%% file: MPQueue.tex
\documentclass[11pt,english,american]{amsart}
\usepackage[T1]{fontenc}
\usepackage[latin9]{inputenc}
\usepackage[letterpaper]{geometry}
\usepackage{array}
\usepackage{amsthm}
\usepackage{amstext}
\usepackage{graphicx}

\makeatletter

\providecommand{\tabularnewline}{\\}

\numberwithin{equation}{section}
\numberwithin{figure}{section}

\usepackage{listings}
\usepackage{color}
\usepackage[all]{xy}
\usepackage{multicol}

\usepackage[labelsep=period,labelfont=sc]{caption}
\usepackage[algo2e,figure,vlined]{algorithm2e}

\numberwithin{table}{section}

\makeatother

\usepackage{babel}

\begin{document}

\numberwithin{lstlisting}{section} 

\title[An MPI Implementation of a Self-Submitting Parallel Job Queue]{An MPI Implementation of a Self-Submitting \\ Parallel Job Queue}

\author{John M. Neuberger}

\author{N\'andor Sieben}

\author{James W. Swift}


\thanks{This research was supported in part by the National Science Foundation
through TeraGrid resources provided by NCSA (TG-DMS090032)}

\date{\today}


\address{Department of Mathematics and Statistics, Northern Arizona University,
Flagstaff, AZ 86011-5717, USA}

\email{john.neuberger@nau.edu, nandor.sieben@nau.edu, jim.swift@nau.edu}
\begin{abstract}
We present a simple and easy to apply methodology for using high-level
self-submitting parallel job queues in an MPI environment. Using C++,
we implemented a library of functions, MPQueue, both for testing our
concepts and for use in real applications. In particular, we have
applied our ideas toward solving computational combinatorics problems
and for finding bifurcation diagrams of solutions of partial differential
equations (PDE). Our method is general and can be applied in many
situations without a lot of programming effort. The key idea is that
workers themselves can easily submit new jobs to the currently running
job queue. Our applications involve complicated data structures, so
we employ serialization to allow data to be effortlessly passed between
nodes. Using our library, one can solve large problems in parallel
without being an expert in MPI. We demonstrate our methodology and
the features of the library with several example programs, and give
some results from our current PDE research. We show that our techniques
are efficient and effective via overhead and scaling experiments.
\end{abstract}
\maketitle

\section{Introduction}

Our methodology is mainly motivated by our work in computational combinatorics
\cite{prooftrees} and partial differential equations (PDE) \cite{NSS2,NSS3,NSS5}.
In combinatorial applications, we have the needs and methods to follow
trees. The bifurcation diagrams we study in PDE are similar in structure
to these trees, and so we can supervise their generation using similar
ideas. Traditional approaches to parallel PDE solvers obtain speed
up by distributing operations on large matrices across nodes. Our
methodology instead uses job queues to solve such problems at a higher
level. Our parallel implementation for creating bifurcation diagrams
for PDE is suggested by the nature of a bifurcation diagram itself.
Such a diagram contains branches and bifurcation points. These objects
spawn each other in a tree hierarchy, so we treat the computation
of these objects as stand-alone jobs. For PDE, this is a new and effective
approach. Processing a single job for one of these applications may
create several new jobs, thus a worker node needs to be able to submit
new jobs to the boss node. These jobs require and produce a lot of
complicated data. For our proof of concept tests and real applications,
we have implemented our ideas in a library of C++ functions, in part
to take advantage of that language's facility with complicated data
structures. In this article, we refer to our library as \emph{MPQueue}.

The Message Passing Interface (MPI) \cite{MPI} is a well-known portable
and efficient communication protocol which has seen extensive use.
In spite of its advantages, MPI is difficult to use because it does
not have a high-level functionality. MPQueue is small, lightweight,
and remedies some of these shortcomings. It uses MPI behind the scenes,
is easy to use, and is Standard Template Library (STL) friendly. MPQueue
has a high-level functionality allowing for the rapid development
of parallel code using our self-submitting job queue technique. In
order to easily pass our data structures between nodes, our library
uses the Boost Serialization Library (BSL) \cite{BoostSerialization}.
It was a design goal of ours to make everything as simple as possible,
so that one does not have to worry about the details of MPI or serialization
when applying our methods. MPQueue is freely available from this paper's
companion website~\cite{companion}.

Section~\ref{sec:Related-work} contains a brief overview of some
popular software packages that facilitate parallel programming. These
fall into roughly three categories, threaded applications for \emph{shared
memory} multi-core processing, message passing for multi-node, \emph{distributed
memory} clusters, and single-system image clusters. Our library uses
message passing and is intended for use on distributed memory clusters.
Sections~\ref{sec:Factoring-example} and~\ref{sec:Matrix-square-example}
explain line-by-line two simple example programs, one for factoring
an integer, and the other for squaring a matrix. The purpose of these
examples is not to provide efficient solutions for these demonstration
applications, but to present our methodology and describe the key
features of MPQueue in full detail. In Section~\ref{sec:Non-attacking-Queens-example}
we give a more serious example, namely that of finding placements
of non-attacking queens on a chessboard. This code is more complicated,
and while not optimal, it does solve the problem on a $20\times20$
board, which would not be possible in a timely manner using serial
code. We also investigate efficiency, scaling, and speedup for this
example. A computationally intensive and mathematically complex example
result for PDE can be found in Section~\ref{sec:Application-to-Nonlinear}.
The full mathematical details of the particular PDE we are interested
in can be found in~\cite{NSS5}, along with our state of the art
numerical results which use MPQueue. 
In Section~\ref{sec:Implementation}
we give an overview of the implementation of MPQueue. This section
also includes an overhead experiment and summarizes the evidence of
our implementation's solid performance. Section~\ref{sec:Conclusions}
contains some concluding remarks, including possible future refinements
to MPQueue.

The companion website~\cite{companion} for this paper contains: 
the MPQueue source code,
the library reference guide (see also Table~\ref{commandsRef}),
all the files needed to compile and execute the three example programs, 
and for the less initiated, line-by-line descriptions of the three example programs.

\section{\label{sec:Related-work}Related work}

In this section we give a short description of some existing systems
offering parallel computing solutions.

\subsection{Shared memory multithreading systems on multicore processors}

Many existing libraries share some features with MPQueue, but unlike
MPQueue, use a shared memory model with multithreading. These systems
are not designed for use with a distributed memory cluster. They could
conceivably be effectively used with some single-system image cluster
software (see Section~\ref{sub:Single-System-Image-(SSI)}).
\begin{itemize}
\item Intel's Cilk++ language \cite{cilkpp} is a linguistic extension of
C++. It is built on the MIT Cilk system \cite{cilk}, which is an
extension of C. In both of these extensions, programs are ordinary
programs with a few additional keywords: cilk\_spawn, cilk\_synk and
silk\_for. A program running on a single processor runs like the original
code without the additional keywords. The Cilk system contains a work-stealing
scheduler.
\item Fastflow \cite{fastflow2} is based on lock-free queues explicitly
designed for programming streaming applications on multi-cores. The
authors report that Fastflow exhibits a substantial speedup against
the state-of-the-art multi-threaded implementation.
\item OpenMP \cite{openmp} is presented as a set of compiler directives
and callable runtime library routines that extend Fortran (and separately,
C and C++) to express shared memory parallelism. It can support pointers
and allocatables, and coarse grain parallelism. It
also includes a callable runtime library with accompanying environment
variables.
\item Intel Threading Building Blocks (TBB) \cite{inteltbb} is a portable
C++ template library for multi-core processors. The library simplifies
the use of lower level threading packages like Pthreads. The library
allocates tasks to cores dynamically using a run-time engine.
\item Pthreads \cite{pthreads} is a POSIX standard for the C programming
language that defines a large number of functions to manage threads.
It uses mutual exclusion (mutex) algorithms to avoid the simultaneous
use of a common resource.
\end{itemize}

\subsection{Message Passing Interface systems}

There are a few existing libraries that are built on top of MPI. These
are C++ interfaces for MPI, without job queues, and mostly without
higher level functionality. It would have been possible to implement
MPQueue on top of some of these packages, in particular, on top of
Boost.MPI.
\begin{itemize}
\item The Boost.MPI library \cite{BoostMPI,modernizing} is a comprehensive
and convenient C++ interface to MPI with some similarity to our library.
Like MPQueue, Boost.MPI uses the BSL to serialize messages.
\item MPI++ \cite{MPIpp} is one of the earliest C++ interfaces to MPI.
The interface is consistent with the C interface.
\item mpi++ \cite{lmpipp} is another C++ interface to MPI. The interface
does not try to be consistent with the C++ interface.
\item Object Oriented MPI (OOMPI) \cite{OOMPI} is a C++ class library for
MPI. It provides MPI functionality though member functions of objects.
\item Para++ is a generic C++ interface for message passing applications.
Like our interface, it is a high level interface built on top of MPI
and is meant to allow for the quick design of parallel applications
without a significant drop in performance. The article \cite{parapp}
describes the package and includes an example application for PDE.
Para++ uses task hierarchies but does not implement job queues.
\end{itemize}

\subsection{\label{sub:Single-System-Image-(SSI)}Single-System Image (SSI) Cluster
software}

A single-system image cluster \cite{SSI,SSIcompare} appears to be
a single system by hiding the distributed nature of resources. Processes
can migrate between nodes for resource balancing purposes. In some
of its implementations, it may be possible to run shared memory applications. 
\begin{itemize}
\item Kerrighed \cite{Kerrighed} is an extension of the Linux operating
system. It allows for applications using OpenMP and Posix multithreading,
presenting an alternative to using MPI on a cluster with distributed
resources. This approach is seemingly simpler than message passing,
but the memory states of the separate nodes must be synchronized during
the execution of a shared memory program. This probably is not as
efficient as a carefully designed message passing solution. 
\item OpenSSI is another SSI clustering system based on Linux. It facilitates
process migration and provides a single process space. OpenSSI can
be used to build a robust, high availability cluster with no single
point of failure. It is a general purpose system that is not specifically
designed for parallel computing. 
\item The Linux Process Migration Infrastructure (LinuxPMI) is a Linux Kernel
extension for single-system image clustering. The project is a continuation
of the abandoned openMosix clustering project which is a fork of MOSIX
\cite{MOSIX}.
\end{itemize}

\subsection{Summary of Alternatives to MPQueue}

As researchers in PDE and combinatorics, we have a need for the computational
power only obtainable through parallel programming. To be practical,
we need a method that uses the distributed memory clusters available
to us. Our key, proven effective programming idea is to use self-submitting
job queues. We did not find an existing package that completely satisfied
all of our requirements, and so wrote our own, MPQueue. Our main design
goal was to make a simple library that is easy to use, allowing effortless,
rapid development of scientific experiments requiring parallel execution,
without sacrificing performance.

Existing shared memory model software for multicore systems have some
similar functionality as MPQueue, but they do not suit all needs because
of the limitations on the number of cores on currently available hardware.
It may be a possibility to use some such systems on top of SSI cluster
software to implement techniques similar to ours over a distributed
memory cluster, but we do not believe that this would be any simpler
than or have as good performance as our simple, more direct message
passing approach.

Other message passing systems are mainly concerned with lower level
tasks. They offer a large variety of ways to send and receive messages
efficiently, but generally do not offer higher level constructs. It
would probably be a reasonable programming alternative to implement
our key idea of the self-submitting job queue on top of some of these
existing systems, but we do not think the result would be as simple
to use or have better performance than our implementation. In the
sequel, we demonstrate with examples how our simple message passing
system handles low level details automatically and offers an easy
to use yet powerful and efficient programming methodology based on
the self-submitting job queue.

\section{\label{sec:Factoring-example}Factoring example}

\begin{figure}
\[
\SelectTips{cm}{}
\xymatrix{
120  \ar@/_.3em/[r] \ar@/_1em/[rr] &
12   \ar@/^.5em/[rr] \ar@/^2em/[rrrrrr] &
10   \ar@/_.5em/[rr] \ar@/_1.8em/[rrrrrr] &
4    \ar@/^1em/[rr] \ar@/^1.5em/[rrr] &
*+[Fo]{5} &
*+[Fo]{2} &
*+[Fo]{2} &
*+[Fo]{3} &
*+[Fo]{2}
\\
}
\]

\caption{\label{fig:factorflow}A possible order of jobs in the job queue during
the factorization of 120. The arrows represent the job submission
process. The circled jobs are the prime number outputs and so they
do not produce new submissions.}

\end{figure}
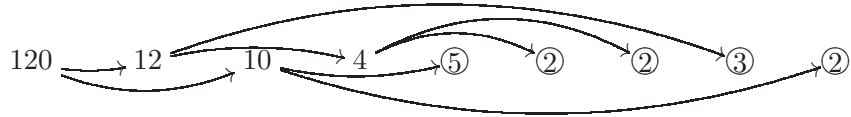

Listing~\ref{lst:factor} shows an example program using the MPQueue
library for factoring an integer. This example demonstrates many of
the features of the library. In particular, it shows the general structure
of a program, including the creation, supervision, and post-processing
of job queues. It also shows the mechanism by which workers themselves
can submit new jobs.
The main idea of the algorithm is to split the input as a product
of two integers that are as large as possible, and then submit these
factors to the job queue for further splitting. These submissions
are done by the workers. The job submission process is visualized
in Figure~\ref{fig:factorflow}. Note that the order of job submissions
is not fully determined.

\lstset{language=C++,basicstyle=\small,identifierstyle=\slshape}
\lstset{morekeywords={string}}
\lstset{emph={MPQswitch,MPQsubmit,MPQstart,MPQinit,MPQtask,MPQsharedata,MPQrunjobs,to_string,from_string,TMPQjobqueue,MPQstop,MPQinfo}, emphstyle=\sffamily}

\begin{lstlisting}[label=lst:factor,float,captionpos=b,caption={Prime factorization example program.},
frame=single,numbers=left,numberstyle=\tiny,lineskip=-1pt,aboveskip=0pt,belowskip=-10pt]
#include "MPQueue.h"
#include "math.h"
using namespace std;

const int SPLIT = 1;		     // only one job type, so no
                                     // cases in MPQswitch 
                                     
void MPQswitch (Tjob & job) {
  int x;
  from_string (x, job.data);           
  int sqt = int (sqrt (double (x)));
  for (int y = sqt; y > 1; y--)      // search for divisors
    if (0 == x % y) {                // found a divisor
      MPQsubmit (Tjob(SPLIT, x/y));  // submit two new jobs
      MPQsubmit (Tjob(SPLIT, y));
      job.data = "";		     // no output to return
      return;
    }
}

int main (int argc, char *argv[]) {
  MPQinit (argc, argv); 
  MPQstart ();                       // only the boss returns
  Tjobqueue inqueue, outqueue;
  int x = 1120581000;		     // factor this number  
  inqueue.push(Tjob(SPLIT,x));       // add one job to the job queue
  MPQrunjobs (inqueue, outqueue);    // supervise queue processing 
  cout << x << " factors as:\n";
  while (!outqueue.empty ()) {       // get the results
    int factor;                      // one factor 
    from_string(factor, outqueue.front().data);  
    cout << factor << " ";
    outqueue.pop();
  }
  MPQstop ();
}
\end{lstlisting}

The companion website~\cite{companion} contains not only the source code for this and the other examples, 
but a more detailed line-by-line explanation of the code.
See also Table~\ref{commandsRef} for a complete list of the commands found in the MPQueue library, 
together with a brief description of their function.  
Here, we include an overview of this example program, in particular the lines of code which 
implement key MPQueue features.
The MPQueue header file included in all our examples contains our new definitions as well
as all header files needed to use MPI and the required parts of the BSL \cite{BoostSerialization}.
In each example, we define a finite list of job types as positive integers, 
although in this first simple example there is only one type of job.

In each of our example programs, 
the main function (found here on line 21) is executed by every node.
First, the MPI is initialized and nodes are split into one boss and several workers. 
Only the boss returns from the \textsf{MPQstart} function call (line 23). 
The workers are then ready to accept jobs.  
The boss creates two job queues, one for storing jobs to do, and another for storing results.
This program computes the prime factorization of the example integer found on line 25.  
The boss places the first splitting job into the job queue and starts the supervision of the workers
via a call to the \textsf{MPQrunjobs} function (line 27), where it will spend the vast majority of its running time.

In this example, the workers split numbers into factors and submit these factors 
(see lines 14--15) to \textsl{inqueue} for further splitting. 
The workers return the prime factors, which the boss collects in the \textsl{outqueue}. 
When there are no more jobs, the boss enters a loop (line 29) and retrieves all the prime factors from the output
queue.

The user-coded \textsf{MPQSwitch} function is called every time a new job needs to be processed.  
There typically are more than the one job type found in this first most simple example.
Here, the workers (and only the workers) spend all their productive time in \textsf{MPQSwitch}, 
recursively finding pairs of factors of an input integer and submitting two new factor jobs for each pair of factors found.
The \textsl{job.data} variable initially contains the serialized input of the job at the time the function is called.  
This variable is replaced by the serialized output of the job. 
It is a key feature for the ease of doing high-level programming with our library that the input and output data can contain arbitrarily complicated data structures.
In this simple example, the output is the same as the input if no factors were found in the loop 
initiated on line 12, whence the input is a prime, otherwise the output is empty.

\section{\label{sec:Matrix-square-example}Matrix square example}

\begin{lstlisting}[label=lst:square,float,captionpos=b,caption={Matrix square example program.},
frame=single,numbers=left,
numberstyle=\tiny,lineskip=-1pt,aboveskip=0pt,belowskip=-10pt]
#include "MPQueue.h"
using namespace std;

enum { NONE, DATA, MULTIPLY, RESULT };
typedef vector < int > Trow;
vector < Trow > matrix, square;
typedef struct {
  int pos;                                     // row index
  Trow result;                                 // output row
} Tdata;

template < class Archive > void		       // needed to serialize Tdata                                                                                                                                                                                 
serialize (Archive & ar, Tdata & data, const unsigned int version) {
  ar & data.pos;
  ar & data.result;
}

void MPQswitch (Tjob & job) {
  Tdata data;
  switch (job.type) {
  case DATA:			               // receive the input matrix
    from_string (matrix, job.data);
    break;
  case MULTIPLY:
    from_string (data.pos, job.data);	       // get the row position
    data.result = Trow (matrix.size (), 0);
    for (int j = 0; j < matrix.size (); j++)   // calutate one row
      for (int k = 0; k < matrix.size (); k++)
	data.result[j] += matrix[data.pos][k] * matrix[k][j];
    job = Tjob(RESULT, data);                  // prepare the result
    MPQtask (job);               	       // send result to boss
    break;
  case RESULT:			               // receive one output row
    from_string (data, job.data);
    square[data.pos] = data.result;
    job.data = "";                             // nothing to return	
  }
}

int main (int argc, char *argv[]) {
  MPQinit (argc, argv);
  MPQstart ();
  Trow row (10, 1);	                  // input matrix containing
  vector < Trow > mat (row.size (), row); // 1 at every entry
  square.resize (row.size ());	          // container for the output
  MPQsharedata (Tjob(DATA, mat));	  // input matrix sent to workers
  Tjobqueue inqueue, outqueue;
  for (int i = 0; i < mat.size (); i++)	  // every row is a separate job
    inqueue.push (Tjob(MULTIPLY, i));
  MPQrunjobs (inqueue, outqueue);	  // run the jobs
  for (int i = 0; i < row.size (); i++) { // print the output matrix
    for (int j = 0; j < row.size (); j++)
      cout << square[i][j] << " ";
    cout << "\n";
  }
  MPQstop ();
}
\end{lstlisting}

Listing~\ref{lst:square} shows an example program for calculating the square of a matrix. 
The example demonstrates several new features of the MPQueue library. 
Namely, it shows how to efficiently share a large amount of data with all the workers using \textsf{MPQsharedata},
how to serialize struct data types using a template function, 
and how to use \textsf{MPQtask} to return results to the boss for immediate processing rather than using the output job queue. 
Unlike the first example, this program uses three job types so that a switch statement is required 
in the \textsf{MPQswitch} function.
Here, we want the job types to be positive integers, so \textsl{NONE} is added to take the unused value of zero.

The variable \textsl{matrix} contains the input matrix (initialized in lines~43--44), 
while the variable \textsl{square} contains the result of the program, the square of the input matrix.
The boss uses our function \textsf{MPQsharedata} on line~46 to painlessly share an STL structure with the nodes.
In lines~48--49 the job queue is filled with \textsl{MULTIPLY} jobs,
each requesting the calculation of one row of the goal matrix.
The boss then starts the supervision of the workers, again spending the bulk of its time in \textsf{MPQrunjobs}.

As in the previous example, the workers execute \textsf{MPQswitch} and spend all their productive time in that function.
Their work, however, this time depends on the job type.  
Also different from the factor example, the \textsf{MPQswitch} function can be executed by the boss, in this case with \textsl{job.type} set to \textsl{RESULT}, as the result of a \textsf{MPQtask} call by a worker.
The variable \textsl{data} contains one row of the result produced by a worker.
When the input matrix is shared by all the workers, 
each worker receives the matrix as a \textsl{DATA} job. 
They deserialize the data and store it locally in the variable \textsl{matrix}.
Lines~24--29 are where a worker calculates one row of the goal matrix.
The input (position within the matrix) is deserialized into \textsl{data.pos}, and the calculated row is stored in \textsl{data.result}.
Workers circumvent the output queue by using \textsf{MPQtask} (line~31) to send their calculated row results to the boss.
Here, the boss receives and deserializes a row and puts it into the appropriate row of the result matrix \textsl{square}.
Thus, in lines~33--36, the boss does some actual work, not just supervising workers.
The boss is usually idle during most of the supervision process.
Thus it can be more efficient for the boss to do tasks such as post-processing during supervision, 
rather than waiting until the workers are all done.

\section{\label{sec:Non-attacking-Queens-example}Non-attacking Queens example}

We now present an application that illustrates how to avoid the \emph{too
many jobs} obstacle to scalability. The example uses local job queues
to avoid excessive communication costs. This technique enhances the
flexibility of our methodology, allowing for the efficient use of
the library in somewhat unexpected situations. 

\begin{figure}
\includegraphics[scale=0.21]{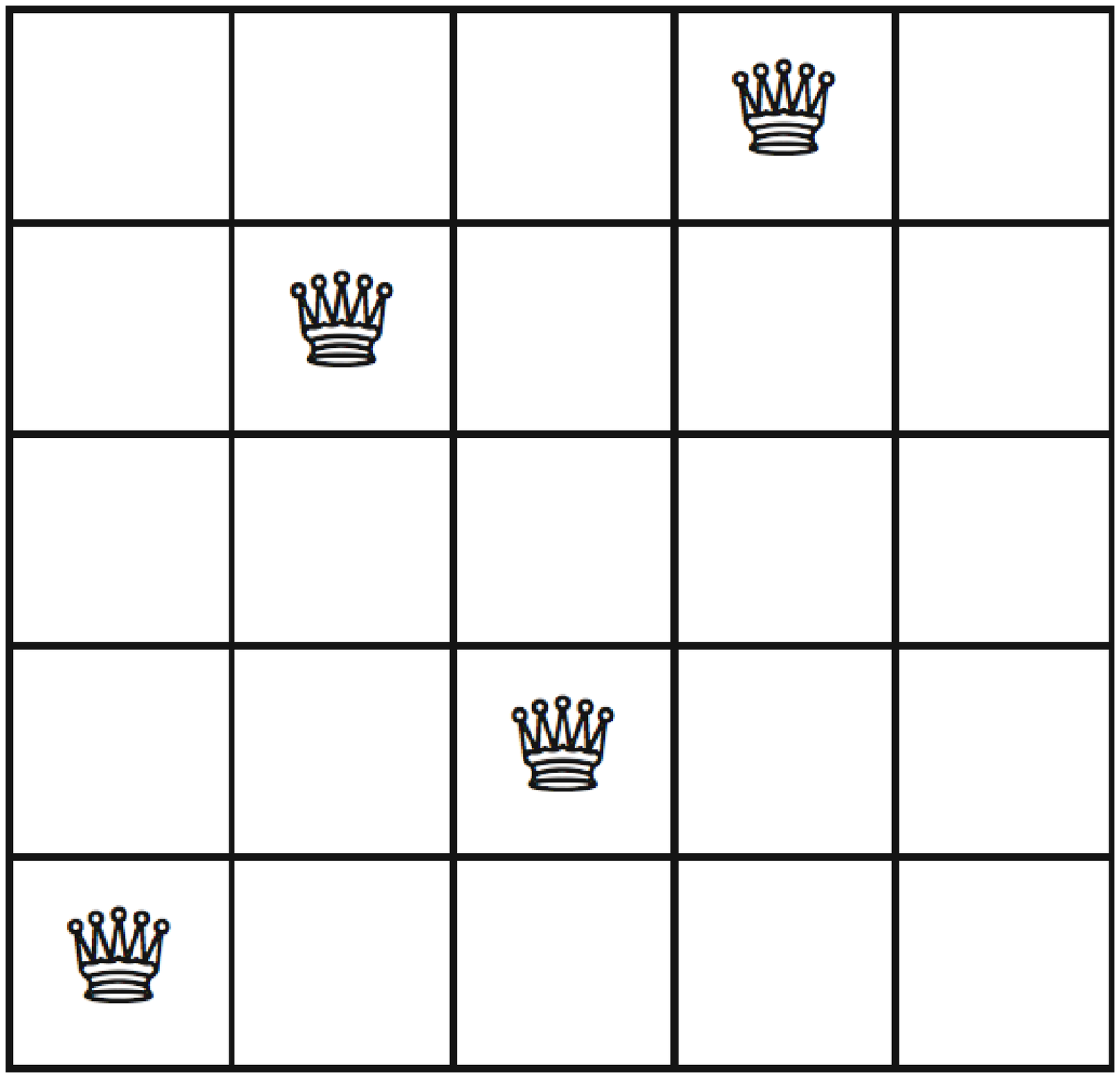}

\caption{\label{fig:partial-placement}A partial placement of queens on a $5\times5$
board. In the example code, this placement is encoded as the vector
$(0,3,1,4)$. This partial placement can be extended to the full placement
$(0,3,1,4,2)$. There are 10 distinct placements on the $5\times5$
board.}

\end{figure}

\begin{figure}
\input{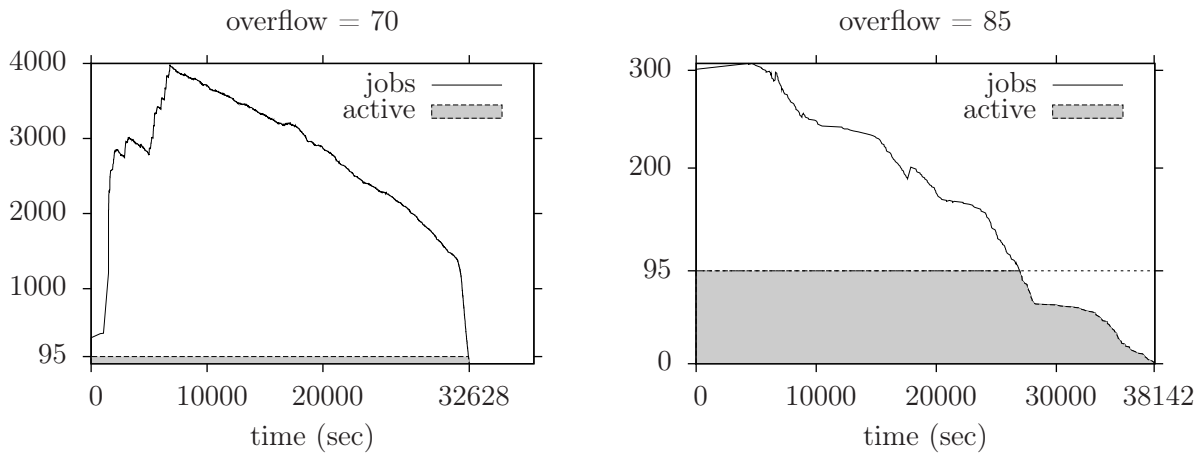}\input{load85.tex}

\caption{\label{fig:Load-diagrams}Load diagrams of the 20-queen placement
example code for two different values of the \textsl{overflow} variable.
In both cases there were 95 available workers. The shaded region shows
the number of workers actively working on jobs. The solid curve shows
the number of jobs available in the global job queue together with
the number of active workers. The unshaded region below 95 represents
idle workers. The pictures show that the choice of 85 for overflow
is too large, since with this choice there are not enough available
jobs and so the program takes longer to finish. The optimal value
of 70 was determined experimentally. From Figures~\ref{fig:speedup-1}
and~\ref{fig:speedup} it appears that the optimal overflow value
is somewhat predictable and depends only on the size of the problem,
not the size of the cluster.}

\end{figure}

The $n$-queen puzzle is a well-known problem in mathematics. It concerns
the placement of $n$ non-attacking queens on an $n\times n$ chess
board. Figure~\ref{fig:partial-placement} shows a valid partial
placement with one missing queen. For a survey of results on the generalizations
of this problem see \cite{queens}. Listing~\ref{lst:queens} shows
an example program which counts the number of solutions on an $n\times n$
board. Figure~\ref{queensPseudo} shows the serial pseudo code. We
were able to run this code for $n\le20$ on a cluster containing 24
Intel(R) Xeon(TM) 2.80GHz dual CPUs with hyper-threading. For $n=20$,
it took $9.0$ hours (see Figure~\ref{fig:Load-diagrams}) using
the $96=24\times2\times2$ cores to find the $39,029,188,884$ solutions.
Note that the state of the art is $n=26,$ that is, the number of
solutions for $n=27$ is not known.

\begin{figure}
\input{speedup15.tex}

\caption{\label{fig:speedup}Speedup as a function of nodes for the 15-queen
placement problem. The bold numbers indicate the overflow values used
for a given curve. It appears that the optimal overflow value does
not depend on the number of nodes, only on the size of the problem.
The dashed line again corresponds to the theoretical maximum speedup.
The number of solutions for this board size is 2,279,184.}

\end{figure}
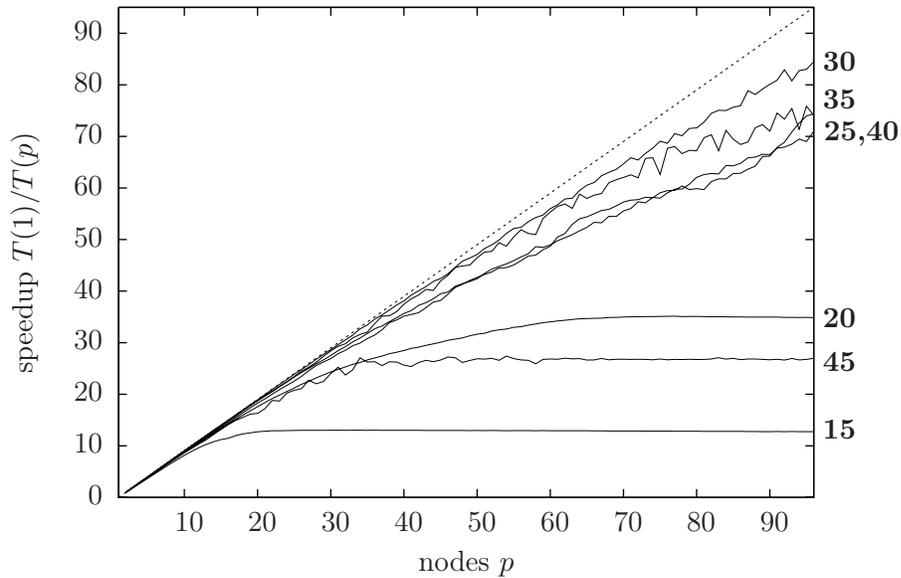

Our code uses a parallel version of a standard backtracking algorithm.
A worker keeps a local job queue containing possible search branches,
i.e., valid partial placements. If this queue grows large enough,
then the worker submits one of the partial placements to the boss
node as a job so that another worker can process it. Submitting jobs
to the boss node too early may result in a very large global job queue
and a lot of unnecessary communication between the nodes. Submitting
jobs too late can starve the workers for jobs. The decision depends
on the number of nodes, the speed of the nodes, and the branching
factor of the search tree. Our code uses a simple heuristic that depends
on an overflow value that we have adjusted experimentally. Figures~\ref{fig:Load-diagrams},
\ref{fig:speedup-1}, and~\ref{fig:speedup} show the effect of different
choices of the value. This very simple submission process could be
fine tuned using the \textsf{MPQinfo} command (see Table~\ref{commandsRef}),
the only MPQueue command not demonstrated in any of our three examples.
Many of the task distribution schemes discussed in the survey paper
\cite{parallelsurvey} could be implemented using the \textsf{MPQinfo}
command.

\input{queensPseudo.tex}

\begin{table}
\begin{centering}
\begin{tabular}{|c|c|c|c|c|}
\hline 
Nodes $p$ & Run time $T(p)$  & Worker usage & Total usage & Efficiency $\frac{T(1)}{pT(p)}$\tabularnewline
\hline
\hline 
1 & 94.08 sec &  & 94.08 CPU$\,$sec & \tabularnewline
\hline 
3 & 47.05 sec & 94.10 CPU$\,$sec & 141.15 CPU$\,$sec & 0.67\tabularnewline
\hline 
6 & 18.77 sec & 93.85 CPU$\,$sec & 112.62 CPU$\,$sec & 0.84\tabularnewline
\hline 
24 & 4.11 sec & 94.53 CPU$\,$sec & 98.64 CPU$\,$sec & 0.95\tabularnewline
\hline 
48  & 2.05 sec & 96.35 CPU$\,$sec & 98.40 CPU$\,$sec & 0.96\tabularnewline
\hline 
72 & 1.44 sec & 102.24 CPU$\,$sec & 103.68 CPU$\,$sec & 0.91\tabularnewline
\hline 
96 & 1.11 sec & 105.45 CPU$\,$sec & 106.56 CPU$\,$sec & 0.88\tabularnewline
\hline
\end{tabular}
\par\end{centering}

\medskip

\caption{\label{tab:scaling}Processing times for the 15-queen placement problem
using the overflow value 30. The \emph{Worker usage }column is the
number of workers multiplied by the run time, while \emph{Total usage}
includes the boss as well. A constant worker usage would indicate
perfect scaling. For larger board sizes, the efficiency increases
up to and beyond our maximum available 96 nodes.}

\end{table}

\begin{lstlisting}[label=lst:queens,float,captionpos=b,caption={Non-attacking queens example program.},
frame=single,numbers=left,numberstyle=\tiny,lineskip=-1pt,aboveskip=0pt,belowskip=-10pt]
#include "MPQueue.h"
#include <deque>
using namespace std;

enum { NONE, PLACE, RESULT };
typedef vector < int > Trow;
unsigned long int allsolutions = 0;             // total number solutions
int overflow = 70;                              // controlls local queue size
const int size = 20;                            // size of the board

bool inline fits (const Trow & row) {
  int j = row.size () - 1;
  for (int i = 0; i < j; i++)
    if ( ( row[i] == row.back () ) || ( abs (row[i] - row[j]) == j - i ) )
      return false;                             // queens interfere
  return true;                                  // last queen fits
}

void MPQswitch (Tjob & job) {
  Trow row;                                     // a partial placement
  deque < Trow > rows;                          // local job queue
  unsigned int solutions = 0;                   // local number of solutions
  switch (job.type) {
  case PLACE:                                   // add queen in next column
    from_string (row, job.data);
    rows.push_back (row);                 	// populate a local job queue 
    while (!rows.empty ()) {              	// still local jobs to do
      row = rows.back ();                     
      rows.pop_back ();
      for (row.push_back (0); row.back () < size; row.back ()++)
	if (fits (row))                         // does the new queen fit? 
	  if (row.size () == size)              // all queens added
	    solutions++;                        // found a new solution
	  else {
	    rows.push_back (row);               // add to local job queue
	    if (rows.size () > overflow) {      // if too many local jobs 
	      MPQsubmit (Tjob(PLACE, rows.front ())); // send one to the boss
	      rows.pop_front ();
	    }
	  }
    }
    job = Tjob(RESULT, solutions);              // prepare the result
    MPQtask (job);          			// send result to the boss
    break;
  case RESULT:
    from_string (solutions, job.data);          // update total solutions
    allsolutions += solutions;                  // with new result
    job.data = "";                              // no result to return 
  }
}

int main (int argc, char *argv[]) {
  MPQinit (argc, argv);
  MPQstart ();
  Tjobqueue inqueue, outqueue;
  inqueue.push (Tjob(PLACE, Trow ()));         // initial empty placement
  MPQrunjobs (inqueue, outqueue);
  cout << allsolutions << " solutions\n";
  MPQstop ();
}
\end{lstlisting} 

In the sequel of this section we provide a description of the more important aspects of the algorithm as they are specifically implemented in the lines of code found in Listing~\ref{lst:queens}.
The \textsl{size} of the board is set to some positive integer, 20 in this example (line 9).
A partial placement of queens is a vector of type \textsl{Trow} of length between $0$ and  \textsl{size}, 
whose integer value $j$ in the $i$-th entry denotes a valid (non-conflicting) 
placement of a queen in the $j$-th row of the $i$-th column of the board.
When the length of a partial placement \textsl{row} equals \textsl{size}, 
it is in fact a full placement and a solution has been found.
After the usual initialization where the boss and worker nodes are created, 
the boss puts an empty placement job on the queue and begins supervision 
by entering \textsf{MPQrunjobs} (line 57), 
where it will stay until the counting is done and its global variable 
\textsl{allsolutions} is output with the final result. 
In the algorithm's counting of numbers of solutions at the ends of branches of partial placements, 
workers return partial sums counting some valid placements to the boss by
submitting a \textsl{RESULTS} task to the boss via a call to \textsf{MPQtask} (line 43).
This causes the boss to update its global variable of the master count (line 47).

In order to control communication costs (and in fact fine-tune performance), 
workers use local job queues named \textsl{rows} (line 27) containing 
up to \textsl{overflow} (line 8) number of partial placement row vectors.
Until the queue is empty, workers pop off local partial placement jobs, 
repeatedly calling the function \textsl{fits} (lines 11 and 31) with a row vector containing 
an attempt to extend to a valid placement to the next column.
A valid extension causes the worker to either 
increment the number of solutions found (line 33) if the edge of the board was reached, 
or otherwise add it as a partial placement job to the local queue for another round of extending by itself (line 35).
Before the next local job is started, the worker checks if the local queue has reached \textsl{overflow},
and if so pops a partial placement job off of the local queue and sends it to the 
boss to be placed on the global queue (lines 36--38).
When a worker's local queue is finally empty, 
the worker returns the partial result via a call to \textsf{MPQtask} and then terminates (lines 43--44).
The worker is of course then available for the boss to give it another partial placement job off of the global queue,
until that queue is itself empty and the final count has been compiled.
\begin{figure}
\input{speedup12.tex}

\caption{\label{fig:speedup-1}Speedup as a function of nodes for the 12-queen
placement problem. The bold numbers indicate the overflow values used
for a given curve. The dashed line is the theoretical maximum speedup
of $p-1$, the number of workers. It appears that adding more than
55 nodes does not increase the speedup. This limitation is the result
of the small board size. The number of solutions is only 14,200.}

\end{figure}
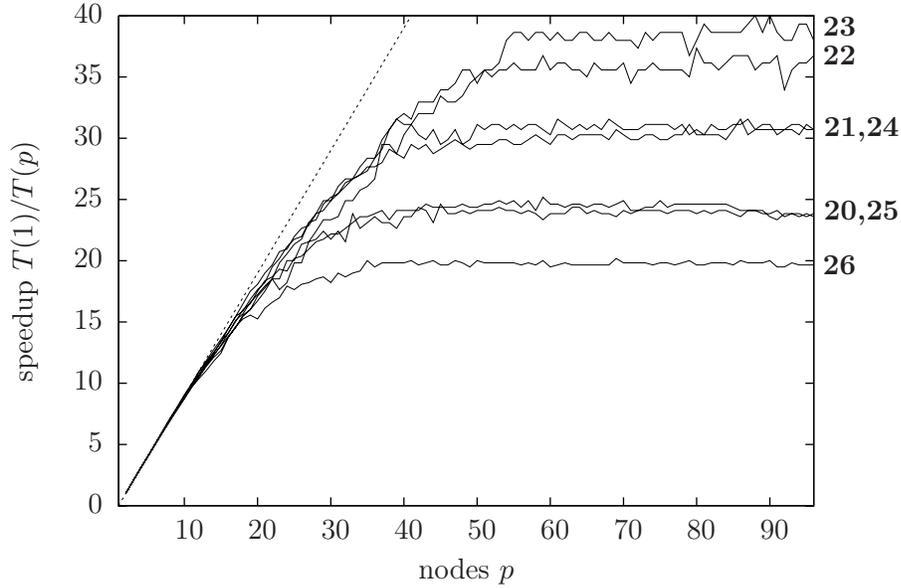

Table~\ref{tab:scaling} and Figures~\ref{fig:speedup-1}~and~\ref{fig:speedup}
show that our approach to the problem scales well. For the size 15
board, we could have used more than the available 96 nodes; to attempt
the unknown $n=27$ case we could make good use of a very large number
of nodes. 

We did not take advantage of the symmetry of the problem, which would
have immediately resulted in a four-fold speedup. This could be done
with minimal work, but our goal here is to present a meaningful yet
simple example demonstrating the use of our library and methodology.
A possible future extension to the MPQueue library that could help
reduce the number of idle workers for problems like this one would
be the implementation of priority job queues.

As a reminder, the companion website~\cite{companion} contains 
the source codes and corresponding detailed line-by-line explanations 
for this and the previous two examples, 
along with the MPQueue library and the few other files needed to compile and execute the programs.
Table~\ref{commandsRef} contains a complete list of MPQueue commands, 
along with a brief description of their function.

\section{\label{sec:Application-to-Nonlinear}Application to Nonlinear Elliptic
PDE}

Our library has provided us with an easy way to port to a parallel
environment our serial code for solving PDE. The resulting increase
in computational power has enabled us to solve the computationally
intensive problem\begin{align}
\Delta u+f_{s}(u) & =0\qquad\text{in }\Omega\label{eq:pde}\\
u & =0\qquad\text{on }\partial\Omega,\nonumber \end{align}
where $\Delta$ is the Laplacian, $\Omega$ is the square or cube,
and $f_{s}$ is the family of nonlinearities $f_{s}(u)=su+u^{3}$,
parameterized by $s\in{\bf R}$. Other regions and nonlinearities
can be handled as well. We had previously developed serial C++ code
for obtaining good approximations of solutions to parameterized PDE
such as~(\ref{eq:pde}), provided that the dimensions involved were
not too big. The reader can refer to \cite{NSS2,NSS3} (serial) and
\cite{NSS5} (parallel) for the mathematical details of our PDE algorithms,
which are based on Newton's method operating in a coefficient space
corresponding to eigenfunction expansions of solution approximations.
Generally, our C++ code follows branches of solutions by starting
with an initial point on a branch and an approximate tangent vector
to the branch, applying Newton's method to find a next point and corresponding
tangent, and repeating until a window is exited. Each time a new bifurcation
point is identified on a branch, our code seeks points on new, bifurcating
branches, which can then also be followed. Processing a bifurcation
point can be very quick but it can also take more time than following
a branch. Thus, it is natural that following a branch and analyzing
a bifurcation point are the two job types we use.

The required input for both job types is complicated. Each input is
a structure with 18 different fields, containing such data as the
discretized solution approximation at $N$ grid points, the corresponding
$M$ eigenfunction expansion coefficients, the parameter $s$, the
current tangent vector, the branch history, and C++ parameter values
that lead to the current state (speed, tolerances, maximum iteration
counts, etc.). For large problems, the embedded vectors may be very
large. The automatic serialization feature of our library makes it
easy to pass such data between nodes.

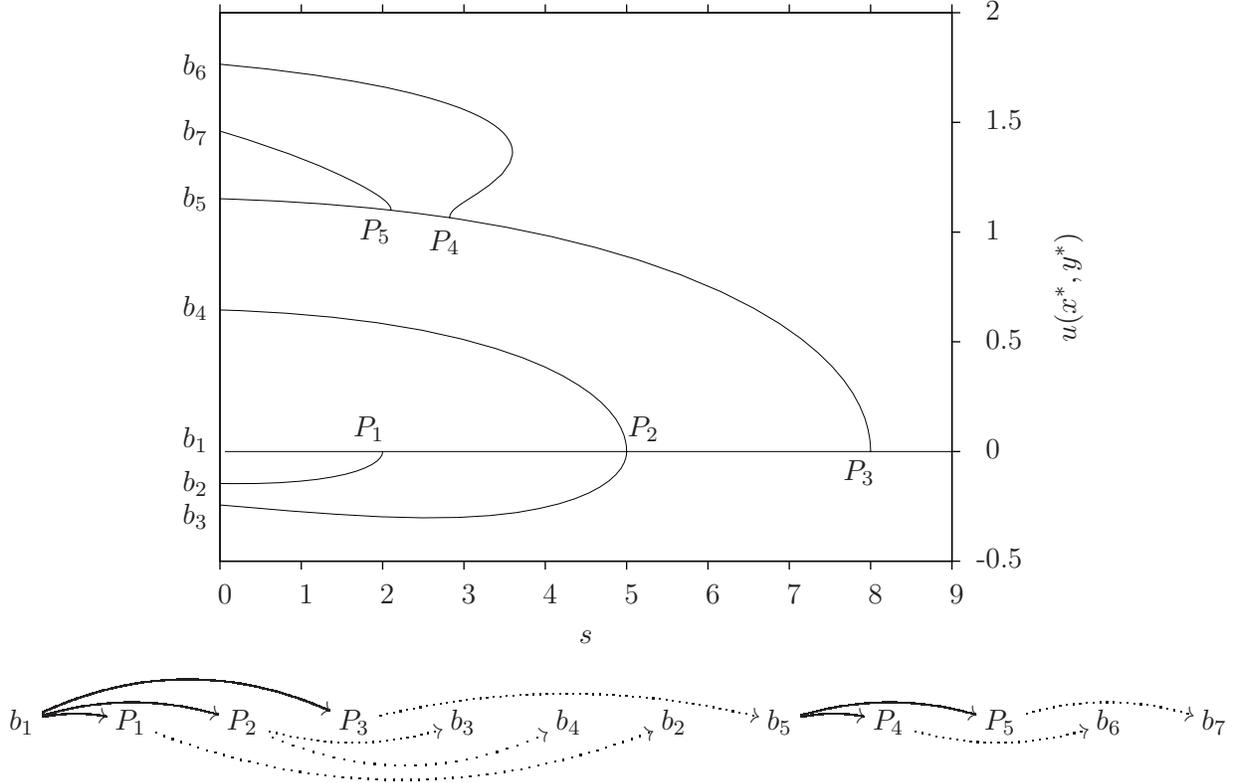
\begin{figure}
\input{nss5_pde_bif.tex}

\[
\SelectTips{cm}{}
\xymatrix{
b_1  \ar@/^.3em/[r]  \ar@/^.7em/[rr] \ar@/^1.5em/[rrr] &
P_1  \ar@{.>}@/_2em/[rrrrr] &
P_2   \ar@{.>}@/_.7em/[rr] \ar@{.>}@/_1.5em/[rrr] &
P_3  \ar@{.>}@/^1em/[rrrr] &
b_3 &
b_4 &
b_2 &
b_5 \ar@/^.3em/[r]  \ar@/^.7em/[rr] &
P_4 \ar@{.>}@/_.7em/[rr]&
P_5 \ar@{.>}@/^.7em/[rr]&
b_6 &
b_7
\\
}
\]

\caption{\label{fig:bif1}The first four primary branches and all the branches
connected to them for PDE~(\ref{eq:pde}) on a square region. For
this example, a total of 12 jobs are placed on the job queue. The
lower diagram represents a possible creation order of these jobs in
the job queue during the generation of the bifurcation diagram presented
in the upper diagram. A solid arrow represents the submission of a
bifurcation analysis job, while a dotted arrow represents a branch
following job. The trivial branch $u=0$ lying on the horizontal axis
is the first branch to be followed. This job, labeled $b_{1}$, is
the first and only job to be placed in the job queue by the boss.
One of the workers follows this trivial branch and encounters the
3 primary bifurcation points and submits 3 bifurcation jobs $P_{1},P_{2,}$
and $P_{3}$. These three bifurcation points are each processed by
a different worker, which leads to the submission of 4 branch jobs.
This process continues until the job queue is empty. The workers do
not send results to the boss using the outqueue; they store all their
results in separate files which are post processed by other scripts
after the termination of the program.}

\end{figure}

Figure~\ref{fig:bif1} shows a partial bifurcation diagram when $\Omega$
is the square $(0,\pi)^{2}$. The horizontal axis is the parameter
$s$ and the vertical axis is the value of the solution at a generic
point $(x^{*},y^{*})$ \cite{NSS3}. The diagram demonstrates how
branch following creates a tree structure whose growth is unpredictable.
Details of the size and speed for this simple example are shown in
the first row of Table~\ref{tab:Processing-times-for}. Using two
nodes is essentially a simulation of our serial code, since there
is only one worker. 

The second row of Table~\ref{tab:Processing-times-for} shows the
same summary for a problem where $\Omega$ is the cube $(0,\pi)^{3}$
and we search for the branches connected to the bifurcation point
on the trivial branch at $s=18$. Parallelization is essential since
it takes about a minute to compute each of the thousands of solutions. 

We conclude the section with a few details of our implementation for
creating bifurcation diagrams. Solution approximations lie in a subspace
spanned by the first $M$ eigenfunctions of the Laplacian, which themselves
have been previously and independently approximated by $M$ vectors
in $R^{N}$, obtained via calls to ARPACK if not known in closed form.
For the example results included in this section, these bases are
well known in terms of sine functions. The construction of the Jacobian
of the object function for Newton's method requires order $M^{2}$
numerical integrations, each requiring order $N$ arithmetic operations.
The number of integrations is reduced somewhat in the presence of
symmetry. The search direction system is solved via a standard LAPACK
subroutine. Each approximated point requires roughly 4 iterations
of Newton's method. Finding bifurcation points requires the computation
of the eigenvalues of the Jacobian, computed by another LAPACK routine. 

\begin{table}
\begin{centering}
\begin{tabular}{|c|c|c|c|c|c|c|c|c|}
\hline 
$\Omega$ & $M$ & $N$ & Branches & Bifurcations & Solutions & Nodes & Run time & Worker usage\tabularnewline
\hline
\hline 
\selectlanguage{english}%
$(0,\pi)^{2}$\selectlanguage{american}
 & $323$ & $41^{2}$ & $7$ & $5$ & $165$ & \begin{tabular}{c}
2\tabularnewline
4\tabularnewline
\end{tabular} & \begin{tabular}{c}
8.17 min\tabularnewline
3.73 min\tabularnewline
\end{tabular} & \begin{tabular}{c}
8.17 CPU$\,$min\tabularnewline
11.19 CPU$\,$min\tabularnewline
\end{tabular}\tabularnewline
\hline 
\selectlanguage{english}%
$(0,\pi)^{3}$\selectlanguage{american}
 & $564$ & $21^{3}$ & $385$ & $364$ & $19724$ & \begin{tabular}{c}
26\tabularnewline
50\tabularnewline
76\tabularnewline
\end{tabular}  & \begin{tabular}{c}
22.4 hour\tabularnewline
11.5 hour\tabularnewline
8.98 hour\tabularnewline
\end{tabular} & \begin{tabular}{c}
560.0 CPU$\,$hour\tabularnewline
563.5 CPU$\,$hour\tabularnewline
673.5 CPU$\,$hour\tabularnewline
\end{tabular}\tabularnewline
\hline
\end{tabular}
\par\end{centering}

\medskip

\caption{\label{tab:Processing-times-for}Processing times for PDE examples.
These results are using NCSA's IA-64 TeraGrid Linux Cluster (Mercury)
with 1.3 GHz nodes \cite{Catlett}. The data for the square region
corresponds to the diagrams found in Figure~\ref{fig:bif1}. The
entries for the cube region are typical of the results found in~\cite{NSS5}.
Note that the cube problem scales well since there are many jobs.}

\end{table}

\section{\label{sec:Implementation}Implementation}

\subsection{Design}

Passing complicated data structures between nodes requires a significant
amount of work in MPI. To avoid this difficulty, we rely on the BSL.
This library encodes every standard STL data structure automatically
as a string, and more complicated data types can be encoded with a
minimal amount of work. The serialization is done mostly automatically
using template functions. A serialized data string is sent between
the nodes in two steps. In the first step, a preliminary message is
sent containing the size of the string together with some additional
information like the job type. In the second step, the string itself
is sent. Most of this is done using point-to-point communication with
\textsf{MPI\_Send} and \textsf{MPI\_Recv.} The exception is the \textsf{MPQsharedata}
command, which uses point-to-point communication for the preliminary
message and uses \textsf{MPI\_Bcast} for the data string.

\input{workers.tex}\input{runjobs.tex}

As a result of calling \textsf{MPQstart}, all the nodes become workers
except node zero, which becomes the boss. The workers go into a waiting
loop, as shown in Figure~\ref{alg2}, and the boss continues its
own work. When the boss requires help, it builds a queue of jobs and
becomes a supervisor using \textsf{MPQrunjobs,} as shown in Figure~\ref{alg1}.
The boss then assigns jobs in the job queue to available workers and
removes these jobs from the job queue. The workers execute \textsf{MPQswitch}
with the job. 

\begin{figure}
\includegraphics[bb=0bp 2.7cm 612bp 710bp,scale=0.8]{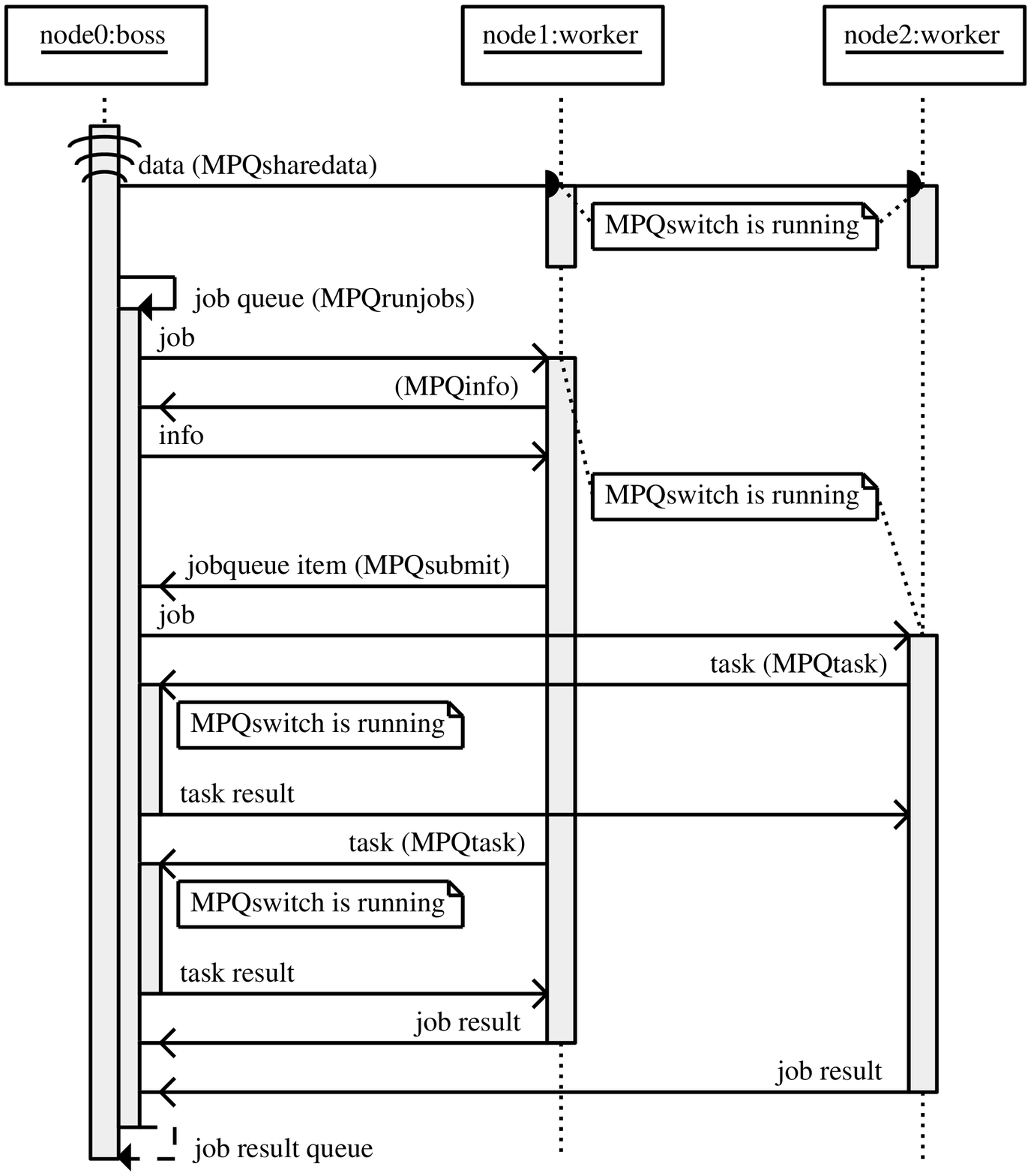}

\caption{\label{fig:Sequence-diagram.}An example sequence diagram for the
communication between the boss and worker nodes. The boss sends out
some global data to the workers using \textsf{MPQsharedata}. The boss
creates a job queue with a single job and then becomes the supervisor
by calling \textsf{MPQrunjobs}. The first worker takes this job and
realizes that an extra job should be created. It uses \textsf{MPQinfo}
to check if there are available workers to take this additional job.
The second worker is not busy, so the first worker submits the new
job. The new job is assigned to the second worker. Both workers need
the value of a global variable so they use \textsf{MPQtask} to get
it from the boss. When the workers are finished, they send the results
to the boss and the supervision phase stops. }

\end{figure}

When a worker finishes a job, it sends a message to the boss, letting
the boss know that the result is available. The boss accepts the result
from the worker and stores it in the output queue. The boss keeps
track of the number of jobs sent out. If the job queue is empty or
every worker is working already, then it only accepts messages and
does not try to assign jobs. The supervision phase ends when the job
queue is empty and the result of every assigned job is received.

During the supervision phase, the workers can create new jobs by sending
a job to the boss using \textsf{MPQsubmit}. The boss puts these jobs
into the currently executing job queue. 

The workers can use \textsf{MPQtask} to ask the boss to execute a
task immediately and return the result to the same worker. To satisfy
this request, the boss executes \textsf{MPQswitch}. This feature can
be used, for example, to set or get the value of a global variable
like a counter. This allows for rudimentary communication between
the workers. The feature can also be used for returning results to
the boss. This allows the boss to post-process the results while the
workers are busy with their jobs. A sequence diagram of the supervision
phase is shown in Figure~\ref{fig:Sequence-diagram.}. At the end
of the program, the boss uses \textsf{MPQstop} to tell the workers
to quit. 

See Table~\ref{commandsRef} for a complete list of the commands found in the MPQueue library, 
together with a brief description of their function.  
The interested reader can consult the companion website~\cite{companion} to find a detailed reference guide.

\begin{table}
\input MPQueueCommandsTable.tex
\caption{
\label{commandsRef}
A complete list of the commands found in the MPQueue library.}
\end{table}

\subsection{Scaling and overhead}

In the non-attacking queens example we presented evidence suggesting
that our methodology scales well. In particular, we show with load
diagrams that all workers can be kept effectively busy if the problem
has very many small jobs that can be managed via a local job queue
to avoid excessive communication. We show that the efficiency of our
parallel solution to this typical application is very good, up to
a certain number of nodes, and that this number of nodes is very large
if the problem is sufficiently big. We show that the speedup for this
example can be close to ideal when using the optimal overflow, that
this optimal overflow value appears to be independent of the number
of nodes and only dependent on the size of the problem, and that for
a sufficiently large problem, an arbitrary number of nodes can be
effectively used. For our PDE application where there are a lesser
number of much larger jobs and local job queues are not needed, we
have observed similarly good performance indicators (see Table~\ref{tab:Processing-times-for}
and~\cite{NSS5}). 

In a final test, we demonstrate that the overhead of our implementation
is reasonable. To observe the cost of bookkeeping, data serialization,
and communication, we ran two experiments where each job was an instruction
to the workers to sleep for one second. In the first experiment, we
sent no additional input or output data, only the type of the job.
In the second experiment, each job was sent with a vector containing
1000 doubles as input, and then the same vector was returned as output.
The addition of data causes some unavoidable delay on our distributed
memory system due to the serialization time and inherent communication
speed of the network. The results are summarized in Table~\ref{tab:overhead}.
It shows that the correspondence between the overhead and the number
of jobs is linear in both experiments.

\begin{table}
\begin{tabular}{|c|c|c|c|c|}
\cline{2-5} 
\multicolumn{1}{c|}{} & \multicolumn{2}{c|}{without data} & \multicolumn{2}{c|}{with data}\tabularnewline
\hline 
Jobs $n$ & Run time $T(p)$ & Overhead $T(p)-n/95$ & Run time $T(p)$ & Overhead $T(p)-n/95$\tabularnewline
\hline
\hline 
760 & 8.02 & 0.02 & 8.73 & 0.73\tabularnewline
\hline 
3040 & 32.07 & 0.07 & 24.88 & 2.88\tabularnewline
\hline 
12160 & 128.19 & 0.19 & 139.50 & 11.50\tabularnewline
\hline 
48640 & 512.76 & 0.76 & 557.90 & 45.90\tabularnewline
\hline 
194560 & 2051.09 & 3.09 & 2231.28 & 183.28\tabularnewline
\hline 
778240 & 8204.51 & 12.51 & 8964.35 & 772.35\tabularnewline
\hline
\end{tabular}\caption{\label{tab:overhead}The time in seconds to execute sleep(1) jobs
on 96 nodes. We used the same cluster specified in Section~\ref{sec:Non-attacking-Queens-example}.
Since there are 95 workers, it takes at least $n/95$ seconds to do
$n$ such jobs. The rest of the time is the overhead. The overhead
is approximately $16$ microseconds per job with no data sent, and
$98$ milliseconds per job with a vector of 1000 doubles serialized,
deserialized and sent both ways for every job. }

\end{table}

\section{\label{sec:Conclusions}Conclusions}

We have demonstrated that parallel self-submitting job queues are
a powerful computational paradigm. Job queues are high-level constructs
not available in other MPI libraries. The ability of workers to submit
jobs to the job queue is the key component that gives our method its
flexibility and power. We implemented our ideas in a light-weight
and easy to use library based on MPI. The library, MPQueue, has been
used successfully in real research applications. Using our interface
saves effort by avoiding low-level coding; scientists without expertise
in parallel programming can rapidly develop code to obtain good results
for serious research problems. Experts in parallel programming can
also take advantage of our methodology for applications that can be
broken into relatively large pieces. In applications such as our PDE
example found in Section~\ref{sec:Application-to-Nonlinear}, this
decomposition is obvious. We also find the approach to be effective
in less obvious situations, as demonstrated in the queens example
in Section~\ref{sec:Non-attacking-Queens-example} which uses local
queues on top of the global job queue to avoid the excessive communication
that would result from transmitting too many small jobs across nodes.
We have supplied evidence in this same example that our approach is
scalable. The overhead associated with our implementation has been
demonstrated to be reasonable. Possibilities for applications are
wide; we have also used MPQueue to find winning strategies in combinatorial
game theory \cite{prooftrees} and in simulating a critical branching
random walk \cite{ES}. 

Further improvements to MPQueue could include developing algorithms
for automatically adjusting the local queue size (i.e., overflow),
or other more sophisticated control procedures for distributing workload,
such as the implementation of priority job queues and the capability
of workers to send jobs directly to workers. We could easily add to
the library the ability for a subset of workers to act as sub-supervisors,
each with their own pool of workers. We believe that the currently
implemented features of MPQueue will suffice for most applications,
and that its simplicity is in many ways an advantage.

In some communication intensive applications, efficiency could be
increased by avoiding the serialization of fixed-length data. The
Boost.MPI library sends fixed length data more efficiently, without
sacrificing convenience. It would be possible to implement MPQueue
on top of Boost.MPI or other MPI interfaces in order to take advantage
of the optimizations offered by those libraries.

\section*{Acknowledgements}

The authors thank the Department of Physics and Astronomy and the
College of Engineering, Forestry and Natural Sciences for providing
access and support in the use of two Linux clusters located on the
Northern Arizona University campus. We also appreciate the TeraGrid
resources provided by NCSA, supported in part by the National Science
Foundation.

\bibliographystyle{plain}
\bibliography{MPQueue,parallel}

\end{document}

%% file: queensPseudo.tex
\incmargin{1em}
\begin{algorithm2e}[h]
\dontprintsemicolon
\linesnumbered
\hrule
\medskip
\SetKwFunction{KwPB}{pushBack}
\SetKwFunction{KwInsert}{insert}
\SetKwFunction{KwRemove}{remove}
\SetKw{KwBreak}{break}
\SetKw{KwWhile}{while}
\SetKwFor{Rep}{repeat}{}{}
\SetKwSwitch{Switch}{Case}{Other}{switch}{}{case}{otherwise}{}

\KwIn{$size$}
\KwOut{$solutions$}
\smallskip
\hrule
\smallskip
add empty row to the $queue$ of partial placements \hfill // no queens placed yet\;
\While{$queue$ is not empty}{
  move the last job from $queue$ into $row$ \;
  increase the size of $row$ \hfill // add room for the next column\;
  \For{$i\in\{0,\ldots,size\}$} {
    set the last coordinate of $row$ to $i$ \hfill // place the next queen into the $i$-th row\;
    \If {$row$ is a good partial placement}{
      \If{$row$ is a full placement}{
        increment $solutions$ \;
      }
      \Else{
        add $row$ to $queue$ \;      
      }
    }    
  }
}
\medskip
\hrule
\caption{\label{queensPseudo} Pseudo code for the serial non-attacking queens placement program. 
The algorithm uses simple back tracking.}

\end{algorithm2e}

%% file: load85.tex
\begingroup%
\makeatletter%
\newcommand{\GNUPLOTspecial}{%
  \@sanitize\catcode`\%=14\relax\special}%
\setlength{\unitlength}{0.0500bp}%
\begin{picture}(4679,3527)(0,0)%
  {\GNUPLOTspecial{"
/gnudict 256 dict def
gnudict begin
%
%
/Color false def
/Blacktext true def
/Solid false def
/Dashlength 1 def
/Landscape false def
/Level1 false def
/Rounded false def
/TransparentPatterns false def
/gnulinewidth 5.000 def
/userlinewidth gnulinewidth def
/vshift -66 def
/dl1 {
  10.0 Dashlength mul mul
  Rounded { currentlinewidth 0.75 mul sub dup 0 le { pop 0.01 } if } if
} def
/dl2 {
  10.0 Dashlength mul mul
  Rounded { currentlinewidth 0.75 mul add } if
} def
/hpt_ 31.5 def
/vpt_ 31.5 def
/hpt hpt_ def
/vpt vpt_ def
Level1 {} {
/SDict 10 dict def
systemdict /pdfmark known not {
  userdict /pdfmark systemdict /cleartomark get put
} if
SDict begin [
  /Title (load85.tex)
  /Subject (gnuplot plot)
  /Creator (gnuplot 4.2 patchlevel 3 )
  /Author (,,,)
  /CreationDate (Thu Dec 17 19:59:23 2009)
  /DOCINFO pdfmark
end
} ifelse
%
%
/M {moveto} bind def
/L {lineto} bind def
/R {rmoveto} bind def
/V {rlineto} bind def
/N {newpath moveto} bind def
/Z {closepath} bind def
/C {setrgbcolor} bind def
/f {rlineto fill} bind def
/vpt2 vpt 2 mul def
/hpt2 hpt 2 mul def
/Lshow {currentpoint stroke M 0 vshift R 
	Blacktext {gsave 0 setgray show grestore} {show} ifelse} def
/Rshow {currentpoint stroke M dup stringwidth pop neg vshift R
	Blacktext {gsave 0 setgray show grestore} {show} ifelse} def
/Cshow {currentpoint stroke M dup stringwidth pop -2 div vshift R 
	Blacktext {gsave 0 setgray show grestore} {show} ifelse} def
/UP {dup vpt_ mul /vpt exch def hpt_ mul /hpt exch def
  /hpt2 hpt 2 mul def /vpt2 vpt 2 mul def} def
/DL {Color {setrgbcolor Solid {pop []} if 0 setdash}
 {pop pop pop 0 setgray Solid {pop []} if 0 setdash} ifelse} def
/BL {stroke userlinewidth 2 mul setlinewidth
	Rounded {1 setlinejoin 1 setlinecap} if} def
/AL {stroke userlinewidth 2 div setlinewidth
	Rounded {1 setlinejoin 1 setlinecap} if} def
/UL {dup gnulinewidth mul /userlinewidth exch def
	dup 1 lt {pop 1} if 10 mul /udl exch def} def
/PL {stroke userlinewidth setlinewidth
	Rounded {1 setlinejoin 1 setlinecap} if} def
/LCw {1 1 1} def
/LCb {0 0 0} def
/LCa {0 0 0} def
/LC0 {1 0 0} def
/LC1 {0 1 0} def
/LC2 {0 0 1} def
/LC3 {1 0 1} def
/LC4 {0 1 1} def
/LC5 {1 1 0} def
/LC6 {0 0 0} def
/LC7 {1 0.3 0} def
/LC8 {0.5 0.5 0.5} def
/LTw {PL [] 1 setgray} def
/LTb {BL [] LCb DL} def
/LTa {AL [1 udl mul 2 udl mul] 0 setdash LCa setrgbcolor} def
/LT0 {PL [] LC0 DL} def
/LT1 {PL [4 dl1 2 dl2] LC1 DL} def
/LT2 {PL [2 dl1 3 dl2] LC2 DL} def
/LT3 {PL [1 dl1 1.5 dl2] LC3 DL} def
/LT4 {PL [6 dl1 2 dl2 1 dl1 2 dl2] LC4 DL} def
/LT5 {PL [3 dl1 3 dl2 1 dl1 3 dl2] LC5 DL} def
/LT6 {PL [2 dl1 2 dl2 2 dl1 6 dl2] LC6 DL} def
/LT7 {PL [1 dl1 2 dl2 6 dl1 2 dl2 1 dl1 2 dl2] LC7 DL} def
/LT8 {PL [2 dl1 2 dl2 2 dl1 2 dl2 2 dl1 2 dl2 2 dl1 4 dl2] LC8 DL} def
/Pnt {stroke [] 0 setdash gsave 1 setlinecap M 0 0 V stroke grestore} def
/Dia {stroke [] 0 setdash 2 copy vpt add M
  hpt neg vpt neg V hpt vpt neg V
  hpt vpt V hpt neg vpt V closepath stroke
  Pnt} def
/Pls {stroke [] 0 setdash vpt sub M 0 vpt2 V
  currentpoint stroke M
  hpt neg vpt neg R hpt2 0 V stroke
 } def
/Box {stroke [] 0 setdash 2 copy exch hpt sub exch vpt add M
  0 vpt2 neg V hpt2 0 V 0 vpt2 V
  hpt2 neg 0 V closepath stroke
  Pnt} def
/Crs {stroke [] 0 setdash exch hpt sub exch vpt add M
  hpt2 vpt2 neg V currentpoint stroke M
  hpt2 neg 0 R hpt2 vpt2 V stroke} def
/TriU {stroke [] 0 setdash 2 copy vpt 1.12 mul add M
  hpt neg vpt -1.62 mul V
  hpt 2 mul 0 V
  hpt neg vpt 1.62 mul V closepath stroke
  Pnt} def
/Star {2 copy Pls Crs} def
/BoxF {stroke [] 0 setdash exch hpt sub exch vpt add M
  0 vpt2 neg V hpt2 0 V 0 vpt2 V
  hpt2 neg 0 V closepath fill} def
/TriUF {stroke [] 0 setdash vpt 1.12 mul add M
  hpt neg vpt -1.62 mul V
  hpt 2 mul 0 V
  hpt neg vpt 1.62 mul V closepath fill} def
/TriD {stroke [] 0 setdash 2 copy vpt 1.12 mul sub M
  hpt neg vpt 1.62 mul V
  hpt 2 mul 0 V
  hpt neg vpt -1.62 mul V closepath stroke
  Pnt} def
/TriDF {stroke [] 0 setdash vpt 1.12 mul sub M
  hpt neg vpt 1.62 mul V
  hpt 2 mul 0 V
  hpt neg vpt -1.62 mul V closepath fill} def
/DiaF {stroke [] 0 setdash vpt add M
  hpt neg vpt neg V hpt vpt neg V
  hpt vpt V hpt neg vpt V closepath fill} def
/Pent {stroke [] 0 setdash 2 copy gsave
  translate 0 hpt M 4 {72 rotate 0 hpt L} repeat
  closepath stroke grestore Pnt} def
/PentF {stroke [] 0 setdash gsave
  translate 0 hpt M 4 {72 rotate 0 hpt L} repeat
  closepath fill grestore} def
/Circle {stroke [] 0 setdash 2 copy
  hpt 0 360 arc stroke Pnt} def
/CircleF {stroke [] 0 setdash hpt 0 360 arc fill} def
/C0 {BL [] 0 setdash 2 copy moveto vpt 90 450 arc} bind def
/C1 {BL [] 0 setdash 2 copy moveto
	2 copy vpt 0 90 arc closepath fill
	vpt 0 360 arc closepath} bind def
/C2 {BL [] 0 setdash 2 copy moveto
	2 copy vpt 90 180 arc closepath fill
	vpt 0 360 arc closepath} bind def
/C3 {BL [] 0 setdash 2 copy moveto
	2 copy vpt 0 180 arc closepath fill
	vpt 0 360 arc closepath} bind def
/C4 {BL [] 0 setdash 2 copy moveto
	2 copy vpt 180 270 arc closepath fill
	vpt 0 360 arc closepath} bind def
/C5 {BL [] 0 setdash 2 copy moveto
	2 copy vpt 0 90 arc
	2 copy moveto
	2 copy vpt 180 270 arc closepath fill
	vpt 0 360 arc} bind def
/C6 {BL [] 0 setdash 2 copy moveto
	2 copy vpt 90 270 arc closepath fill
	vpt 0 360 arc closepath} bind def
/C7 {BL [] 0 setdash 2 copy moveto
	2 copy vpt 0 270 arc closepath fill
	vpt 0 360 arc closepath} bind def
/C8 {BL [] 0 setdash 2 copy moveto
	2 copy vpt 270 360 arc closepath fill
	vpt 0 360 arc closepath} bind def
/C9 {BL [] 0 setdash 2 copy moveto
	2 copy vpt 270 450 arc closepath fill
	vpt 0 360 arc closepath} bind def
/C10 {BL [] 0 setdash 2 copy 2 copy moveto vpt 270 360 arc closepath fill
	2 copy moveto
	2 copy vpt 90 180 arc closepath fill
	vpt 0 360 arc closepath} bind def
/C11 {BL [] 0 setdash 2 copy moveto
	2 copy vpt 0 180 arc closepath fill
	2 copy moveto
	2 copy vpt 270 360 arc closepath fill
	vpt 0 360 arc closepath} bind def
/C12 {BL [] 0 setdash 2 copy moveto
	2 copy vpt 180 360 arc closepath fill
	vpt 0 360 arc closepath} bind def
/C13 {BL [] 0 setdash 2 copy moveto
	2 copy vpt 0 90 arc closepath fill
	2 copy moveto
	2 copy vpt 180 360 arc closepath fill
	vpt 0 360 arc closepath} bind def
/C14 {BL [] 0 setdash 2 copy moveto
	2 copy vpt 90 360 arc closepath fill
	vpt 0 360 arc} bind def
/C15 {BL [] 0 setdash 2 copy vpt 0 360 arc closepath fill
	vpt 0 360 arc closepath} bind def
/Rec {newpath 4 2 roll moveto 1 index 0 rlineto 0 exch rlineto
	neg 0 rlineto closepath} bind def
/Square {dup Rec} bind def
/Bsquare {vpt sub exch vpt sub exch vpt2 Square} bind def
/S0 {BL [] 0 setdash 2 copy moveto 0 vpt rlineto BL Bsquare} bind def
/S1 {BL [] 0 setdash 2 copy vpt Square fill Bsquare} bind def
/S2 {BL [] 0 setdash 2 copy exch vpt sub exch vpt Square fill Bsquare} bind def
/S3 {BL [] 0 setdash 2 copy exch vpt sub exch vpt2 vpt Rec fill Bsquare} bind def
/S4 {BL [] 0 setdash 2 copy exch vpt sub exch vpt sub vpt Square fill Bsquare} bind def
/S5 {BL [] 0 setdash 2 copy 2 copy vpt Square fill
	exch vpt sub exch vpt sub vpt Square fill Bsquare} bind def
/S6 {BL [] 0 setdash 2 copy exch vpt sub exch vpt sub vpt vpt2 Rec fill Bsquare} bind def
/S7 {BL [] 0 setdash 2 copy exch vpt sub exch vpt sub vpt vpt2 Rec fill
	2 copy vpt Square fill Bsquare} bind def
/S8 {BL [] 0 setdash 2 copy vpt sub vpt Square fill Bsquare} bind def
/S9 {BL [] 0 setdash 2 copy vpt sub vpt vpt2 Rec fill Bsquare} bind def
/S10 {BL [] 0 setdash 2 copy vpt sub vpt Square fill 2 copy exch vpt sub exch vpt Square fill
	Bsquare} bind def
/S11 {BL [] 0 setdash 2 copy vpt sub vpt Square fill 2 copy exch vpt sub exch vpt2 vpt Rec fill
	Bsquare} bind def
/S12 {BL [] 0 setdash 2 copy exch vpt sub exch vpt sub vpt2 vpt Rec fill Bsquare} bind def
/S13 {BL [] 0 setdash 2 copy exch vpt sub exch vpt sub vpt2 vpt Rec fill
	2 copy vpt Square fill Bsquare} bind def
/S14 {BL [] 0 setdash 2 copy exch vpt sub exch vpt sub vpt2 vpt Rec fill
	2 copy exch vpt sub exch vpt Square fill Bsquare} bind def
/S15 {BL [] 0 setdash 2 copy Bsquare fill Bsquare} bind def
/D0 {gsave translate 45 rotate 0 0 S0 stroke grestore} bind def
/D1 {gsave translate 45 rotate 0 0 S1 stroke grestore} bind def
/D2 {gsave translate 45 rotate 0 0 S2 stroke grestore} bind def
/D3 {gsave translate 45 rotate 0 0 S3 stroke grestore} bind def
/D4 {gsave translate 45 rotate 0 0 S4 stroke grestore} bind def
/D5 {gsave translate 45 rotate 0 0 S5 stroke grestore} bind def
/D6 {gsave translate 45 rotate 0 0 S6 stroke grestore} bind def
/D7 {gsave translate 45 rotate 0 0 S7 stroke grestore} bind def
/D8 {gsave translate 45 rotate 0 0 S8 stroke grestore} bind def
/D9 {gsave translate 45 rotate 0 0 S9 stroke grestore} bind def
/D10 {gsave translate 45 rotate 0 0 S10 stroke grestore} bind def
/D11 {gsave translate 45 rotate 0 0 S11 stroke grestore} bind def
/D12 {gsave translate 45 rotate 0 0 S12 stroke grestore} bind def
/D13 {gsave translate 45 rotate 0 0 S13 stroke grestore} bind def
/D14 {gsave translate 45 rotate 0 0 S14 stroke grestore} bind def
/D15 {gsave translate 45 rotate 0 0 S15 stroke grestore} bind def
/DiaE {stroke [] 0 setdash vpt add M
  hpt neg vpt neg V hpt vpt neg V
  hpt vpt V hpt neg vpt V closepath stroke} def
/BoxE {stroke [] 0 setdash exch hpt sub exch vpt add M
  0 vpt2 neg V hpt2 0 V 0 vpt2 V
  hpt2 neg 0 V closepath stroke} def
/TriUE {stroke [] 0 setdash vpt 1.12 mul add M
  hpt neg vpt -1.62 mul V
  hpt 2 mul 0 V
  hpt neg vpt 1.62 mul V closepath stroke} def
/TriDE {stroke [] 0 setdash vpt 1.12 mul sub M
  hpt neg vpt 1.62 mul V
  hpt 2 mul 0 V
  hpt neg vpt -1.62 mul V closepath stroke} def
/PentE {stroke [] 0 setdash gsave
  translate 0 hpt M 4 {72 rotate 0 hpt L} repeat
  closepath stroke grestore} def
/CircE {stroke [] 0 setdash 
  hpt 0 360 arc stroke} def
/Opaque {gsave closepath 1 setgray fill grestore 0 setgray closepath} def
/DiaW {stroke [] 0 setdash vpt add M
  hpt neg vpt neg V hpt vpt neg V
  hpt vpt V hpt neg vpt V Opaque stroke} def
/BoxW {stroke [] 0 setdash exch hpt sub exch vpt add M
  0 vpt2 neg V hpt2 0 V 0 vpt2 V
  hpt2 neg 0 V Opaque stroke} def
/TriUW {stroke [] 0 setdash vpt 1.12 mul add M
  hpt neg vpt -1.62 mul V
  hpt 2 mul 0 V
  hpt neg vpt 1.62 mul V Opaque stroke} def
/TriDW {stroke [] 0 setdash vpt 1.12 mul sub M
  hpt neg vpt 1.62 mul V
  hpt 2 mul 0 V
  hpt neg vpt -1.62 mul V Opaque stroke} def
/PentW {stroke [] 0 setdash gsave
  translate 0 hpt M 4 {72 rotate 0 hpt L} repeat
  Opaque stroke grestore} def
/CircW {stroke [] 0 setdash 
  hpt 0 360 arc Opaque stroke} def
/BoxFill {gsave Rec 1 setgray fill grestore} def
/Density {
  /Fillden exch def
  currentrgbcolor
  /ColB exch def /ColG exch def /ColR exch def
  /ColR ColR Fillden mul Fillden sub 1 add def
  /ColG ColG Fillden mul Fillden sub 1 add def
  /ColB ColB Fillden mul Fillden sub 1 add def
  ColR ColG ColB setrgbcolor} def
/BoxColFill {gsave Rec PolyFill} def
/PolyFill {gsave Density fill grestore grestore} def
/h {rlineto rlineto rlineto gsave closepath fill grestore} bind def
%
%
/PatternFill {gsave /PFa [ 9 2 roll ] def
  PFa 0 get PFa 2 get 2 div add PFa 1 get PFa 3 get 2 div add translate
  PFa 2 get -2 div PFa 3 get -2 div PFa 2 get PFa 3 get Rec
  gsave 1 setgray fill grestore clip
  currentlinewidth 0.5 mul setlinewidth
  /PFs PFa 2 get dup mul PFa 3 get dup mul add sqrt def
  0 0 M PFa 5 get rotate PFs -2 div dup translate
  0 1 PFs PFa 4 get div 1 add floor cvi
	{PFa 4 get mul 0 M 0 PFs V} for
  0 PFa 6 get ne {
	0 1 PFs PFa 4 get div 1 add floor cvi
	{PFa 4 get mul 0 2 1 roll M PFs 0 V} for
 } if
  stroke grestore} def
/languagelevel where
 {pop languagelevel} {1} ifelse
 2 lt
	{/InterpretLevel1 true def}
	{/InterpretLevel1 Level1 def}
 ifelse
%
%
/Level2PatternFill {
/Tile8x8 {/PaintType 2 /PatternType 1 /TilingType 1 /BBox [0 0 8 8] /XStep 8 /YStep 8}
	bind def
/KeepColor {currentrgbcolor [/Pattern /DeviceRGB] setcolorspace} bind def
<< Tile8x8
 /PaintProc {0.5 setlinewidth pop 0 0 M 8 8 L 0 8 M 8 0 L stroke} 
>> matrix makepattern
/Pat1 exch def
<< Tile8x8
 /PaintProc {0.5 setlinewidth pop 0 0 M 8 8 L 0 8 M 8 0 L stroke
	0 4 M 4 8 L 8 4 L 4 0 L 0 4 L stroke}
>> matrix makepattern
/Pat2 exch def
<< Tile8x8
 /PaintProc {0.5 setlinewidth pop 0 0 M 0 8 L
	8 8 L 8 0 L 0 0 L fill}
>> matrix makepattern
/Pat3 exch def
<< Tile8x8
 /PaintProc {0.5 setlinewidth pop -4 8 M 8 -4 L
	0 12 M 12 0 L stroke}
>> matrix makepattern
/Pat4 exch def
<< Tile8x8
 /PaintProc {0.5 setlinewidth pop -4 0 M 8 12 L
	0 -4 M 12 8 L stroke}
>> matrix makepattern
/Pat5 exch def
<< Tile8x8
 /PaintProc {0.5 setlinewidth pop -2 8 M 4 -4 L
	0 12 M 8 -4 L 4 12 M 10 0 L stroke}
>> matrix makepattern
/Pat6 exch def
<< Tile8x8
 /PaintProc {0.5 setlinewidth pop -2 0 M 4 12 L
	0 -4 M 8 12 L 4 -4 M 10 8 L stroke}
>> matrix makepattern
/Pat7 exch def
<< Tile8x8
 /PaintProc {0.5 setlinewidth pop 8 -2 M -4 4 L
	12 0 M -4 8 L 12 4 M 0 10 L stroke}
>> matrix makepattern
/Pat8 exch def
<< Tile8x8
 /PaintProc {0.5 setlinewidth pop 0 -2 M 12 4 L
	-4 0 M 12 8 L -4 4 M 8 10 L stroke}
>> matrix makepattern
/Pat9 exch def
/Pattern1 {PatternBgnd KeepColor Pat1 setpattern} bind def
/Pattern2 {PatternBgnd KeepColor Pat2 setpattern} bind def
/Pattern3 {PatternBgnd KeepColor Pat3 setpattern} bind def
/Pattern4 {PatternBgnd KeepColor Landscape {Pat5} {Pat4} ifelse setpattern} bind def
/Pattern5 {PatternBgnd KeepColor Landscape {Pat4} {Pat5} ifelse setpattern} bind def
/Pattern6 {PatternBgnd KeepColor Landscape {Pat9} {Pat6} ifelse setpattern} bind def
/Pattern7 {PatternBgnd KeepColor Landscape {Pat8} {Pat7} ifelse setpattern} bind def
} def
%
%
%
/PatternBgnd {
  TransparentPatterns {} {gsave 1 setgray fill grestore} ifelse
} def
%
%
/Level1PatternFill {
/Pattern1 {0.250 Density} bind def
/Pattern2 {0.500 Density} bind def
/Pattern3 {0.750 Density} bind def
/Pattern4 {0.125 Density} bind def
/Pattern5 {0.375 Density} bind def
/Pattern6 {0.625 Density} bind def
/Pattern7 {0.875 Density} bind def
} def
%
%
Level1 {Level1PatternFill} {Level2PatternFill} ifelse
/Symbol-Oblique /Symbol findfont [1 0 .167 1 0 0] makefont
dup length dict begin {1 index /FID eq {pop pop} {def} ifelse} forall
currentdict end definefont pop
end
gnudict begin
gsave
0 0 translate
0.050 0.050 scale
0 setgray
newpath
1.000 UL
LTb
883 663 M
-63 0 V
3520 0 R
63 0 V
883 1364 M
-63 0 V
3520 0 R
63 0 V
883 2139 M
-63 0 V
3520 0 R
63 0 V
883 2876 M
-63 0 V
3520 0 R
63 0 V
883 663 M
0 -63 V
0 2328 R
0 63 V
1788 663 M
0 -63 V
0 2328 R
0 63 V
2693 663 M
0 -63 V
0 2328 R
0 63 V
3598 663 M
0 -63 V
0 2328 R
0 63 V
4335 663 M
0 -63 V
0 2328 R
0 63 V
stroke
883 2928 N
883 663 L
3457 0 V
0 2265 V
-3457 0 V
Z stroke
LCb setrgbcolor
LTb
0.800 UP
1.000 UL
LTb
1.000 UL
LT0
LTb
LT0
3570 2765 M
530 0 V
883 663 M
0 944 V
0 1247 V
0 30 V
393 44 V
1 -7 V
25 7 V
17 -7 V
13 -8 V
19 -7 V
22 -8 V
2 -7 V
5 0 V
4 -7 V
23 -8 V
3 -7 V
0 -7 V
24 7 V
4 -7 V
9 -8 V
0 -15 V
1 8 V
1 -8 V
2 -7 V
2 -7 V
0 -8 V
11 -7 V
3 -7 V
0 -8 V
11 8 V
3 59 V
10 -8 V
0 -7 V
2 -8 V
1 -7 V
5 -7 V
1 -8 V
0 -7 V
1 -7 V
5 -8 V
1 -7 V
3 -8 V
5 -7 V
2 -7 V
6 -8 V
2 -7 V
1 -7 V
3 -8 V
0 -7 V
2 7 V
6 -7 V
4 -8 V
0 -7 V
5 -7 V
2 -8 V
21 -7 V
2 -7 V
3 -8 V
1 -7 V
7 -8 V
3 -7 V
7 -7 V
1 -8 V
10 -7 V
1 0 V
12 -7 V
6 -8 V
1 -7 V
7 -8 V
3 -7 V
2 -7 V
0 -8 V
13 -7 V
6 -8 V
5 -7 V
3 -7 V
1 -8 V
2 -7 V
9 -7 V
6 -8 V
1 15 V
10 15 V
3 -8 V
8 -7 V
1 -7 V
5 -8 V
20 -7 V
3 -8 V
46 -7 V
13 -7 V
5 -8 V
3 -7 V
1 -7 V
17 -8 V
16 -7 V
149 -8 V
5 -7 V
14 -7 V
1 0 V
24 0 V
17 -8 V
26 -7 V
38 -7 V
2138 2404 L
29 -7 V
31 -8 V
20 -7 V
15 -7 V
7 -8 V
18 -7 V
8 -7 V
6 -8 V
3 -7 V
5 -8 V
2 -7 V
16 -7 V
2 -8 V
1 -7 V
11 -7 V
1 -8 V
3 -7 V
1 -8 V
5 -7 V
19 -7 V
8 -8 V
7 -15 V
1 -7 V
4 -7 V
6 -8 V
3 -7 V
10 -7 V
3 -8 V
7 -7 V
11 -8 V
5 -7 V
5 -7 V
9 -8 V
2 -7 V
5 -7 V
6 -8 V
1 -7 V
12 -8 V
1 -7 V
2 -7 V
3 -8 V
7 -7 V
5 -7 V
2 -8 V
7 -7 V
3 -8 V
25 89 V
17 -7 V
22 -8 V
2 -7 V
2 -8 V
5 -7 V
21 -7 V
2 -8 V
11 -7 V
1 -7 V
6 -8 V
15 -7 V
1 -8 V
12 -7 V
7 -7 V
0 -8 V
7 -7 V
6 -7 V
4 -8 V
14 -7 V
2 -8 V
19 -7 V
3 -7 V
1 -8 V
6 -7 V
2 -7 V
1 -8 V
5 -7 V
5 -8 V
3 -7 V
4 -7 V
13 -8 V
9 -7 V
5 -8 V
36 -7 V
0 -7 V
22 7 V
23 -7 V
1 -8 V
11 8 V
68 -8 V
1 -7 V
45 -7 V
27 -8 V
3 -7 V
11 -8 V
23 -7 V
21 -7 V
4 -8 V
1 -7 V
18 -7 V
16 -8 V
1 -7 V
5 -8 V
1 -7 V
1 -7 V
9 -8 V
7 -7 V
3092 1748 L
3 -8 V
4 -7 V
4 -8 V
2 -7 V
0 -7 V
2 -8 V
2 -7 V
1 -7 V
10 -8 V
15 -7 V
4 -8 V
2 -7 V
3 -7 V
0 -8 V
3 -7 V
5 -8 V
1 -7 V
2 -7 V
0 -8 V
10 -7 V
13 -7 V
1 -8 V
5 -7 V
8 -8 V
0 -7 V
12 -7 V
5 -8 V
1 -7 V
4 -7 V
1 -8 V
4 -7 V
0 -8 V
5 -7 V
3 -7 V
11 -8 V
12 -7 V
1 -7 V
3 -8 V
3 -7 V
2 -8 V
17 -7 V
7 -7 V
5 -8 V
1 -7 V
4 -7 V
2 7 V
2 -7 V
5 -8 V
5 -7 V
2 -8 V
4 -7 V
1 -7 V
5 -8 V
3 -7 V
2 -7 V
2 -8 V
2 -7 V
0 -8 V
5 -7 V
5 -7 V
0 -8 V
1 -7 V
0 -8 V
3 -7 V
8 -7 V
4 -8 V
8 -7 V
14 -7 V
1 -8 V
3 -7 V
0 -8 V
4 -7 V
2 -7 V
3 -8 V
5 -7 V
2 -7 V
4 -8 V
2 -7 V
3 -8 V
1 -14 V
4 -8 V
4 -7 V
0 -7 V
5 -8 V
6 -7 V
4 -8 V
12 -7 V
156 -7 V
32 -8 V
85 -7 V
52 -7 V
13 -8 V
25 -7 V
21 -8 V
41 -7 V
24 -7 V
5 -8 V
0 -7 V
13 -7 V
12 -8 V
10 -7 V
6 -8 V
6 -7 V
2 -7 V
3982 980 L
0 -7 V
1 -8 V
12 -7 V
4 -7 V
6 -8 V
3 -7 V
8 -7 V
9 -8 V
10 -7 V
1 -8 V
8 -7 V
9 -7 V
0 -8 V
2 -7 V
4 -7 V
3 -8 V
1 -7 V
9 -8 V
2 -7 V
11 -7 V
3 -8 V
1 -7 V
24 -7 V
5 -8 V
1 -7 V
30 -8 V
1 -7 V
1 -7 V
11 -8 V
17 -7 V
17 -7 V
5 -8 V
47 -7 V
8 -8 V
7 -7 V
3 -7 V
12 -8 V
6 -7 V
7 -7 V
25 -8 V
11 -7 V
8 -8 V
stroke
LT1
LTb
LT1
0.200 3570 2515 530 100 BoxColFill
3570 2515 N
530 0 V
0 100 V
-530 0 V
0 -100 V
Z stroke
gsave 883 663 N 0 0 V 0 701 V 0 0 V 0 0 V 393 0 V 1 0 V 25 0 V 17 0 V 13 0 V 19 0 V 22 0 V 2 0 V 0 0 V 5 0 V 4 0 V 23 0 V 3 0 V 0 0 V 24 0 V 4 0 V 9 0 V 0 0 V 1 0 V 1 0 V 2 0 V 2 0 V 0 0 V 11 0 V 3 0 V 0 0 V 11 0 V 3 0 V 10 0 V 0 0 V 2 0 V 1 0 V 5 0 V 1 0 V 0 0 V 1 0 V 5 0 V 1 0 V 3 0 V 5 0 V 2 0 V 6 0 V 2 0 V 1 0 V 3 0 V 0 0 V 2 0 V 6 0 V 4 0 V 0 0 V 5 0 V 2 0 V 21 0 V 2 0 V 3 0 V 1 0 V 7 0 V 3 0 V 7 0 V 1 0 V 10 0 V 1 0 V 12 0 V 6 0 V 1 0 V 7 0 V 3 0 V 2 0 V 0 0 V 13 0 V 6 0 V 5 0 V 3 0 V 1 0 V 2 0 V 9 0 V 6 0 V 1 0 V 10 0 V 3 0 V 8 0 V 1 0 V 5 0 V 20 0 V 3 0 V 46 0 V 13 0 V 5 0 V 3 0 V 1 0 V 17 0 V 16 0 V 149 0 V 5 0 V 14 0 V 1 0 V 24 0 V 17 0 V 26 0 V 38 0 V 32 0 V 29 0 V 31 0 V 20 0 V 15 0 V 7 0 V 18 0 V 8 0 V 6 0 V 3 0 V 5 0 V 2 0 V 16 0 V 2 0 V 1 0 V 11 0 V 1 0 V 3 0 V 1 0 V 5 0 V 19 0 V 8 0 V 7 0 V 1 0 V 4 0 V 6 0 V 3 0 V 10 0 V 3 0 V 7 0 V 11 0 V 5 0 V 5 0 V 9 0 V 2 0 V 5 0 V 6 0 V 1 0 V 12 0 V 1 0 V 2 0 V 3 0 V 7 0 V 5 0 V 2 0 V 7 0 V 3 0 V 25 0 V 17 0 V 22 0 V 2 0 V 2 0 V 5 0 V 21 0 V 2 0 V 11 0 V 1 0 V 6 0 V 15 0 V 1 0 V 12 0 V 7 0 V 0 0 V 7 0 V 6 0 V 4 0 V 14 0 V 2 0 V 19 0 V 3 0 V 1 0 V 6 0 V 2 0 V 1 0 V 5 0 V 5 0 V 3 0 V 4 0 V 13 0 V 9 0 V 5 0 V 36 0 V 0 0 V 22 0 V 23 0 V 1 0 V 11 0 V 68 0 V 1 0 V 45 0 V 27 0 V 3 0 V 11 0 V 23 0 V 21 0 V 4 0 V 1 0 V 18 0 V 16 0 V 1 0 V 5 0 V 1 0 V 1 0 V 9 0 V 7 0 V 3 0 V 3 0 V 4 0 V 4 0 V 2 0 V 0 0 V 2 0 V 2 0 V 1 0 V 10 0 V 15 0 V 4 0 V 2 0 V 3 0 V 0 0 V 3 0 V 5 0 V 1 0 V 2 0 V 0 0 V 10 0 V 13 0 V 1 0 V 5 0 V 8 0 V 0 0 V 12 0 V 5 0 V 1 0 V 4 0 V 1 0 V 4 0 V 0 0 V 5 0 V 3 0 V 11 0 V 12 0 V 1 0 V 3 0 V 3 0 V 2 0 V 17 0 V 7 0 V 5 0 V 1 0 V 4 0 V 2 0 V 2 0 V 5 0 V 5 0 V 2 0 V 4 0 V 1 0 V 5 0 V 3 0 V 2 -7 V 2 -8 V 2 -7 V 0 -8 V 5 -7 V 5 -7 V 0 -8 V 1 -7 V 0 -8 V 3 -7 V 8 -7 V 4 -8 V 8 -7 V 14 -7 V 1 -8 V 3 -7 V 0 -8 V 4 -7 V 2 -7 V 3 -8 V 5 -7 V 2 -7 V 4 -8 V 2 -7 V 3 -8 V 1 -14 V 4 -8 V 4 -7 V 0 -7 V 5 -8 V 6 -7 V 4 -8 V 12 -7 V 156 -7 V 32 -8 V 85 -7 V 52 -7 V 13 -8 V 25 -7 V 21 -8 V 41 -7 V 24 -7 V 5 -8 V 0 -7 V 13 -7 V 12 -8 V 10 -7 V 6 -8 V 6 -7 V 2 -7 V 38 -8 V 0 -7 V 1 -8 V 12 -7 V 4 -7 V 6 -8 V 3 -7 V 8 -7 V 9 -8 V 10 -7 V 1 -8 V 8 -7 V 9 -7 V 0 -8 V 2 -7 V 4 -7 V 3 -8 V 1 -7 V 9 -8 V 2 -7 V 11 -7 V 3 -8 V 1 -7 V 24 -7 V 5 -8 V 1 -7 V 30 -8 V 1 -7 V 1 -7 V 11 -8 V 17 -7 V 17 -7 V 5 -8 V 47 -7 V 8 -8 V 7 -7 V 3 -7 V 12 -8 V 6 -7 V 7 -7 V 25 -8 V 11 -7 V 8 -8 V 0 -7 V -3452 0 V 0.20 PolyFill
883 663 M
0 701 V
393 0 V
1 0 V
25 0 V
17 0 V
13 0 V
19 0 V
22 0 V
2 0 V
5 0 V
4 0 V
23 0 V
3 0 V
24 0 V
4 0 V
9 0 V
1 0 V
1 0 V
2 0 V
2 0 V
11 0 V
3 0 V
11 0 V
3 0 V
10 0 V
2 0 V
1 0 V
5 0 V
1 0 V
1 0 V
5 0 V
1 0 V
3 0 V
5 0 V
2 0 V
6 0 V
2 0 V
1 0 V
3 0 V
2 0 V
6 0 V
4 0 V
5 0 V
2 0 V
21 0 V
2 0 V
3 0 V
1 0 V
7 0 V
3 0 V
7 0 V
1 0 V
10 0 V
1 0 V
12 0 V
6 0 V
1 0 V
7 0 V
3 0 V
2 0 V
13 0 V
6 0 V
5 0 V
3 0 V
1 0 V
2 0 V
9 0 V
6 0 V
1 0 V
10 0 V
3 0 V
8 0 V
1 0 V
5 0 V
20 0 V
3 0 V
46 0 V
13 0 V
5 0 V
3 0 V
1 0 V
17 0 V
16 0 V
149 0 V
5 0 V
14 0 V
1 0 V
24 0 V
17 0 V
26 0 V
38 0 V
32 0 V
29 0 V
31 0 V
20 0 V
15 0 V
7 0 V
18 0 V
8 0 V
6 0 V
3 0 V
5 0 V
2 0 V
16 0 V
2300 1364 L
1 0 V
11 0 V
1 0 V
3 0 V
1 0 V
5 0 V
19 0 V
8 0 V
7 0 V
1 0 V
4 0 V
6 0 V
3 0 V
10 0 V
3 0 V
7 0 V
11 0 V
5 0 V
5 0 V
9 0 V
2 0 V
5 0 V
6 0 V
1 0 V
12 0 V
1 0 V
2 0 V
3 0 V
7 0 V
5 0 V
2 0 V
7 0 V
3 0 V
25 0 V
17 0 V
22 0 V
2 0 V
2 0 V
5 0 V
21 0 V
2 0 V
11 0 V
1 0 V
6 0 V
15 0 V
1 0 V
12 0 V
7 0 V
7 0 V
6 0 V
4 0 V
14 0 V
2 0 V
19 0 V
3 0 V
1 0 V
6 0 V
2 0 V
1 0 V
5 0 V
5 0 V
3 0 V
4 0 V
13 0 V
9 0 V
5 0 V
36 0 V
22 0 V
23 0 V
1 0 V
11 0 V
68 0 V
1 0 V
45 0 V
27 0 V
3 0 V
11 0 V
23 0 V
21 0 V
4 0 V
1 0 V
18 0 V
16 0 V
1 0 V
5 0 V
1 0 V
1 0 V
9 0 V
7 0 V
3 0 V
3 0 V
4 0 V
4 0 V
2 0 V
2 0 V
2 0 V
1 0 V
10 0 V
15 0 V
4 0 V
2 0 V
3 0 V
3 0 V
5 0 V
3153 1364 L
2 0 V
10 0 V
13 0 V
1 0 V
5 0 V
8 0 V
12 0 V
5 0 V
1 0 V
4 0 V
1 0 V
4 0 V
5 0 V
3 0 V
11 0 V
12 0 V
1 0 V
3 0 V
3 0 V
2 0 V
17 0 V
7 0 V
5 0 V
1 0 V
4 0 V
2 0 V
2 0 V
5 0 V
5 0 V
2 0 V
4 0 V
1 0 V
5 0 V
3 0 V
2 -7 V
2 -8 V
2 -7 V
0 -8 V
5 -7 V
5 -7 V
0 -8 V
1 -7 V
0 -8 V
3 -7 V
8 -7 V
4 -8 V
8 -7 V
14 -7 V
1 -8 V
3 -7 V
0 -8 V
4 -7 V
2 -7 V
3 -8 V
5 -7 V
2 -7 V
4 -8 V
2 -7 V
3 -8 V
1 -14 V
4 -8 V
4 -7 V
0 -7 V
5 -8 V
6 -7 V
4 -8 V
12 -7 V
156 -7 V
32 -8 V
85 -7 V
52 -7 V
13 -8 V
25 -7 V
21 -8 V
41 -7 V
24 -7 V
5 -8 V
0 -7 V
13 -7 V
12 -8 V
10 -7 V
6 -8 V
6 -7 V
2 -7 V
38 -8 V
0 -7 V
1 -8 V
12 -7 V
4 -7 V
6 -8 V
3 -7 V
8 -7 V
9 -8 V
10 -7 V
1 -8 V
8 -7 V
9 -7 V
0 -8 V
2 -7 V
4 -7 V
3 -8 V
1 -7 V
9 -8 V
2 -7 V
4085 833 L
3 -8 V
1 -7 V
24 -7 V
5 -8 V
1 -7 V
30 -8 V
1 -7 V
1 -7 V
11 -8 V
17 -7 V
17 -7 V
5 -8 V
47 -7 V
8 -8 V
7 -7 V
3 -7 V
12 -8 V
6 -7 V
7 -7 V
25 -8 V
11 -7 V
8 -8 V
stroke
1.000 UL
LT2
883 1364 M
35 0 V
35 0 V
35 0 V
35 0 V
35 0 V
35 0 V
34 0 V
35 0 V
35 0 V
35 0 V
35 0 V
35 0 V
35 0 V
35 0 V
35 0 V
35 0 V
35 0 V
35 0 V
34 0 V
35 0 V
35 0 V
35 0 V
35 0 V
35 0 V
35 0 V
35 0 V
35 0 V
35 0 V
35 0 V
35 0 V
34 0 V
35 0 V
35 0 V
35 0 V
35 0 V
35 0 V
35 0 V
35 0 V
35 0 V
35 0 V
35 0 V
35 0 V
35 0 V
34 0 V
35 0 V
35 0 V
35 0 V
35 0 V
35 0 V
35 0 V
35 0 V
35 0 V
35 0 V
35 0 V
35 0 V
34 0 V
35 0 V
35 0 V
35 0 V
35 0 V
35 0 V
35 0 V
35 0 V
35 0 V
35 0 V
35 0 V
35 0 V
35 0 V
34 0 V
35 0 V
35 0 V
35 0 V
35 0 V
35 0 V
35 0 V
35 0 V
35 0 V
35 0 V
35 0 V
35 0 V
34 0 V
35 0 V
35 0 V
35 0 V
35 0 V
35 0 V
35 0 V
35 0 V
35 0 V
35 0 V
35 0 V
35 0 V
34 0 V
35 0 V
35 0 V
35 0 V
35 0 V
35 0 V
35 0 V
stroke
LTb
883 2928 N
883 663 L
3457 0 V
0 2265 V
-3457 0 V
Z stroke
0.800 UP
1.000 UL
LTb
stroke
grestore
end
showpage
  }}%
  \put(3450,2565){\makebox(0,0)[r]{\strut{}active}}%
  \put(3450,2765){\makebox(0,0)[r]{\strut{}jobs}}%
  \put(2611,3228){\makebox(0,0){\strut{}overflow = 85}}%
  \put(2611,100){\makebox(0,0){\strut{}time (sec)}}%
  \put(4335,400){\makebox(0,0){\strut{} 38142}}%
  \put(3598,400){\makebox(0,0){\strut{} 30000}}%
  \put(2693,400){\makebox(0,0){\strut{} 20000}}%
  \put(1788,400){\makebox(0,0){\strut{} 10000}}%
  \put(883,400){\makebox(0,0){\strut{} 0}}%
  \put(700,2876){\makebox(0,0)[r]{\strut{} 300}}%
  \put(700,2139){\makebox(0,0)[r]{\strut{} 200}}%
  \put(700,1364){\makebox(0,0)[r]{\strut{} 95}}%
  \put(700,663){\makebox(0,0)[r]{\strut{} 0}}%
\end{picture}%
\endgroup
 

%% file: speedup12.tex
\begingroup%
\makeatletter%
\newcommand{\GNUPLOTspecial}{%
  \@sanitize\catcode`\%=14\relax\special}%
\setlength{\unitlength}{0.0500bp}%
\begin{picture}(6479,4535)(0,0)%
  {\GNUPLOTspecial{"
/gnudict 256 dict def
gnudict begin
%
%
/Color false def
/Blacktext true def
/Solid false def
/Dashlength 1 def
/Landscape false def
/Level1 false def
/Rounded false def
/TransparentPatterns false def
/gnulinewidth 5.000 def
/userlinewidth gnulinewidth def
/vshift -66 def
/dl1 {
  10.0 Dashlength mul mul
  Rounded { currentlinewidth 0.75 mul sub dup 0 le { pop 0.01 } if } if
} def
/dl2 {
  10.0 Dashlength mul mul
  Rounded { currentlinewidth 0.75 mul add } if
} def
/hpt_ 31.5 def
/vpt_ 31.5 def
/hpt hpt_ def
/vpt vpt_ def
Level1 {} {
/SDict 10 dict def
systemdict /pdfmark known not {
  userdict /pdfmark systemdict /cleartomark get put
} if
SDict begin [
  /Title (speedup12.tex)
  /Subject (gnuplot plot)
  /Creator (gnuplot 4.2 patchlevel 2 )
  /Author (n,,,)
  /CreationDate (Wed Sep 22 15:37:03 2010)
  /DOCINFO pdfmark
end
} ifelse
%
%
/M {moveto} bind def
/L {lineto} bind def
/R {rmoveto} bind def
/V {rlineto} bind def
/N {newpath moveto} bind def
/Z {closepath} bind def
/C {setrgbcolor} bind def
/f {rlineto fill} bind def
/vpt2 vpt 2 mul def
/hpt2 hpt 2 mul def
/Lshow {currentpoint stroke M 0 vshift R 
	Blacktext {gsave 0 setgray show grestore} {show} ifelse} def
/Rshow {currentpoint stroke M dup stringwidth pop neg vshift R
	Blacktext {gsave 0 setgray show grestore} {show} ifelse} def
/Cshow {currentpoint stroke M dup stringwidth pop -2 div vshift R 
	Blacktext {gsave 0 setgray show grestore} {show} ifelse} def
/UP {dup vpt_ mul /vpt exch def hpt_ mul /hpt exch def
  /hpt2 hpt 2 mul def /vpt2 vpt 2 mul def} def
/DL {Color {setrgbcolor Solid {pop []} if 0 setdash}
 {pop pop pop 0 setgray Solid {pop []} if 0 setdash} ifelse} def
/BL {stroke userlinewidth 2 mul setlinewidth
	Rounded {1 setlinejoin 1 setlinecap} if} def
/AL {stroke userlinewidth 2 div setlinewidth
	Rounded {1 setlinejoin 1 setlinecap} if} def
/UL {dup gnulinewidth mul /userlinewidth exch def
	dup 1 lt {pop 1} if 10 mul /udl exch def} def
/PL {stroke userlinewidth setlinewidth
	Rounded {1 setlinejoin 1 setlinecap} if} def
/LCw {1 1 1} def
/LCb {0 0 0} def
/LCa {0 0 0} def
/LC0 {1 0 0} def
/LC1 {0 1 0} def
/LC2 {0 0 1} def
/LC3 {1 0 1} def
/LC4 {0 1 1} def
/LC5 {1 1 0} def
/LC6 {0 0 0} def
/LC7 {1 0.3 0} def
/LC8 {0.5 0.5 0.5} def
/LTw {PL [] 1 setgray} def
/LTb {BL [] LCb DL} def
/LTa {AL [1 udl mul 2 udl mul] 0 setdash LCa setrgbcolor} def
/LT0 {PL [] LC0 DL} def
/LT1 {PL [4 dl1 2 dl2] LC1 DL} def
/LT2 {PL [2 dl1 3 dl2] LC2 DL} def
/LT3 {PL [1 dl1 1.5 dl2] LC3 DL} def
/LT4 {PL [6 dl1 2 dl2 1 dl1 2 dl2] LC4 DL} def
/LT5 {PL [3 dl1 3 dl2 1 dl1 3 dl2] LC5 DL} def
/LT6 {PL [2 dl1 2 dl2 2 dl1 6 dl2] LC6 DL} def
/LT7 {PL [1 dl1 2 dl2 6 dl1 2 dl2 1 dl1 2 dl2] LC7 DL} def
/LT8 {PL [2 dl1 2 dl2 2 dl1 2 dl2 2 dl1 2 dl2 2 dl1 4 dl2] LC8 DL} def
/Pnt {stroke [] 0 setdash gsave 1 setlinecap M 0 0 V stroke grestore} def
/Dia {stroke [] 0 setdash 2 copy vpt add M
  hpt neg vpt neg V hpt vpt neg V
  hpt vpt V hpt neg vpt V closepath stroke
  Pnt} def
/Pls {stroke [] 0 setdash vpt sub M 0 vpt2 V
  currentpoint stroke M
  hpt neg vpt neg R hpt2 0 V stroke
 } def
/Box {stroke [] 0 setdash 2 copy exch hpt sub exch vpt add M
  0 vpt2 neg V hpt2 0 V 0 vpt2 V
  hpt2 neg 0 V closepath stroke
  Pnt} def
/Crs {stroke [] 0 setdash exch hpt sub exch vpt add M
  hpt2 vpt2 neg V currentpoint stroke M
  hpt2 neg 0 R hpt2 vpt2 V stroke} def
/TriU {stroke [] 0 setdash 2 copy vpt 1.12 mul add M
  hpt neg vpt -1.62 mul V
  hpt 2 mul 0 V
  hpt neg vpt 1.62 mul V closepath stroke
  Pnt} def
/Star {2 copy Pls Crs} def
/BoxF {stroke [] 0 setdash exch hpt sub exch vpt add M
  0 vpt2 neg V hpt2 0 V 0 vpt2 V
  hpt2 neg 0 V closepath fill} def
/TriUF {stroke [] 0 setdash vpt 1.12 mul add M
  hpt neg vpt -1.62 mul V
  hpt 2 mul 0 V
  hpt neg vpt 1.62 mul V closepath fill} def
/TriD {stroke [] 0 setdash 2 copy vpt 1.12 mul sub M
  hpt neg vpt 1.62 mul V
  hpt 2 mul 0 V
  hpt neg vpt -1.62 mul V closepath stroke
  Pnt} def
/TriDF {stroke [] 0 setdash vpt 1.12 mul sub M
  hpt neg vpt 1.62 mul V
  hpt 2 mul 0 V
  hpt neg vpt -1.62 mul V closepath fill} def
/DiaF {stroke [] 0 setdash vpt add M
  hpt neg vpt neg V hpt vpt neg V
  hpt vpt V hpt neg vpt V closepath fill} def
/Pent {stroke [] 0 setdash 2 copy gsave
  translate 0 hpt M 4 {72 rotate 0 hpt L} repeat
  closepath stroke grestore Pnt} def
/PentF {stroke [] 0 setdash gsave
  translate 0 hpt M 4 {72 rotate 0 hpt L} repeat
  closepath fill grestore} def
/Circle {stroke [] 0 setdash 2 copy
  hpt 0 360 arc stroke Pnt} def
/CircleF {stroke [] 0 setdash hpt 0 360 arc fill} def
/C0 {BL [] 0 setdash 2 copy moveto vpt 90 450 arc} bind def
/C1 {BL [] 0 setdash 2 copy moveto
	2 copy vpt 0 90 arc closepath fill
	vpt 0 360 arc closepath} bind def
/C2 {BL [] 0 setdash 2 copy moveto
	2 copy vpt 90 180 arc closepath fill
	vpt 0 360 arc closepath} bind def
/C3 {BL [] 0 setdash 2 copy moveto
	2 copy vpt 0 180 arc closepath fill
	vpt 0 360 arc closepath} bind def
/C4 {BL [] 0 setdash 2 copy moveto
	2 copy vpt 180 270 arc closepath fill
	vpt 0 360 arc closepath} bind def
/C5 {BL [] 0 setdash 2 copy moveto
	2 copy vpt 0 90 arc
	2 copy moveto
	2 copy vpt 180 270 arc closepath fill
	vpt 0 360 arc} bind def
/C6 {BL [] 0 setdash 2 copy moveto
	2 copy vpt 90 270 arc closepath fill
	vpt 0 360 arc closepath} bind def
/C7 {BL [] 0 setdash 2 copy moveto
	2 copy vpt 0 270 arc closepath fill
	vpt 0 360 arc closepath} bind def
/C8 {BL [] 0 setdash 2 copy moveto
	2 copy vpt 270 360 arc closepath fill
	vpt 0 360 arc closepath} bind def
/C9 {BL [] 0 setdash 2 copy moveto
	2 copy vpt 270 450 arc closepath fill
	vpt 0 360 arc closepath} bind def
/C10 {BL [] 0 setdash 2 copy 2 copy moveto vpt 270 360 arc closepath fill
	2 copy moveto
	2 copy vpt 90 180 arc closepath fill
	vpt 0 360 arc closepath} bind def
/C11 {BL [] 0 setdash 2 copy moveto
	2 copy vpt 0 180 arc closepath fill
	2 copy moveto
	2 copy vpt 270 360 arc closepath fill
	vpt 0 360 arc closepath} bind def
/C12 {BL [] 0 setdash 2 copy moveto
	2 copy vpt 180 360 arc closepath fill
	vpt 0 360 arc closepath} bind def
/C13 {BL [] 0 setdash 2 copy moveto
	2 copy vpt 0 90 arc closepath fill
	2 copy moveto
	2 copy vpt 180 360 arc closepath fill
	vpt 0 360 arc closepath} bind def
/C14 {BL [] 0 setdash 2 copy moveto
	2 copy vpt 90 360 arc closepath fill
	vpt 0 360 arc} bind def
/C15 {BL [] 0 setdash 2 copy vpt 0 360 arc closepath fill
	vpt 0 360 arc closepath} bind def
/Rec {newpath 4 2 roll moveto 1 index 0 rlineto 0 exch rlineto
	neg 0 rlineto closepath} bind def
/Square {dup Rec} bind def
/Bsquare {vpt sub exch vpt sub exch vpt2 Square} bind def
/S0 {BL [] 0 setdash 2 copy moveto 0 vpt rlineto BL Bsquare} bind def
/S1 {BL [] 0 setdash 2 copy vpt Square fill Bsquare} bind def
/S2 {BL [] 0 setdash 2 copy exch vpt sub exch vpt Square fill Bsquare} bind def
/S3 {BL [] 0 setdash 2 copy exch vpt sub exch vpt2 vpt Rec fill Bsquare} bind def
/S4 {BL [] 0 setdash 2 copy exch vpt sub exch vpt sub vpt Square fill Bsquare} bind def
/S5 {BL [] 0 setdash 2 copy 2 copy vpt Square fill
	exch vpt sub exch vpt sub vpt Square fill Bsquare} bind def
/S6 {BL [] 0 setdash 2 copy exch vpt sub exch vpt sub vpt vpt2 Rec fill Bsquare} bind def
/S7 {BL [] 0 setdash 2 copy exch vpt sub exch vpt sub vpt vpt2 Rec fill
	2 copy vpt Square fill Bsquare} bind def
/S8 {BL [] 0 setdash 2 copy vpt sub vpt Square fill Bsquare} bind def
/S9 {BL [] 0 setdash 2 copy vpt sub vpt vpt2 Rec fill Bsquare} bind def
/S10 {BL [] 0 setdash 2 copy vpt sub vpt Square fill 2 copy exch vpt sub exch vpt Square fill
	Bsquare} bind def
/S11 {BL [] 0 setdash 2 copy vpt sub vpt Square fill 2 copy exch vpt sub exch vpt2 vpt Rec fill
	Bsquare} bind def
/S12 {BL [] 0 setdash 2 copy exch vpt sub exch vpt sub vpt2 vpt Rec fill Bsquare} bind def
/S13 {BL [] 0 setdash 2 copy exch vpt sub exch vpt sub vpt2 vpt Rec fill
	2 copy vpt Square fill Bsquare} bind def
/S14 {BL [] 0 setdash 2 copy exch vpt sub exch vpt sub vpt2 vpt Rec fill
	2 copy exch vpt sub exch vpt Square fill Bsquare} bind def
/S15 {BL [] 0 setdash 2 copy Bsquare fill Bsquare} bind def
/D0 {gsave translate 45 rotate 0 0 S0 stroke grestore} bind def
/D1 {gsave translate 45 rotate 0 0 S1 stroke grestore} bind def
/D2 {gsave translate 45 rotate 0 0 S2 stroke grestore} bind def
/D3 {gsave translate 45 rotate 0 0 S3 stroke grestore} bind def
/D4 {gsave translate 45 rotate 0 0 S4 stroke grestore} bind def
/D5 {gsave translate 45 rotate 0 0 S5 stroke grestore} bind def
/D6 {gsave translate 45 rotate 0 0 S6 stroke grestore} bind def
/D7 {gsave translate 45 rotate 0 0 S7 stroke grestore} bind def
/D8 {gsave translate 45 rotate 0 0 S8 stroke grestore} bind def
/D9 {gsave translate 45 rotate 0 0 S9 stroke grestore} bind def
/D10 {gsave translate 45 rotate 0 0 S10 stroke grestore} bind def
/D11 {gsave translate 45 rotate 0 0 S11 stroke grestore} bind def
/D12 {gsave translate 45 rotate 0 0 S12 stroke grestore} bind def
/D13 {gsave translate 45 rotate 0 0 S13 stroke grestore} bind def
/D14 {gsave translate 45 rotate 0 0 S14 stroke grestore} bind def
/D15 {gsave translate 45 rotate 0 0 S15 stroke grestore} bind def
/DiaE {stroke [] 0 setdash vpt add M
  hpt neg vpt neg V hpt vpt neg V
  hpt vpt V hpt neg vpt V closepath stroke} def
/BoxE {stroke [] 0 setdash exch hpt sub exch vpt add M
  0 vpt2 neg V hpt2 0 V 0 vpt2 V
  hpt2 neg 0 V closepath stroke} def
/TriUE {stroke [] 0 setdash vpt 1.12 mul add M
  hpt neg vpt -1.62 mul V
  hpt 2 mul 0 V
  hpt neg vpt 1.62 mul V closepath stroke} def
/TriDE {stroke [] 0 setdash vpt 1.12 mul sub M
  hpt neg vpt 1.62 mul V
  hpt 2 mul 0 V
  hpt neg vpt -1.62 mul V closepath stroke} def
/PentE {stroke [] 0 setdash gsave
  translate 0 hpt M 4 {72 rotate 0 hpt L} repeat
  closepath stroke grestore} def
/CircE {stroke [] 0 setdash 
  hpt 0 360 arc stroke} def
/Opaque {gsave closepath 1 setgray fill grestore 0 setgray closepath} def
/DiaW {stroke [] 0 setdash vpt add M
  hpt neg vpt neg V hpt vpt neg V
  hpt vpt V hpt neg vpt V Opaque stroke} def
/BoxW {stroke [] 0 setdash exch hpt sub exch vpt add M
  0 vpt2 neg V hpt2 0 V 0 vpt2 V
  hpt2 neg 0 V Opaque stroke} def
/TriUW {stroke [] 0 setdash vpt 1.12 mul add M
  hpt neg vpt -1.62 mul V
  hpt 2 mul 0 V
  hpt neg vpt 1.62 mul V Opaque stroke} def
/TriDW {stroke [] 0 setdash vpt 1.12 mul sub M
  hpt neg vpt 1.62 mul V
  hpt 2 mul 0 V
  hpt neg vpt -1.62 mul V Opaque stroke} def
/PentW {stroke [] 0 setdash gsave
  translate 0 hpt M 4 {72 rotate 0 hpt L} repeat
  Opaque stroke grestore} def
/CircW {stroke [] 0 setdash 
  hpt 0 360 arc Opaque stroke} def
/BoxFill {gsave Rec 1 setgray fill grestore} def
/Density {
  /Fillden exch def
  currentrgbcolor
  /ColB exch def /ColG exch def /ColR exch def
  /ColR ColR Fillden mul Fillden sub 1 add def
  /ColG ColG Fillden mul Fillden sub 1 add def
  /ColB ColB Fillden mul Fillden sub 1 add def
  ColR ColG ColB setrgbcolor} def
/BoxColFill {gsave Rec PolyFill} def
/PolyFill {gsave Density fill grestore grestore} def
/h {rlineto rlineto rlineto gsave fill grestore} bind def
%
%
/PatternFill {gsave /PFa [ 9 2 roll ] def
  PFa 0 get PFa 2 get 2 div add PFa 1 get PFa 3 get 2 div add translate
  PFa 2 get -2 div PFa 3 get -2 div PFa 2 get PFa 3 get Rec
  gsave 1 setgray fill grestore clip
  currentlinewidth 0.5 mul setlinewidth
  /PFs PFa 2 get dup mul PFa 3 get dup mul add sqrt def
  0 0 M PFa 5 get rotate PFs -2 div dup translate
  0 1 PFs PFa 4 get div 1 add floor cvi
	{PFa 4 get mul 0 M 0 PFs V} for
  0 PFa 6 get ne {
	0 1 PFs PFa 4 get div 1 add floor cvi
	{PFa 4 get mul 0 2 1 roll M PFs 0 V} for
 } if
  stroke grestore} def
/languagelevel where
 {pop languagelevel} {1} ifelse
 2 lt
	{/InterpretLevel1 true def}
	{/InterpretLevel1 Level1 def}
 ifelse
%
%
/Level2PatternFill {
/Tile8x8 {/PaintType 2 /PatternType 1 /TilingType 1 /BBox [0 0 8 8] /XStep 8 /YStep 8}
	bind def
/KeepColor {currentrgbcolor [/Pattern /DeviceRGB] setcolorspace} bind def
<< Tile8x8
 /PaintProc {0.5 setlinewidth pop 0 0 M 8 8 L 0 8 M 8 0 L stroke} 
>> matrix makepattern
/Pat1 exch def
<< Tile8x8
 /PaintProc {0.5 setlinewidth pop 0 0 M 8 8 L 0 8 M 8 0 L stroke
	0 4 M 4 8 L 8 4 L 4 0 L 0 4 L stroke}
>> matrix makepattern
/Pat2 exch def
<< Tile8x8
 /PaintProc {0.5 setlinewidth pop 0 0 M 0 8 L
	8 8 L 8 0 L 0 0 L fill}
>> matrix makepattern
/Pat3 exch def
<< Tile8x8
 /PaintProc {0.5 setlinewidth pop -4 8 M 8 -4 L
	0 12 M 12 0 L stroke}
>> matrix makepattern
/Pat4 exch def
<< Tile8x8
 /PaintProc {0.5 setlinewidth pop -4 0 M 8 12 L
	0 -4 M 12 8 L stroke}
>> matrix makepattern
/Pat5 exch def
<< Tile8x8
 /PaintProc {0.5 setlinewidth pop -2 8 M 4 -4 L
	0 12 M 8 -4 L 4 12 M 10 0 L stroke}
>> matrix makepattern
/Pat6 exch def
<< Tile8x8
 /PaintProc {0.5 setlinewidth pop -2 0 M 4 12 L
	0 -4 M 8 12 L 4 -4 M 10 8 L stroke}
>> matrix makepattern
/Pat7 exch def
<< Tile8x8
 /PaintProc {0.5 setlinewidth pop 8 -2 M -4 4 L
	12 0 M -4 8 L 12 4 M 0 10 L stroke}
>> matrix makepattern
/Pat8 exch def
<< Tile8x8
 /PaintProc {0.5 setlinewidth pop 0 -2 M 12 4 L
	-4 0 M 12 8 L -4 4 M 8 10 L stroke}
>> matrix makepattern
/Pat9 exch def
/Pattern1 {PatternBgnd KeepColor Pat1 setpattern} bind def
/Pattern2 {PatternBgnd KeepColor Pat2 setpattern} bind def
/Pattern3 {PatternBgnd KeepColor Pat3 setpattern} bind def
/Pattern4 {PatternBgnd KeepColor Landscape {Pat5} {Pat4} ifelse setpattern} bind def
/Pattern5 {PatternBgnd KeepColor Landscape {Pat4} {Pat5} ifelse setpattern} bind def
/Pattern6 {PatternBgnd KeepColor Landscape {Pat9} {Pat6} ifelse setpattern} bind def
/Pattern7 {PatternBgnd KeepColor Landscape {Pat8} {Pat7} ifelse setpattern} bind def
} def
%
%
%
/PatternBgnd {
  TransparentPatterns {} {gsave 1 setgray fill grestore} ifelse
} def
%
%
/Level1PatternFill {
/Pattern1 {0.250 Density} bind def
/Pattern2 {0.500 Density} bind def
/Pattern3 {0.750 Density} bind def
/Pattern4 {0.125 Density} bind def
/Pattern5 {0.375 Density} bind def
/Pattern6 {0.625 Density} bind def
/Pattern7 {0.875 Density} bind def
} def
%
%
Level1 {Level1PatternFill} {Level2PatternFill} ifelse
/Symbol-Oblique /Symbol findfont [1 0 .167 1 0 0] makefont
dup length dict begin {1 index /FID eq {pop pop} {def} ifelse} forall
currentdict end definefont pop
end
gnudict begin
gsave
0 0 translate
0.050 0.050 scale
0 setgray
newpath
1.000 UL
LTb
900 600 M
63 0 V
5177 0 R
-63 0 V
900 1062 M
63 0 V
5177 0 R
-63 0 V
900 1524 M
63 0 V
5177 0 R
-63 0 V
900 1986 M
63 0 V
5177 0 R
-63 0 V
900 2448 M
63 0 V
5177 0 R
-63 0 V
900 2910 M
63 0 V
5177 0 R
-63 0 V
900 3372 M
63 0 V
5177 0 R
-63 0 V
900 3834 M
63 0 V
5177 0 R
-63 0 V
900 4296 M
63 0 V
5177 0 R
-63 0 V
1396 600 M
0 63 V
0 3633 R
0 -63 V
1948 600 M
0 63 V
0 3633 R
0 -63 V
2500 600 M
0 63 V
0 3633 R
0 -63 V
3051 600 M
0 63 V
0 3633 R
0 -63 V
3603 600 M
0 63 V
0 3633 R
0 -63 V
4154 600 M
0 63 V
0 3633 R
0 -63 V
4706 600 M
0 63 V
0 3633 R
0 -63 V
5257 600 M
0 63 V
0 3633 R
0 -63 V
5809 600 M
0 63 V
0 3633 R
0 -63 V
900 4296 M
900 600 L
5240 0 V
0 3696 V
-5240 0 V
stroke
LCb setrgbcolor
LTb
LCb setrgbcolor
LTb
0.800 UP
1.000 UL
LTb
1.000 UL
LT2
900 600 M
53 89 V
53 88 V
53 89 V
53 89 V
53 88 V
53 89 V
53 89 V
52 88 V
53 89 V
53 89 V
53 88 V
53 89 V
53 89 V
53 88 V
53 89 V
53 89 V
53 88 V
53 89 V
53 89 V
53 88 V
53 89 V
52 89 V
53 88 V
53 89 V
53 89 V
53 88 V
53 89 V
53 89 V
53 88 V
53 89 V
53 89 V
53 88 V
53 89 V
53 89 V
53 88 V
52 89 V
53 89 V
53 88 V
53 89 V
53 89 V
53 88 V
36 61 V
stroke
LT0
955 692 M
55 95 V
55 92 V
56 89 V
55 92 V
55 92 V
55 91 V
55 91 V
55 88 V
56 86 V
55 83 V
55 87 V
55 72 V
55 90 V
55 62 V
56 88 V
55 89 V
55 44 V
55 95 V
55 79 V
55 29 V
55 59 V
56 -15 V
55 95 V
55 17 V
55 53 V
55 56 V
55 19 V
56 40 V
55 0 V
55 42 V
55 43 V
55 45 V
55 0 V
56 23 V
55 24 V
55 -70 V
55 70 V
55 0 V
55 0 V
55 0 V
56 48 V
55 -24 V
55 0 V
55 0 V
55 0 V
55 24 V
56 -24 V
55 0 V
55 24 V
55 0 V
55 26 V
55 -26 V
56 26 V
55 -50 V
55 50 V
55 -74 V
55 100 V
55 -52 V
55 0 V
56 -24 V
55 24 V
55 0 V
55 -24 V
55 24 V
55 0 V
56 -24 V
55 24 V
55 26 V
55 -74 V
55 0 V
55 24 V
56 0 V
55 50 V
55 -50 V
55 24 V
55 0 V
55 0 V
55 0 V
56 0 V
55 0 V
55 0 V
55 0 V
55 -24 V
55 -24 V
56 0 V
55 0 V
55 0 V
55 -47 V
55 -23 V
55 46 V
56 0 V
55 0 V
55 -23 V
55 23 V
955 693 M
55 95 V
55 94 V
56 90 V
55 93 V
55 92 V
55 98 V
55 87 V
55 90 V
56 88 V
1507 1610 L
55 86 V
55 81 V
55 78 V
55 81 V
56 92 V
55 95 V
55 95 V
55 52 V
55 85 V
55 61 V
55 101 V
56 18 V
55 95 V
55 62 V
55 21 V
55 90 V
55 24 V
56 74 V
55 52 V
55 55 V
55 57 V
55 30 V
55 61 V
56 0 V
55 32 V
55 101 V
55 -34 V
55 -34 V
55 104 V
55 -36 V
56 36 V
55 -70 V
55 34 V
55 36 V
55 36 V
55 -36 V
56 -36 V
55 36 V
55 0 V
55 0 V
55 36 V
55 0 V
56 -36 V
55 0 V
55 36 V
55 37 V
55 -37 V
55 -36 V
55 73 V
56 0 V
55 -37 V
55 37 V
55 0 V
55 0 V
55 0 V
56 -37 V
55 76 V
55 -39 V
55 -37 V
55 0 V
55 37 V
56 0 V
55 -37 V
55 0 V
55 37 V
55 0 V
55 78 V
55 -78 V
56 0 V
55 0 V
55 0 V
55 -37 V
55 76 V
55 79 V
56 -118 V
55 39 V
55 39 V
55 0 V
55 -78 V
55 39 V
56 0 V
55 -39 V
55 78 V
55 -39 V
955 694 M
55 95 V
55 93 V
56 93 V
55 93 V
55 93 V
55 92 V
55 87 V
55 95 V
56 81 V
55 99 V
55 69 V
55 73 V
55 76 V
55 103 V
56 92 V
55 41 V
55 88 V
55 73 V
55 53 V
2058 2401 L
55 64 V
56 88 V
55 57 V
55 20 V
55 105 V
55 68 V
55 98 V
56 0 V
55 107 V
55 57 V
55 0 V
55 91 V
55 65 V
56 0 V
55 140 V
55 76 V
55 79 V
55 42 V
55 -42 V
55 129 V
56 0 V
55 0 V
55 92 V
55 0 V
55 49 V
55 101 V
56 0 V
55 -101 V
55 101 V
55 -52 V
55 52 V
55 0 V
56 53 V
55 -53 V
55 0 V
55 0 V
55 108 V
55 -108 V
55 53 V
56 0 V
55 -53 V
55 0 V
55 -52 V
55 105 V
55 -53 V
56 0 V
55 0 V
55 53 V
55 -154 V
55 101 V
55 0 V
56 53 V
55 -53 V
55 0 V
55 0 V
55 0 V
55 -52 V
55 216 V
56 -111 V
55 0 V
55 -53 V
55 53 V
55 55 V
55 0 V
56 0 V
55 -108 V
55 53 V
55 0 V
55 55 V
55 -258 V
56 150 V
55 53 V
55 0 V
55 55 V
955 694 M
55 95 V
55 93 V
56 92 V
55 90 V
55 88 V
55 93 V
55 92 V
55 78 V
56 93 V
55 73 V
55 81 V
55 70 V
55 79 V
55 91 V
56 97 V
55 49 V
55 86 V
55 59 V
55 64 V
55 54 V
55 121 V
56 68 V
55 72 V
55 38 V
55 125 V
55 22 V
55 118 V
56 52 V
55 26 V
2610 3036 L
55 29 V
55 30 V
55 30 V
56 96 V
55 68 V
55 148 V
55 79 V
55 -40 V
55 0 V
55 82 V
56 0 V
55 0 V
55 133 V
55 -46 V
55 0 V
55 46 V
56 95 V
55 49 V
55 52 V
55 0 V
55 53 V
55 170 V
56 60 V
55 0 V
55 -60 V
55 60 V
55 0 V
55 -60 V
55 60 V
56 -60 V
55 0 V
55 60 V
55 0 V
55 -60 V
55 0 V
56 60 V
55 -60 V
55 0 V
55 0 V
55 -59 V
55 119 V
56 -60 V
55 60 V
55 0 V
55 0 V
55 0 V
55 -175 V
55 115 V
56 123 V
55 -63 V
55 0 V
55 0 V
55 0 V
55 0 V
56 63 V
53 63 V
2 0 R
55 -126 V
54 126 V
2 0 R
54 -126 V
55 -60 V
56 0 V
55 123 V
55 0 V
55 -123 V
955 694 M
55 95 V
55 92 V
56 92 V
55 89 V
55 98 V
55 91 V
55 78 V
55 89 V
56 82 V
55 77 V
55 85 V
55 76 V
55 66 V
55 74 V
56 67 V
55 83 V
55 51 V
55 66 V
55 73 V
55 93 V
55 0 V
56 138 V
55 0 V
55 123 V
55 19 V
55 80 V
55 86 V
56 0 V
55 23 V
55 121 V
55 0 V
55 79 V
55 28 V
56 57 V
55 260 V
55 0 V
55 -70 V
3051 3398 L
55 78 V
55 -78 V
56 -37 V
55 76 V
55 -112 V
55 36 V
55 76 V
55 -112 V
56 36 V
55 115 V
55 -39 V
55 39 V
55 0 V
55 -39 V
56 39 V
55 -39 V
55 39 V
55 0 V
55 -39 V
55 -39 V
55 118 V
56 -40 V
55 -39 V
55 39 V
55 -39 V
55 39 V
55 -39 V
56 79 V
55 -40 V
55 -39 V
55 0 V
55 0 V
55 39 V
56 0 V
55 0 V
55 0 V
55 -39 V
55 39 V
55 -78 V
55 39 V
56 -39 V
55 78 V
55 -39 V
55 0 V
55 39 V
55 0 V
56 40 V
55 -79 V
55 0 V
55 0 V
55 0 V
55 0 V
56 39 V
55 0 V
55 -39 V
55 0 V
955 694 M
55 94 V
55 91 V
56 92 V
55 92 V
55 89 V
55 87 V
55 82 V
55 85 V
56 87 V
55 79 V
55 79 V
55 68 V
55 51 V
55 100 V
56 75 V
55 93 V
55 41 V
55 102 V
55 62 V
55 68 V
55 -81 V
56 53 V
55 149 V
55 0 V
55 159 V
55 19 V
55 61 V
56 -61 V
55 61 V
55 -80 V
55 212 V
55 -111 V
55 43 V
56 45 V
55 -45 V
55 0 V
55 -43 V
55 88 V
55 0 V
55 71 V
56 -24 V
55 -92 V
55 140 V
55 -72 V
55 0 V
55 24 V
56 -24 V
3603 2827 L
55 0 V
55 0 V
55 24 V
55 -24 V
56 0 V
55 -24 V
55 24 V
55 -24 V
55 -46 V
55 46 V
55 0 V
56 0 V
55 48 V
55 0 V
55 -48 V
55 24 V
55 -24 V
56 0 V
55 -23 V
55 47 V
55 -24 V
55 24 V
55 0 V
56 0 V
55 0 V
55 -24 V
55 24 V
55 0 V
55 0 V
55 -24 V
56 24 V
55 -24 V
55 24 V
55 0 V
55 -24 V
55 -23 V
56 47 V
55 0 V
55 -24 V
55 0 V
55 24 V
55 -47 V
56 23 V
55 -23 V
55 23 V
55 -23 V
955 694 M
55 95 V
55 92 V
56 91 V
55 90 V
55 90 V
55 85 V
55 92 V
55 80 V
56 87 V
55 63 V
55 61 V
55 70 V
55 60 V
55 120 V
56 83 V
55 56 V
55 29 V
55 -29 V
55 81 V
55 44 V
55 35 V
56 88 V
55 -27 V
55 40 V
55 13 V
55 28 V
55 15 V
56 -43 V
55 72 V
55 -29 V
55 29 V
55 15 V
55 79 V
56 -48 V
55 31 V
55 0 V
55 0 V
55 0 V
55 17 V
55 -33 V
56 16 V
55 17 V
55 -33 V
55 0 V
55 33 V
55 -17 V
56 -31 V
55 48 V
55 -17 V
55 0 V
55 0 V
55 0 V
56 0 V
55 -16 V
55 16 V
55 -16 V
55 33 V
4154 2401 L
55 15 V
56 16 V
55 -16 V
55 0 V
55 0 V
55 0 V
55 0 V
56 0 V
55 49 V
55 -33 V
55 0 V
55 0 V
55 0 V
56 -16 V
55 16 V
55 0 V
55 -16 V
55 16 V
55 17 V
55 -33 V
56 0 V
55 16 V
55 0 V
55 0 V
55 -16 V
55 33 V
56 -17 V
55 0 V
55 0 V
55 17 V
55 -48 V
55 0 V
56 31 V
55 -16 V
55 0 V
55 0 V
stroke
LTb
900 4296 M
900 600 L
5240 0 V
0 3696 V
-5240 0 V
0.800 UP
stroke
grestore
end
showpage
  }}%
  \put(6333,2416){\makebox(0,0){\bf 26}}%
  \put(6333,4213){\makebox(0,0){\bf 23}}%
  \put(6333,3995){\makebox(0,0){\bf 22}}%
  \put(6499,3437){\makebox(0,0){\bf 21,24}}%
  \put(6499,2803){\makebox(0,0){\bf 20,25}}%
  \put(3520,100){\makebox(0,0){\strut{}nodes $p$}}%
  \put(200,2448){%
  \special{ps: gsave currentpoint currentpoint translate
270 rotate neg exch neg exch translate}%
  \makebox(0,0){\strut{}speedup $T(1)/T(p)$}%
  \special{ps: currentpoint grestore moveto}%
  }%
  \put(5809,400){\makebox(0,0){\strut{} 90}}%
  \put(5257,400){\makebox(0,0){\strut{} 80}}%
  \put(4706,400){\makebox(0,0){\strut{} 70}}%
  \put(4154,400){\makebox(0,0){\strut{} 60}}%
  \put(3603,400){\makebox(0,0){\strut{} 50}}%
  \put(3051,400){\makebox(0,0){\strut{} 40}}%
  \put(2500,400){\makebox(0,0){\strut{} 30}}%
  \put(1948,400){\makebox(0,0){\strut{} 20}}%
  \put(1396,400){\makebox(0,0){\strut{} 10}}%
  \put(780,4296){\makebox(0,0)[r]{\strut{} 40}}%
  \put(780,3834){\makebox(0,0)[r]{\strut{} 35}}%
  \put(780,3372){\makebox(0,0)[r]{\strut{} 30}}%
  \put(780,2910){\makebox(0,0)[r]{\strut{} 25}}%
  \put(780,2448){\makebox(0,0)[r]{\strut{} 20}}%
  \put(780,1986){\makebox(0,0)[r]{\strut{} 15}}%
  \put(780,1524){\makebox(0,0)[r]{\strut{} 10}}%
  \put(780,1062){\makebox(0,0)[r]{\strut{} 5}}%
  \put(780,600){\makebox(0,0)[r]{\strut{} 0}}%
\end{picture}%
\endgroup
 

%% file: speedup15.tex
\begingroup%
\makeatletter%
\newcommand{\GNUPLOTspecial}{%
  \@sanitize\catcode`\%=14\relax\special}%
\setlength{\unitlength}{0.0500bp}%
\begin{picture}(6479,4535)(0,0)%
  {\GNUPLOTspecial{"
/gnudict 256 dict def
gnudict begin
%
%
/Color false def
/Blacktext true def
/Solid false def
/Dashlength 1 def
/Landscape false def
/Level1 false def
/Rounded false def
/TransparentPatterns false def
/gnulinewidth 5.000 def
/userlinewidth gnulinewidth def
/vshift -66 def
/dl1 {
  10.0 Dashlength mul mul
  Rounded { currentlinewidth 0.75 mul sub dup 0 le { pop 0.01 } if } if
} def
/dl2 {
  10.0 Dashlength mul mul
  Rounded { currentlinewidth 0.75 mul add } if
} def
/hpt_ 31.5 def
/vpt_ 31.5 def
/hpt hpt_ def
/vpt vpt_ def
Level1 {} {
/SDict 10 dict def
systemdict /pdfmark known not {
  userdict /pdfmark systemdict /cleartomark get put
} if
SDict begin [
  /Title (speedup15.tex)
  /Subject (gnuplot plot)
  /Creator (gnuplot 4.2 patchlevel 2 )
  /Author (n,,,)
  /CreationDate (Wed Sep 22 15:37:11 2010)
  /DOCINFO pdfmark
end
} ifelse
%
%
/M {moveto} bind def
/L {lineto} bind def
/R {rmoveto} bind def
/V {rlineto} bind def
/N {newpath moveto} bind def
/Z {closepath} bind def
/C {setrgbcolor} bind def
/f {rlineto fill} bind def
/vpt2 vpt 2 mul def
/hpt2 hpt 2 mul def
/Lshow {currentpoint stroke M 0 vshift R 
	Blacktext {gsave 0 setgray show grestore} {show} ifelse} def
/Rshow {currentpoint stroke M dup stringwidth pop neg vshift R
	Blacktext {gsave 0 setgray show grestore} {show} ifelse} def
/Cshow {currentpoint stroke M dup stringwidth pop -2 div vshift R 
	Blacktext {gsave 0 setgray show grestore} {show} ifelse} def
/UP {dup vpt_ mul /vpt exch def hpt_ mul /hpt exch def
  /hpt2 hpt 2 mul def /vpt2 vpt 2 mul def} def
/DL {Color {setrgbcolor Solid {pop []} if 0 setdash}
 {pop pop pop 0 setgray Solid {pop []} if 0 setdash} ifelse} def
/BL {stroke userlinewidth 2 mul setlinewidth
	Rounded {1 setlinejoin 1 setlinecap} if} def
/AL {stroke userlinewidth 2 div setlinewidth
	Rounded {1 setlinejoin 1 setlinecap} if} def
/UL {dup gnulinewidth mul /userlinewidth exch def
	dup 1 lt {pop 1} if 10 mul /udl exch def} def
/PL {stroke userlinewidth setlinewidth
	Rounded {1 setlinejoin 1 setlinecap} if} def
/LCw {1 1 1} def
/LCb {0 0 0} def
/LCa {0 0 0} def
/LC0 {1 0 0} def
/LC1 {0 1 0} def
/LC2 {0 0 1} def
/LC3 {1 0 1} def
/LC4 {0 1 1} def
/LC5 {1 1 0} def
/LC6 {0 0 0} def
/LC7 {1 0.3 0} def
/LC8 {0.5 0.5 0.5} def
/LTw {PL [] 1 setgray} def
/LTb {BL [] LCb DL} def
/LTa {AL [1 udl mul 2 udl mul] 0 setdash LCa setrgbcolor} def
/LT0 {PL [] LC0 DL} def
/LT1 {PL [4 dl1 2 dl2] LC1 DL} def
/LT2 {PL [2 dl1 3 dl2] LC2 DL} def
/LT3 {PL [1 dl1 1.5 dl2] LC3 DL} def
/LT4 {PL [6 dl1 2 dl2 1 dl1 2 dl2] LC4 DL} def
/LT5 {PL [3 dl1 3 dl2 1 dl1 3 dl2] LC5 DL} def
/LT6 {PL [2 dl1 2 dl2 2 dl1 6 dl2] LC6 DL} def
/LT7 {PL [1 dl1 2 dl2 6 dl1 2 dl2 1 dl1 2 dl2] LC7 DL} def
/LT8 {PL [2 dl1 2 dl2 2 dl1 2 dl2 2 dl1 2 dl2 2 dl1 4 dl2] LC8 DL} def
/Pnt {stroke [] 0 setdash gsave 1 setlinecap M 0 0 V stroke grestore} def
/Dia {stroke [] 0 setdash 2 copy vpt add M
  hpt neg vpt neg V hpt vpt neg V
  hpt vpt V hpt neg vpt V closepath stroke
  Pnt} def
/Pls {stroke [] 0 setdash vpt sub M 0 vpt2 V
  currentpoint stroke M
  hpt neg vpt neg R hpt2 0 V stroke
 } def
/Box {stroke [] 0 setdash 2 copy exch hpt sub exch vpt add M
  0 vpt2 neg V hpt2 0 V 0 vpt2 V
  hpt2 neg 0 V closepath stroke
  Pnt} def
/Crs {stroke [] 0 setdash exch hpt sub exch vpt add M
  hpt2 vpt2 neg V currentpoint stroke M
  hpt2 neg 0 R hpt2 vpt2 V stroke} def
/TriU {stroke [] 0 setdash 2 copy vpt 1.12 mul add M
  hpt neg vpt -1.62 mul V
  hpt 2 mul 0 V
  hpt neg vpt 1.62 mul V closepath stroke
  Pnt} def
/Star {2 copy Pls Crs} def
/BoxF {stroke [] 0 setdash exch hpt sub exch vpt add M
  0 vpt2 neg V hpt2 0 V 0 vpt2 V
  hpt2 neg 0 V closepath fill} def
/TriUF {stroke [] 0 setdash vpt 1.12 mul add M
  hpt neg vpt -1.62 mul V
  hpt 2 mul 0 V
  hpt neg vpt 1.62 mul V closepath fill} def
/TriD {stroke [] 0 setdash 2 copy vpt 1.12 mul sub M
  hpt neg vpt 1.62 mul V
  hpt 2 mul 0 V
  hpt neg vpt -1.62 mul V closepath stroke
  Pnt} def
/TriDF {stroke [] 0 setdash vpt 1.12 mul sub M
  hpt neg vpt 1.62 mul V
  hpt 2 mul 0 V
  hpt neg vpt -1.62 mul V closepath fill} def
/DiaF {stroke [] 0 setdash vpt add M
  hpt neg vpt neg V hpt vpt neg V
  hpt vpt V hpt neg vpt V closepath fill} def
/Pent {stroke [] 0 setdash 2 copy gsave
  translate 0 hpt M 4 {72 rotate 0 hpt L} repeat
  closepath stroke grestore Pnt} def
/PentF {stroke [] 0 setdash gsave
  translate 0 hpt M 4 {72 rotate 0 hpt L} repeat
  closepath fill grestore} def
/Circle {stroke [] 0 setdash 2 copy
  hpt 0 360 arc stroke Pnt} def
/CircleF {stroke [] 0 setdash hpt 0 360 arc fill} def
/C0 {BL [] 0 setdash 2 copy moveto vpt 90 450 arc} bind def
/C1 {BL [] 0 setdash 2 copy moveto
	2 copy vpt 0 90 arc closepath fill
	vpt 0 360 arc closepath} bind def
/C2 {BL [] 0 setdash 2 copy moveto
	2 copy vpt 90 180 arc closepath fill
	vpt 0 360 arc closepath} bind def
/C3 {BL [] 0 setdash 2 copy moveto
	2 copy vpt 0 180 arc closepath fill
	vpt 0 360 arc closepath} bind def
/C4 {BL [] 0 setdash 2 copy moveto
	2 copy vpt 180 270 arc closepath fill
	vpt 0 360 arc closepath} bind def
/C5 {BL [] 0 setdash 2 copy moveto
	2 copy vpt 0 90 arc
	2 copy moveto
	2 copy vpt 180 270 arc closepath fill
	vpt 0 360 arc} bind def
/C6 {BL [] 0 setdash 2 copy moveto
	2 copy vpt 90 270 arc closepath fill
	vpt 0 360 arc closepath} bind def
/C7 {BL [] 0 setdash 2 copy moveto
	2 copy vpt 0 270 arc closepath fill
	vpt 0 360 arc closepath} bind def
/C8 {BL [] 0 setdash 2 copy moveto
	2 copy vpt 270 360 arc closepath fill
	vpt 0 360 arc closepath} bind def
/C9 {BL [] 0 setdash 2 copy moveto
	2 copy vpt 270 450 arc closepath fill
	vpt 0 360 arc closepath} bind def
/C10 {BL [] 0 setdash 2 copy 2 copy moveto vpt 270 360 arc closepath fill
	2 copy moveto
	2 copy vpt 90 180 arc closepath fill
	vpt 0 360 arc closepath} bind def
/C11 {BL [] 0 setdash 2 copy moveto
	2 copy vpt 0 180 arc closepath fill
	2 copy moveto
	2 copy vpt 270 360 arc closepath fill
	vpt 0 360 arc closepath} bind def
/C12 {BL [] 0 setdash 2 copy moveto
	2 copy vpt 180 360 arc closepath fill
	vpt 0 360 arc closepath} bind def
/C13 {BL [] 0 setdash 2 copy moveto
	2 copy vpt 0 90 arc closepath fill
	2 copy moveto
	2 copy vpt 180 360 arc closepath fill
	vpt 0 360 arc closepath} bind def
/C14 {BL [] 0 setdash 2 copy moveto
	2 copy vpt 90 360 arc closepath fill
	vpt 0 360 arc} bind def
/C15 {BL [] 0 setdash 2 copy vpt 0 360 arc closepath fill
	vpt 0 360 arc closepath} bind def
/Rec {newpath 4 2 roll moveto 1 index 0 rlineto 0 exch rlineto
	neg 0 rlineto closepath} bind def
/Square {dup Rec} bind def
/Bsquare {vpt sub exch vpt sub exch vpt2 Square} bind def
/S0 {BL [] 0 setdash 2 copy moveto 0 vpt rlineto BL Bsquare} bind def
/S1 {BL [] 0 setdash 2 copy vpt Square fill Bsquare} bind def
/S2 {BL [] 0 setdash 2 copy exch vpt sub exch vpt Square fill Bsquare} bind def
/S3 {BL [] 0 setdash 2 copy exch vpt sub exch vpt2 vpt Rec fill Bsquare} bind def
/S4 {BL [] 0 setdash 2 copy exch vpt sub exch vpt sub vpt Square fill Bsquare} bind def
/S5 {BL [] 0 setdash 2 copy 2 copy vpt Square fill
	exch vpt sub exch vpt sub vpt Square fill Bsquare} bind def
/S6 {BL [] 0 setdash 2 copy exch vpt sub exch vpt sub vpt vpt2 Rec fill Bsquare} bind def
/S7 {BL [] 0 setdash 2 copy exch vpt sub exch vpt sub vpt vpt2 Rec fill
	2 copy vpt Square fill Bsquare} bind def
/S8 {BL [] 0 setdash 2 copy vpt sub vpt Square fill Bsquare} bind def
/S9 {BL [] 0 setdash 2 copy vpt sub vpt vpt2 Rec fill Bsquare} bind def
/S10 {BL [] 0 setdash 2 copy vpt sub vpt Square fill 2 copy exch vpt sub exch vpt Square fill
	Bsquare} bind def
/S11 {BL [] 0 setdash 2 copy vpt sub vpt Square fill 2 copy exch vpt sub exch vpt2 vpt Rec fill
	Bsquare} bind def
/S12 {BL [] 0 setdash 2 copy exch vpt sub exch vpt sub vpt2 vpt Rec fill Bsquare} bind def
/S13 {BL [] 0 setdash 2 copy exch vpt sub exch vpt sub vpt2 vpt Rec fill
	2 copy vpt Square fill Bsquare} bind def
/S14 {BL [] 0 setdash 2 copy exch vpt sub exch vpt sub vpt2 vpt Rec fill
	2 copy exch vpt sub exch vpt Square fill Bsquare} bind def
/S15 {BL [] 0 setdash 2 copy Bsquare fill Bsquare} bind def
/D0 {gsave translate 45 rotate 0 0 S0 stroke grestore} bind def
/D1 {gsave translate 45 rotate 0 0 S1 stroke grestore} bind def
/D2 {gsave translate 45 rotate 0 0 S2 stroke grestore} bind def
/D3 {gsave translate 45 rotate 0 0 S3 stroke grestore} bind def
/D4 {gsave translate 45 rotate 0 0 S4 stroke grestore} bind def
/D5 {gsave translate 45 rotate 0 0 S5 stroke grestore} bind def
/D6 {gsave translate 45 rotate 0 0 S6 stroke grestore} bind def
/D7 {gsave translate 45 rotate 0 0 S7 stroke grestore} bind def
/D8 {gsave translate 45 rotate 0 0 S8 stroke grestore} bind def
/D9 {gsave translate 45 rotate 0 0 S9 stroke grestore} bind def
/D10 {gsave translate 45 rotate 0 0 S10 stroke grestore} bind def
/D11 {gsave translate 45 rotate 0 0 S11 stroke grestore} bind def
/D12 {gsave translate 45 rotate 0 0 S12 stroke grestore} bind def
/D13 {gsave translate 45 rotate 0 0 S13 stroke grestore} bind def
/D14 {gsave translate 45 rotate 0 0 S14 stroke grestore} bind def
/D15 {gsave translate 45 rotate 0 0 S15 stroke grestore} bind def
/DiaE {stroke [] 0 setdash vpt add M
  hpt neg vpt neg V hpt vpt neg V
  hpt vpt V hpt neg vpt V closepath stroke} def
/BoxE {stroke [] 0 setdash exch hpt sub exch vpt add M
  0 vpt2 neg V hpt2 0 V 0 vpt2 V
  hpt2 neg 0 V closepath stroke} def
/TriUE {stroke [] 0 setdash vpt 1.12 mul add M
  hpt neg vpt -1.62 mul V
  hpt 2 mul 0 V
  hpt neg vpt 1.62 mul V closepath stroke} def
/TriDE {stroke [] 0 setdash vpt 1.12 mul sub M
  hpt neg vpt 1.62 mul V
  hpt 2 mul 0 V
  hpt neg vpt -1.62 mul V closepath stroke} def
/PentE {stroke [] 0 setdash gsave
  translate 0 hpt M 4 {72 rotate 0 hpt L} repeat
  closepath stroke grestore} def
/CircE {stroke [] 0 setdash 
  hpt 0 360 arc stroke} def
/Opaque {gsave closepath 1 setgray fill grestore 0 setgray closepath} def
/DiaW {stroke [] 0 setdash vpt add M
  hpt neg vpt neg V hpt vpt neg V
  hpt vpt V hpt neg vpt V Opaque stroke} def
/BoxW {stroke [] 0 setdash exch hpt sub exch vpt add M
  0 vpt2 neg V hpt2 0 V 0 vpt2 V
  hpt2 neg 0 V Opaque stroke} def
/TriUW {stroke [] 0 setdash vpt 1.12 mul add M
  hpt neg vpt -1.62 mul V
  hpt 2 mul 0 V
  hpt neg vpt 1.62 mul V Opaque stroke} def
/TriDW {stroke [] 0 setdash vpt 1.12 mul sub M
  hpt neg vpt 1.62 mul V
  hpt 2 mul 0 V
  hpt neg vpt -1.62 mul V Opaque stroke} def
/PentW {stroke [] 0 setdash gsave
  translate 0 hpt M 4 {72 rotate 0 hpt L} repeat
  Opaque stroke grestore} def
/CircW {stroke [] 0 setdash 
  hpt 0 360 arc Opaque stroke} def
/BoxFill {gsave Rec 1 setgray fill grestore} def
/Density {
  /Fillden exch def
  currentrgbcolor
  /ColB exch def /ColG exch def /ColR exch def
  /ColR ColR Fillden mul Fillden sub 1 add def
  /ColG ColG Fillden mul Fillden sub 1 add def
  /ColB ColB Fillden mul Fillden sub 1 add def
  ColR ColG ColB setrgbcolor} def
/BoxColFill {gsave Rec PolyFill} def
/PolyFill {gsave Density fill grestore grestore} def
/h {rlineto rlineto rlineto gsave fill grestore} bind def
%
%
/PatternFill {gsave /PFa [ 9 2 roll ] def
  PFa 0 get PFa 2 get 2 div add PFa 1 get PFa 3 get 2 div add translate
  PFa 2 get -2 div PFa 3 get -2 div PFa 2 get PFa 3 get Rec
  gsave 1 setgray fill grestore clip
  currentlinewidth 0.5 mul setlinewidth
  /PFs PFa 2 get dup mul PFa 3 get dup mul add sqrt def
  0 0 M PFa 5 get rotate PFs -2 div dup translate
  0 1 PFs PFa 4 get div 1 add floor cvi
	{PFa 4 get mul 0 M 0 PFs V} for
  0 PFa 6 get ne {
	0 1 PFs PFa 4 get div 1 add floor cvi
	{PFa 4 get mul 0 2 1 roll M PFs 0 V} for
 } if
  stroke grestore} def
/languagelevel where
 {pop languagelevel} {1} ifelse
 2 lt
	{/InterpretLevel1 true def}
	{/InterpretLevel1 Level1 def}
 ifelse
%
%
/Level2PatternFill {
/Tile8x8 {/PaintType 2 /PatternType 1 /TilingType 1 /BBox [0 0 8 8] /XStep 8 /YStep 8}
	bind def
/KeepColor {currentrgbcolor [/Pattern /DeviceRGB] setcolorspace} bind def
<< Tile8x8
 /PaintProc {0.5 setlinewidth pop 0 0 M 8 8 L 0 8 M 8 0 L stroke} 
>> matrix makepattern
/Pat1 exch def
<< Tile8x8
 /PaintProc {0.5 setlinewidth pop 0 0 M 8 8 L 0 8 M 8 0 L stroke
	0 4 M 4 8 L 8 4 L 4 0 L 0 4 L stroke}
>> matrix makepattern
/Pat2 exch def
<< Tile8x8
 /PaintProc {0.5 setlinewidth pop 0 0 M 0 8 L
	8 8 L 8 0 L 0 0 L fill}
>> matrix makepattern
/Pat3 exch def
<< Tile8x8
 /PaintProc {0.5 setlinewidth pop -4 8 M 8 -4 L
	0 12 M 12 0 L stroke}
>> matrix makepattern
/Pat4 exch def
<< Tile8x8
 /PaintProc {0.5 setlinewidth pop -4 0 M 8 12 L
	0 -4 M 12 8 L stroke}
>> matrix makepattern
/Pat5 exch def
<< Tile8x8
 /PaintProc {0.5 setlinewidth pop -2 8 M 4 -4 L
	0 12 M 8 -4 L 4 12 M 10 0 L stroke}
>> matrix makepattern
/Pat6 exch def
<< Tile8x8
 /PaintProc {0.5 setlinewidth pop -2 0 M 4 12 L
	0 -4 M 8 12 L 4 -4 M 10 8 L stroke}
>> matrix makepattern
/Pat7 exch def
<< Tile8x8
 /PaintProc {0.5 setlinewidth pop 8 -2 M -4 4 L
	12 0 M -4 8 L 12 4 M 0 10 L stroke}
>> matrix makepattern
/Pat8 exch def
<< Tile8x8
 /PaintProc {0.5 setlinewidth pop 0 -2 M 12 4 L
	-4 0 M 12 8 L -4 4 M 8 10 L stroke}
>> matrix makepattern
/Pat9 exch def
/Pattern1 {PatternBgnd KeepColor Pat1 setpattern} bind def
/Pattern2 {PatternBgnd KeepColor Pat2 setpattern} bind def
/Pattern3 {PatternBgnd KeepColor Pat3 setpattern} bind def
/Pattern4 {PatternBgnd KeepColor Landscape {Pat5} {Pat4} ifelse setpattern} bind def
/Pattern5 {PatternBgnd KeepColor Landscape {Pat4} {Pat5} ifelse setpattern} bind def
/Pattern6 {PatternBgnd KeepColor Landscape {Pat9} {Pat6} ifelse setpattern} bind def
/Pattern7 {PatternBgnd KeepColor Landscape {Pat8} {Pat7} ifelse setpattern} bind def
} def
%
%
%
/PatternBgnd {
  TransparentPatterns {} {gsave 1 setgray fill grestore} ifelse
} def
%
%
/Level1PatternFill {
/Pattern1 {0.250 Density} bind def
/Pattern2 {0.500 Density} bind def
/Pattern3 {0.750 Density} bind def
/Pattern4 {0.125 Density} bind def
/Pattern5 {0.375 Density} bind def
/Pattern6 {0.625 Density} bind def
/Pattern7 {0.875 Density} bind def
} def
%
%
Level1 {Level1PatternFill} {Level2PatternFill} ifelse
/Symbol-Oblique /Symbol findfont [1 0 .167 1 0 0] makefont
dup length dict begin {1 index /FID eq {pop pop} {def} ifelse} forall
currentdict end definefont pop
end
gnudict begin
gsave
0 0 translate
0.050 0.050 scale
0 setgray
newpath
1.000 UL
LTb
900 600 M
63 0 V
5177 0 R
-63 0 V
900 989 M
63 0 V
5177 0 R
-63 0 V
900 1378 M
63 0 V
5177 0 R
-63 0 V
900 1767 M
63 0 V
5177 0 R
-63 0 V
900 2156 M
63 0 V
5177 0 R
-63 0 V
900 2545 M
63 0 V
5177 0 R
-63 0 V
900 2934 M
63 0 V
5177 0 R
-63 0 V
900 3323 M
63 0 V
5177 0 R
-63 0 V
900 3712 M
63 0 V
5177 0 R
-63 0 V
900 4101 M
63 0 V
5177 0 R
-63 0 V
1396 600 M
0 63 V
0 3633 R
0 -63 V
1948 600 M
0 63 V
0 3633 R
0 -63 V
2500 600 M
0 63 V
0 3633 R
0 -63 V
3051 600 M
0 63 V
0 3633 R
0 -63 V
3603 600 M
0 63 V
0 3633 R
0 -63 V
4154 600 M
0 63 V
0 3633 R
0 -63 V
4706 600 M
0 63 V
0 3633 R
0 -63 V
5257 600 M
0 63 V
0 3633 R
0 -63 V
5809 600 M
0 63 V
0 3633 R
0 -63 V
900 4296 M
900 600 L
5240 0 V
0 3696 V
-5240 0 V
stroke
LCb setrgbcolor
LTb
LCb setrgbcolor
LTb
0.800 UP
1.000 UL
LTb
1.000 UL
LT2
900 600 M
53 37 V
53 38 V
53 37 V
53 37 V
53 38 V
53 37 V
53 37 V
52 38 V
53 37 V
53 37 V
53 38 V
53 37 V
53 37 V
53 38 V
53 37 V
53 37 V
53 38 V
53 37 V
53 37 V
53 38 V
53 37 V
52 37 V
53 38 V
53 37 V
53 37 V
53 38 V
53 37 V
53 37 V
53 38 V
53 37 V
53 37 V
53 38 V
53 37 V
53 37 V
53 38 V
52 37 V
53 37 V
53 38 V
53 37 V
53 37 V
53 38 V
53 37 V
53 37 V
53 38 V
53 37 V
53 37 V
53 38 V
53 37 V
53 37 V
52 38 V
53 37 V
53 37 V
53 38 V
53 37 V
53 37 V
53 38 V
53 37 V
53 37 V
53 38 V
53 37 V
53 37 V
53 38 V
53 37 V
52 37 V
53 38 V
53 37 V
53 37 V
53 38 V
53 37 V
53 37 V
53 38 V
53 37 V
53 37 V
53 38 V
53 37 V
53 37 V
53 38 V
52 37 V
53 37 V
53 38 V
53 37 V
53 37 V
53 38 V
53 37 V
53 37 V
53 38 V
53 37 V
53 37 V
53 38 V
53 37 V
53 37 V
52 38 V
53 37 V
53 37 V
53 38 V
53 37 V
53 37 V
53 38 V
53 37 V
stroke
LT0
955 636 M
55 36 V
55 36 V
56 36 V
55 36 V
55 35 V
55 36 V
55 33 V
55 33 V
56 31 V
55 28 V
55 24 V
55 20 V
55 16 V
55 10 V
56 20 V
55 13 V
55 8 V
55 8 V
55 4 V
55 4 V
55 0 V
56 1 V
55 1 V
55 1 V
55 0 V
55 0 V
55 0 V
56 1 V
55 0 V
55 -1 V
55 1 V
55 0 V
55 -1 V
56 1 V
55 -1 V
55 0 V
55 0 V
55 0 V
55 0 V
55 0 V
56 -1 V
55 0 V
55 0 V
55 0 V
55 0 V
55 0 V
56 -1 V
55 1 V
55 -1 V
55 1 V
55 -1 V
55 -1 V
56 1 V
55 0 V
55 0 V
55 -1 V
55 1 V
55 -1 V
55 0 V
56 0 V
55 0 V
55 0 V
55 0 V
55 -1 V
55 0 V
56 -1 V
55 0 V
55 0 V
55 0 V
55 0 V
55 -1 V
56 0 V
55 0 V
55 0 V
55 -1 V
55 0 V
55 0 V
55 0 V
56 -1 V
55 0 V
55 0 V
55 0 V
55 0 V
55 0 V
56 -1 V
55 0 V
55 -1 V
55 0 V
55 1 V
55 -1 V
56 1 V
55 -1 V
55 0 V
55 0 V
955 638 M
55 38 V
55 39 V
56 38 V
55 39 V
55 38 V
55 39 V
55 37 V
55 37 V
56 38 V
1507 1017 L
55 36 V
55 34 V
55 34 V
55 32 V
56 34 V
55 34 V
55 32 V
55 31 V
55 31 V
55 30 V
55 27 V
56 27 V
55 26 V
55 28 V
55 23 V
55 25 V
55 23 V
56 22 V
55 20 V
55 18 V
55 20 V
55 13 V
55 20 V
56 15 V
55 14 V
55 18 V
55 11 V
55 13 V
55 16 V
55 9 V
56 14 V
55 11 V
55 12 V
55 15 V
55 9 V
55 9 V
56 16 V
55 12 V
55 7 V
55 17 V
55 9 V
55 6 V
56 13 V
55 5 V
55 11 V
55 9 V
55 10 V
55 6 V
55 7 V
56 5 V
55 6 V
55 5 V
55 2 V
55 2 V
55 1 V
56 5 V
55 1 V
55 2 V
55 2 V
55 1 V
55 1 V
56 0 V
55 1 V
55 1 V
55 1 V
55 -3 V
55 1 V
55 0 V
56 -1 V
55 -2 V
55 1 V
55 -1 V
55 -1 V
55 0 V
56 -1 V
55 0 V
55 0 V
55 0 V
55 0 V
55 -2 V
56 0 V
55 -1 V
55 1 V
55 -2 V
955 639 M
55 39 V
55 40 V
56 39 V
55 39 V
55 39 V
55 39 V
55 39 V
55 39 V
56 39 V
55 39 V
55 38 V
55 37 V
55 37 V
55 38 V
56 40 V
55 38 V
55 37 V
55 36 V
55 35 V
2058 1401 L
55 36 V
56 34 V
55 37 V
55 34 V
55 31 V
55 34 V
55 30 V
56 34 V
55 36 V
55 30 V
55 36 V
55 29 V
55 34 V
56 36 V
55 21 V
55 32 V
55 30 V
55 32 V
55 22 V
55 39 V
56 21 V
55 29 V
55 38 V
55 13 V
55 43 V
55 20 V
56 17 V
55 20 V
55 40 V
55 33 V
55 20 V
55 19 V
56 25 V
55 28 V
55 24 V
55 53 V
55 -5 V
55 18 V
55 32 V
56 23 V
55 23 V
55 29 V
55 26 V
55 38 V
55 -4 V
56 49 V
55 -1 V
55 42 V
55 18 V
55 51 V
55 -8 V
56 41 V
55 -6 V
55 40 V
55 28 V
55 1 V
55 36 V
55 52 V
56 21 V
55 8 V
55 -3 V
55 63 V
55 -2 V
55 19 V
56 41 V
55 6 V
55 21 V
55 -2 V
55 48 V
55 8 V
56 23 V
55 38 V
55 -19 V
55 77 V
955 639 M
55 39 V
55 39 V
56 39 V
55 39 V
55 39 V
55 39 V
55 39 V
55 40 V
56 39 V
55 38 V
55 38 V
55 41 V
55 37 V
55 37 V
56 40 V
55 38 V
55 39 V
55 37 V
55 40 V
55 38 V
55 38 V
56 38 V
55 39 V
55 37 V
55 35 V
55 40 V
55 37 V
56 38 V
55 36 V
2610 1788 L
55 36 V
55 47 V
55 29 V
56 45 V
55 31 V
55 37 V
55 38 V
55 31 V
55 44 V
55 32 V
56 32 V
55 39 V
55 22 V
55 35 V
55 49 V
55 47 V
56 24 V
55 26 V
55 45 V
55 28 V
55 36 V
55 50 V
56 38 V
55 8 V
55 43 V
55 31 V
55 19 V
55 48 V
55 29 V
56 44 V
55 6 V
55 58 V
55 40 V
55 20 V
55 51 V
56 26 V
55 22 V
55 42 V
55 46 V
55 -17 V
55 49 V
56 41 V
55 52 V
55 -16 V
55 73 V
55 -16 V
55 55 V
55 4 V
56 44 V
55 66 V
55 23 V
55 -2 V
55 25 V
55 11 V
56 114 V
55 -29 V
55 41 V
55 36 V
55 31 V
55 79 V
56 -84 V
55 77 V
55 10 V
55 57 V
955 639 M
55 38 V
55 39 V
56 38 V
55 39 V
55 39 V
55 38 V
55 38 V
55 39 V
56 39 V
55 35 V
55 36 V
55 40 V
55 34 V
55 41 V
56 39 V
55 42 V
55 38 V
55 38 V
55 35 V
55 34 V
55 38 V
56 37 V
55 38 V
55 42 V
55 29 V
55 36 V
55 46 V
56 41 V
55 35 V
55 19 V
55 44 V
55 26 V
55 18 V
56 64 V
55 56 V
55 3 V
55 31 V
55 56 V
55 41 V
3161 2126 L
56 43 V
55 -8 V
55 71 V
55 42 V
55 61 V
55 11 V
56 7 V
55 55 V
55 42 V
55 6 V
55 65 V
55 -26 V
56 71 V
55 53 V
55 -28 V
55 -9 V
55 121 V
55 49 V
55 43 V
56 18 V
55 -33 V
55 108 V
55 -33 V
55 25 V
55 12 V
56 82 V
55 22 V
55 18 V
55 19 V
55 -22 V
55 146 V
56 5 V
55 -126 V
55 197 V
55 14 V
55 -54 V
55 25 V
55 -3 V
56 98 V
55 23 V
55 -121 V
55 154 V
55 -67 V
55 -40 V
56 19 V
55 107 V
55 24 V
55 -50 V
55 137 V
55 -46 V
56 81 V
55 -158 V
55 175 V
55 -69 V
955 638 M
55 37 V
55 38 V
56 38 V
55 38 V
55 38 V
55 38 V
55 37 V
55 37 V
56 38 V
55 33 V
55 36 V
55 38 V
55 37 V
55 39 V
56 39 V
55 39 V
55 38 V
55 36 V
55 31 V
55 36 V
55 34 V
56 32 V
55 32 V
55 36 V
55 43 V
55 38 V
55 27 V
56 26 V
55 37 V
55 26 V
55 34 V
55 20 V
55 34 V
56 27 V
55 39 V
55 54 V
55 20 V
55 29 V
55 13 V
55 13 V
56 38 V
55 42 V
55 12 V
55 47 V
55 51 V
55 25 V
56 29 V
55 23 V
55 21 V
3713 2301 L
55 30 V
55 2 V
56 20 V
55 33 V
55 35 V
55 6 V
55 42 V
55 25 V
55 70 V
56 27 V
55 69 V
55 42 V
55 14 V
55 24 V
55 24 V
56 9 V
55 26 V
55 29 V
55 14 V
55 4 V
55 18 V
56 -4 V
55 38 V
55 16 V
55 1 V
55 34 V
55 -27 V
55 7 V
56 -7 V
55 60 V
55 21 V
55 -1 V
55 41 V
55 2 V
56 31 V
55 22 V
55 73 V
55 3 V
55 66 V
55 39 V
56 78 V
55 66 V
55 54 V
55 16 V
955 638 M
55 38 V
55 38 V
56 38 V
55 37 V
55 38 V
55 37 V
55 35 V
55 37 V
56 34 V
55 34 V
55 31 V
55 40 V
55 37 V
55 37 V
56 32 V
55 23 V
55 24 V
55 6 V
55 35 V
55 62 V
55 -4 V
56 60 V
55 21 V
55 12 V
55 64 V
55 -27 V
55 17 V
56 60 V
55 48 V
55 -59 V
55 39 V
55 91 V
55 -32 V
56 3 V
55 -30 V
55 29 V
55 -2 V
55 -23 V
55 -11 V
55 30 V
56 17 V
55 -10 V
55 12 V
55 23 V
55 -5 V
55 -36 V
56 15 V
55 13 V
55 -3 V
55 6 V
55 -15 V
55 32 V
56 -19 V
55 -10 V
55 -3 V
55 -26 V
55 33 V
55 3 V
55 -2 V
4265 1650 L
55 -5 V
55 2 V
55 -11 V
55 1 V
55 11 V
56 -10 V
55 4 V
55 0 V
55 -1 V
55 1 V
55 -1 V
56 -3 V
55 6 V
55 -6 V
55 3 V
55 -1 V
55 2 V
55 -2 V
56 1 V
55 3 V
55 10 V
55 -11 V
55 6 V
55 -10 V
56 5 V
55 0 V
55 -3 V
55 -3 V
55 4 V
55 -6 V
56 12 V
55 -11 V
55 4 V
55 8 V
stroke
LTb
900 4296 M
900 600 L
5240 0 V
0 3696 V
-5240 0 V
0.800 UP
stroke
grestore
end
showpage
  }}%
  \put(6333,1625){\makebox(0,0){\bf 45}}%
  \put(6333,3601){\makebox(0,0){\bf 35}}%
  \put(6333,3887){\makebox(0,0){\bf 30}}%
  \put(6499,3345){\makebox(0,0){\bf 25,40}}%
  \put(6333,1954){\makebox(0,0){\bf 20}}%
  \put(6333,1112){\makebox(0,0){\bf 15}}%
  \put(3520,100){\makebox(0,0){\strut{}nodes $p$}}%
  \put(200,2448){%
  \special{ps: gsave currentpoint currentpoint translate
270 rotate neg exch neg exch translate}%
  \makebox(0,0){\strut{}speedup $T(1)/T(p)$}%
  \special{ps: currentpoint grestore moveto}%
  }%
  \put(5809,400){\makebox(0,0){\strut{} 90}}%
  \put(5257,400){\makebox(0,0){\strut{} 80}}%
  \put(4706,400){\makebox(0,0){\strut{} 70}}%
  \put(4154,400){\makebox(0,0){\strut{} 60}}%
  \put(3603,400){\makebox(0,0){\strut{} 50}}%
  \put(3051,400){\makebox(0,0){\strut{} 40}}%
  \put(2500,400){\makebox(0,0){\strut{} 30}}%
  \put(1948,400){\makebox(0,0){\strut{} 20}}%
  \put(1396,400){\makebox(0,0){\strut{} 10}}%
  \put(780,4101){\makebox(0,0)[r]{\strut{} 90}}%
  \put(780,3712){\makebox(0,0)[r]{\strut{} 80}}%
  \put(780,3323){\makebox(0,0)[r]{\strut{} 70}}%
  \put(780,2934){\makebox(0,0)[r]{\strut{} 60}}%
  \put(780,2545){\makebox(0,0)[r]{\strut{} 50}}%
  \put(780,2156){\makebox(0,0)[r]{\strut{} 40}}%
  \put(780,1767){\makebox(0,0)[r]{\strut{} 30}}%
  \put(780,1378){\makebox(0,0)[r]{\strut{} 20}}%
  \put(780,989){\makebox(0,0)[r]{\strut{} 10}}%
  \put(780,600){\makebox(0,0)[r]{\strut{} 0}}%
\end{picture}%
\endgroup
 

%% file: nss5_pde_bif.tex
\begingroup%
\makeatletter%
\newcommand{\GNUPLOTspecial}{%
  \@sanitize\catcode`\%=14\relax\special}%
\setlength{\unitlength}{0.0500bp}%
\begin{picture}(7200,5040)(0,0)%
  {\GNUPLOTspecial{"
/gnudict 256 dict def
gnudict begin
%
%
/Color false def
/Blacktext true def
/Solid false def
/Dashlength 1 def
/Landscape false def
/Level1 false def
/Rounded false def
/TransparentPatterns false def
/gnulinewidth 5.000 def
/userlinewidth gnulinewidth def
/vshift -66 def
/dl1 {
  10.0 Dashlength mul mul
  Rounded { currentlinewidth 0.75 mul sub dup 0 le { pop 0.01 } if } if
} def
/dl2 {
  10.0 Dashlength mul mul
  Rounded { currentlinewidth 0.75 mul add } if
} def
/hpt_ 31.5 def
/vpt_ 31.5 def
/hpt hpt_ def
/vpt vpt_ def
Level1 {} {
/SDict 10 dict def
systemdict /pdfmark known not {
  userdict /pdfmark systemdict /cleartomark get put
} if
SDict begin [
  /Title (nss5_pde_bif.tex)
  /Subject (gnuplot plot)
  /Creator (gnuplot 4.2 patchlevel 3 )
  /Author (,,,)
  /CreationDate (Wed Dec 16 15:01:33 2009)
  /DOCINFO pdfmark
end
} ifelse
%
%
/M {moveto} bind def
/L {lineto} bind def
/R {rmoveto} bind def
/V {rlineto} bind def
/N {newpath moveto} bind def
/Z {closepath} bind def
/C {setrgbcolor} bind def
/f {rlineto fill} bind def
/vpt2 vpt 2 mul def
/hpt2 hpt 2 mul def
/Lshow {currentpoint stroke M 0 vshift R 
	Blacktext {gsave 0 setgray show grestore} {show} ifelse} def
/Rshow {currentpoint stroke M dup stringwidth pop neg vshift R
	Blacktext {gsave 0 setgray show grestore} {show} ifelse} def
/Cshow {currentpoint stroke M dup stringwidth pop -2 div vshift R 
	Blacktext {gsave 0 setgray show grestore} {show} ifelse} def
/UP {dup vpt_ mul /vpt exch def hpt_ mul /hpt exch def
  /hpt2 hpt 2 mul def /vpt2 vpt 2 mul def} def
/DL {Color {setrgbcolor Solid {pop []} if 0 setdash}
 {pop pop pop 0 setgray Solid {pop []} if 0 setdash} ifelse} def
/BL {stroke userlinewidth 2 mul setlinewidth
	Rounded {1 setlinejoin 1 setlinecap} if} def
/AL {stroke userlinewidth 2 div setlinewidth
	Rounded {1 setlinejoin 1 setlinecap} if} def
/UL {dup gnulinewidth mul /userlinewidth exch def
	dup 1 lt {pop 1} if 10 mul /udl exch def} def
/PL {stroke userlinewidth setlinewidth
	Rounded {1 setlinejoin 1 setlinecap} if} def
/LCw {1 1 1} def
/LCb {0 0 0} def
/LCa {0 0 0} def
/LC0 {1 0 0} def
/LC1 {0 1 0} def
/LC2 {0 0 1} def
/LC3 {1 0 1} def
/LC4 {0 1 1} def
/LC5 {1 1 0} def
/LC6 {0 0 0} def
/LC7 {1 0.3 0} def
/LC8 {0.5 0.5 0.5} def
/LTw {PL [] 1 setgray} def
/LTb {BL [] LCb DL} def
/LTa {AL [1 udl mul 2 udl mul] 0 setdash LCa setrgbcolor} def
/LT0 {PL [] LC0 DL} def
/LT1 {PL [4 dl1 2 dl2] LC1 DL} def
/LT2 {PL [2 dl1 3 dl2] LC2 DL} def
/LT3 {PL [1 dl1 1.5 dl2] LC3 DL} def
/LT4 {PL [6 dl1 2 dl2 1 dl1 2 dl2] LC4 DL} def
/LT5 {PL [3 dl1 3 dl2 1 dl1 3 dl2] LC5 DL} def
/LT6 {PL [2 dl1 2 dl2 2 dl1 6 dl2] LC6 DL} def
/LT7 {PL [1 dl1 2 dl2 6 dl1 2 dl2 1 dl1 2 dl2] LC7 DL} def
/LT8 {PL [2 dl1 2 dl2 2 dl1 2 dl2 2 dl1 2 dl2 2 dl1 4 dl2] LC8 DL} def
/Pnt {stroke [] 0 setdash gsave 1 setlinecap M 0 0 V stroke grestore} def
/Dia {stroke [] 0 setdash 2 copy vpt add M
  hpt neg vpt neg V hpt vpt neg V
  hpt vpt V hpt neg vpt V closepath stroke
  Pnt} def
/Pls {stroke [] 0 setdash vpt sub M 0 vpt2 V
  currentpoint stroke M
  hpt neg vpt neg R hpt2 0 V stroke
 } def
/Box {stroke [] 0 setdash 2 copy exch hpt sub exch vpt add M
  0 vpt2 neg V hpt2 0 V 0 vpt2 V
  hpt2 neg 0 V closepath stroke
  Pnt} def
/Crs {stroke [] 0 setdash exch hpt sub exch vpt add M
  hpt2 vpt2 neg V currentpoint stroke M
  hpt2 neg 0 R hpt2 vpt2 V stroke} def
/TriU {stroke [] 0 setdash 2 copy vpt 1.12 mul add M
  hpt neg vpt -1.62 mul V
  hpt 2 mul 0 V
  hpt neg vpt 1.62 mul V closepath stroke
  Pnt} def
/Star {2 copy Pls Crs} def
/BoxF {stroke [] 0 setdash exch hpt sub exch vpt add M
  0 vpt2 neg V hpt2 0 V 0 vpt2 V
  hpt2 neg 0 V closepath fill} def
/TriUF {stroke [] 0 setdash vpt 1.12 mul add M
  hpt neg vpt -1.62 mul V
  hpt 2 mul 0 V
  hpt neg vpt 1.62 mul V closepath fill} def
/TriD {stroke [] 0 setdash 2 copy vpt 1.12 mul sub M
  hpt neg vpt 1.62 mul V
  hpt 2 mul 0 V
  hpt neg vpt -1.62 mul V closepath stroke
  Pnt} def
/TriDF {stroke [] 0 setdash vpt 1.12 mul sub M
  hpt neg vpt 1.62 mul V
  hpt 2 mul 0 V
  hpt neg vpt -1.62 mul V closepath fill} def
/DiaF {stroke [] 0 setdash vpt add M
  hpt neg vpt neg V hpt vpt neg V
  hpt vpt V hpt neg vpt V closepath fill} def
/Pent {stroke [] 0 setdash 2 copy gsave
  translate 0 hpt M 4 {72 rotate 0 hpt L} repeat
  closepath stroke grestore Pnt} def
/PentF {stroke [] 0 setdash gsave
  translate 0 hpt M 4 {72 rotate 0 hpt L} repeat
  closepath fill grestore} def
/Circle {stroke [] 0 setdash 2 copy
  hpt 0 360 arc stroke Pnt} def
/CircleF {stroke [] 0 setdash hpt 0 360 arc fill} def
/C0 {BL [] 0 setdash 2 copy moveto vpt 90 450 arc} bind def
/C1 {BL [] 0 setdash 2 copy moveto
	2 copy vpt 0 90 arc closepath fill
	vpt 0 360 arc closepath} bind def
/C2 {BL [] 0 setdash 2 copy moveto
	2 copy vpt 90 180 arc closepath fill
	vpt 0 360 arc closepath} bind def
/C3 {BL [] 0 setdash 2 copy moveto
	2 copy vpt 0 180 arc closepath fill
	vpt 0 360 arc closepath} bind def
/C4 {BL [] 0 setdash 2 copy moveto
	2 copy vpt 180 270 arc closepath fill
	vpt 0 360 arc closepath} bind def
/C5 {BL [] 0 setdash 2 copy moveto
	2 copy vpt 0 90 arc
	2 copy moveto
	2 copy vpt 180 270 arc closepath fill
	vpt 0 360 arc} bind def
/C6 {BL [] 0 setdash 2 copy moveto
	2 copy vpt 90 270 arc closepath fill
	vpt 0 360 arc closepath} bind def
/C7 {BL [] 0 setdash 2 copy moveto
	2 copy vpt 0 270 arc closepath fill
	vpt 0 360 arc closepath} bind def
/C8 {BL [] 0 setdash 2 copy moveto
	2 copy vpt 270 360 arc closepath fill
	vpt 0 360 arc closepath} bind def
/C9 {BL [] 0 setdash 2 copy moveto
	2 copy vpt 270 450 arc closepath fill
	vpt 0 360 arc closepath} bind def
/C10 {BL [] 0 setdash 2 copy 2 copy moveto vpt 270 360 arc closepath fill
	2 copy moveto
	2 copy vpt 90 180 arc closepath fill
	vpt 0 360 arc closepath} bind def
/C11 {BL [] 0 setdash 2 copy moveto
	2 copy vpt 0 180 arc closepath fill
	2 copy moveto
	2 copy vpt 270 360 arc closepath fill
	vpt 0 360 arc closepath} bind def
/C12 {BL [] 0 setdash 2 copy moveto
	2 copy vpt 180 360 arc closepath fill
	vpt 0 360 arc closepath} bind def
/C13 {BL [] 0 setdash 2 copy moveto
	2 copy vpt 0 90 arc closepath fill
	2 copy moveto
	2 copy vpt 180 360 arc closepath fill
	vpt 0 360 arc closepath} bind def
/C14 {BL [] 0 setdash 2 copy moveto
	2 copy vpt 90 360 arc closepath fill
	vpt 0 360 arc} bind def
/C15 {BL [] 0 setdash 2 copy vpt 0 360 arc closepath fill
	vpt 0 360 arc closepath} bind def
/Rec {newpath 4 2 roll moveto 1 index 0 rlineto 0 exch rlineto
	neg 0 rlineto closepath} bind def
/Square {dup Rec} bind def
/Bsquare {vpt sub exch vpt sub exch vpt2 Square} bind def
/S0 {BL [] 0 setdash 2 copy moveto 0 vpt rlineto BL Bsquare} bind def
/S1 {BL [] 0 setdash 2 copy vpt Square fill Bsquare} bind def
/S2 {BL [] 0 setdash 2 copy exch vpt sub exch vpt Square fill Bsquare} bind def
/S3 {BL [] 0 setdash 2 copy exch vpt sub exch vpt2 vpt Rec fill Bsquare} bind def
/S4 {BL [] 0 setdash 2 copy exch vpt sub exch vpt sub vpt Square fill Bsquare} bind def
/S5 {BL [] 0 setdash 2 copy 2 copy vpt Square fill
	exch vpt sub exch vpt sub vpt Square fill Bsquare} bind def
/S6 {BL [] 0 setdash 2 copy exch vpt sub exch vpt sub vpt vpt2 Rec fill Bsquare} bind def
/S7 {BL [] 0 setdash 2 copy exch vpt sub exch vpt sub vpt vpt2 Rec fill
	2 copy vpt Square fill Bsquare} bind def
/S8 {BL [] 0 setdash 2 copy vpt sub vpt Square fill Bsquare} bind def
/S9 {BL [] 0 setdash 2 copy vpt sub vpt vpt2 Rec fill Bsquare} bind def
/S10 {BL [] 0 setdash 2 copy vpt sub vpt Square fill 2 copy exch vpt sub exch vpt Square fill
	Bsquare} bind def
/S11 {BL [] 0 setdash 2 copy vpt sub vpt Square fill 2 copy exch vpt sub exch vpt2 vpt Rec fill
	Bsquare} bind def
/S12 {BL [] 0 setdash 2 copy exch vpt sub exch vpt sub vpt2 vpt Rec fill Bsquare} bind def
/S13 {BL [] 0 setdash 2 copy exch vpt sub exch vpt sub vpt2 vpt Rec fill
	2 copy vpt Square fill Bsquare} bind def
/S14 {BL [] 0 setdash 2 copy exch vpt sub exch vpt sub vpt2 vpt Rec fill
	2 copy exch vpt sub exch vpt Square fill Bsquare} bind def
/S15 {BL [] 0 setdash 2 copy Bsquare fill Bsquare} bind def
/D0 {gsave translate 45 rotate 0 0 S0 stroke grestore} bind def
/D1 {gsave translate 45 rotate 0 0 S1 stroke grestore} bind def
/D2 {gsave translate 45 rotate 0 0 S2 stroke grestore} bind def
/D3 {gsave translate 45 rotate 0 0 S3 stroke grestore} bind def
/D4 {gsave translate 45 rotate 0 0 S4 stroke grestore} bind def
/D5 {gsave translate 45 rotate 0 0 S5 stroke grestore} bind def
/D6 {gsave translate 45 rotate 0 0 S6 stroke grestore} bind def
/D7 {gsave translate 45 rotate 0 0 S7 stroke grestore} bind def
/D8 {gsave translate 45 rotate 0 0 S8 stroke grestore} bind def
/D9 {gsave translate 45 rotate 0 0 S9 stroke grestore} bind def
/D10 {gsave translate 45 rotate 0 0 S10 stroke grestore} bind def
/D11 {gsave translate 45 rotate 0 0 S11 stroke grestore} bind def
/D12 {gsave translate 45 rotate 0 0 S12 stroke grestore} bind def
/D13 {gsave translate 45 rotate 0 0 S13 stroke grestore} bind def
/D14 {gsave translate 45 rotate 0 0 S14 stroke grestore} bind def
/D15 {gsave translate 45 rotate 0 0 S15 stroke grestore} bind def
/DiaE {stroke [] 0 setdash vpt add M
  hpt neg vpt neg V hpt vpt neg V
  hpt vpt V hpt neg vpt V closepath stroke} def
/BoxE {stroke [] 0 setdash exch hpt sub exch vpt add M
  0 vpt2 neg V hpt2 0 V 0 vpt2 V
  hpt2 neg 0 V closepath stroke} def
/TriUE {stroke [] 0 setdash vpt 1.12 mul add M
  hpt neg vpt -1.62 mul V
  hpt 2 mul 0 V
  hpt neg vpt 1.62 mul V closepath stroke} def
/TriDE {stroke [] 0 setdash vpt 1.12 mul sub M
  hpt neg vpt 1.62 mul V
  hpt 2 mul 0 V
  hpt neg vpt -1.62 mul V closepath stroke} def
/PentE {stroke [] 0 setdash gsave
  translate 0 hpt M 4 {72 rotate 0 hpt L} repeat
  closepath stroke grestore} def
/CircE {stroke [] 0 setdash 
  hpt 0 360 arc stroke} def
/Opaque {gsave closepath 1 setgray fill grestore 0 setgray closepath} def
/DiaW {stroke [] 0 setdash vpt add M
  hpt neg vpt neg V hpt vpt neg V
  hpt vpt V hpt neg vpt V Opaque stroke} def
/BoxW {stroke [] 0 setdash exch hpt sub exch vpt add M
  0 vpt2 neg V hpt2 0 V 0 vpt2 V
  hpt2 neg 0 V Opaque stroke} def
/TriUW {stroke [] 0 setdash vpt 1.12 mul add M
  hpt neg vpt -1.62 mul V
  hpt 2 mul 0 V
  hpt neg vpt 1.62 mul V Opaque stroke} def
/TriDW {stroke [] 0 setdash vpt 1.12 mul sub M
  hpt neg vpt 1.62 mul V
  hpt 2 mul 0 V
  hpt neg vpt -1.62 mul V Opaque stroke} def
/PentW {stroke [] 0 setdash gsave
  translate 0 hpt M 4 {72 rotate 0 hpt L} repeat
  Opaque stroke grestore} def
/CircW {stroke [] 0 setdash 
  hpt 0 360 arc Opaque stroke} def
/BoxFill {gsave Rec 1 setgray fill grestore} def
/Density {
  /Fillden exch def
  currentrgbcolor
  /ColB exch def /ColG exch def /ColR exch def
  /ColR ColR Fillden mul Fillden sub 1 add def
  /ColG ColG Fillden mul Fillden sub 1 add def
  /ColB ColB Fillden mul Fillden sub 1 add def
  ColR ColG ColB setrgbcolor} def
/BoxColFill {gsave Rec PolyFill} def
/PolyFill {gsave Density fill grestore grestore} def
/h {rlineto rlineto rlineto gsave closepath fill grestore} bind def
%
%
/PatternFill {gsave /PFa [ 9 2 roll ] def
  PFa 0 get PFa 2 get 2 div add PFa 1 get PFa 3 get 2 div add translate
  PFa 2 get -2 div PFa 3 get -2 div PFa 2 get PFa 3 get Rec
  gsave 1 setgray fill grestore clip
  currentlinewidth 0.5 mul setlinewidth
  /PFs PFa 2 get dup mul PFa 3 get dup mul add sqrt def
  0 0 M PFa 5 get rotate PFs -2 div dup translate
  0 1 PFs PFa 4 get div 1 add floor cvi
	{PFa 4 get mul 0 M 0 PFs V} for
  0 PFa 6 get ne {
	0 1 PFs PFa 4 get div 1 add floor cvi
	{PFa 4 get mul 0 2 1 roll M PFs 0 V} for
 } if
  stroke grestore} def
/languagelevel where
 {pop languagelevel} {1} ifelse
 2 lt
	{/InterpretLevel1 true def}
	{/InterpretLevel1 Level1 def}
 ifelse
%
%
/Level2PatternFill {
/Tile8x8 {/PaintType 2 /PatternType 1 /TilingType 1 /BBox [0 0 8 8] /XStep 8 /YStep 8}
	bind def
/KeepColor {currentrgbcolor [/Pattern /DeviceRGB] setcolorspace} bind def
<< Tile8x8
 /PaintProc {0.5 setlinewidth pop 0 0 M 8 8 L 0 8 M 8 0 L stroke} 
>> matrix makepattern
/Pat1 exch def
<< Tile8x8
 /PaintProc {0.5 setlinewidth pop 0 0 M 8 8 L 0 8 M 8 0 L stroke
	0 4 M 4 8 L 8 4 L 4 0 L 0 4 L stroke}
>> matrix makepattern
/Pat2 exch def
<< Tile8x8
 /PaintProc {0.5 setlinewidth pop 0 0 M 0 8 L
	8 8 L 8 0 L 0 0 L fill}
>> matrix makepattern
/Pat3 exch def
<< Tile8x8
 /PaintProc {0.5 setlinewidth pop -4 8 M 8 -4 L
	0 12 M 12 0 L stroke}
>> matrix makepattern
/Pat4 exch def
<< Tile8x8
 /PaintProc {0.5 setlinewidth pop -4 0 M 8 12 L
	0 -4 M 12 8 L stroke}
>> matrix makepattern
/Pat5 exch def
<< Tile8x8
 /PaintProc {0.5 setlinewidth pop -2 8 M 4 -4 L
	0 12 M 8 -4 L 4 12 M 10 0 L stroke}
>> matrix makepattern
/Pat6 exch def
<< Tile8x8
 /PaintProc {0.5 setlinewidth pop -2 0 M 4 12 L
	0 -4 M 8 12 L 4 -4 M 10 8 L stroke}
>> matrix makepattern
/Pat7 exch def
<< Tile8x8
 /PaintProc {0.5 setlinewidth pop 8 -2 M -4 4 L
	12 0 M -4 8 L 12 4 M 0 10 L stroke}
>> matrix makepattern
/Pat8 exch def
<< Tile8x8
 /PaintProc {0.5 setlinewidth pop 0 -2 M 12 4 L
	-4 0 M 12 8 L -4 4 M 8 10 L stroke}
>> matrix makepattern
/Pat9 exch def
/Pattern1 {PatternBgnd KeepColor Pat1 setpattern} bind def
/Pattern2 {PatternBgnd KeepColor Pat2 setpattern} bind def
/Pattern3 {PatternBgnd KeepColor Pat3 setpattern} bind def
/Pattern4 {PatternBgnd KeepColor Landscape {Pat5} {Pat4} ifelse setpattern} bind def
/Pattern5 {PatternBgnd KeepColor Landscape {Pat4} {Pat5} ifelse setpattern} bind def
/Pattern6 {PatternBgnd KeepColor Landscape {Pat9} {Pat6} ifelse setpattern} bind def
/Pattern7 {PatternBgnd KeepColor Landscape {Pat8} {Pat7} ifelse setpattern} bind def
} def
%
%
%
/PatternBgnd {
  TransparentPatterns {} {gsave 1 setgray fill grestore} ifelse
} def
%
%
/Level1PatternFill {
/Pattern1 {0.250 Density} bind def
/Pattern2 {0.500 Density} bind def
/Pattern3 {0.750 Density} bind def
/Pattern4 {0.125 Density} bind def
/Pattern5 {0.375 Density} bind def
/Pattern6 {0.625 Density} bind def
/Pattern7 {0.875 Density} bind def
} def
%
%
Level1 {Level1PatternFill} {Level2PatternFill} ifelse
/Symbol-Oblique /Symbol findfont [1 0 .167 1 0 0] makefont
dup length dict begin {1 index /FID eq {pop pop} {def} ifelse} forall
currentdict end definefont pop
end
gnudict begin
gsave
0 0 translate
0.050 0.050 scale
0 setgray
newpath
1.000 UL
LTb
600 663 M
0 -63 V
0 4200 R
0 63 V
1213 663 M
0 -63 V
0 4200 R
0 63 V
1826 663 M
0 -63 V
0 4200 R
0 63 V
2439 663 M
0 -63 V
0 4200 R
0 63 V
3052 663 M
0 -63 V
0 4200 R
0 63 V
3665 663 M
0 -63 V
0 4200 R
0 63 V
4278 663 M
0 -63 V
0 4200 R
0 63 V
4891 663 M
0 -63 V
0 4200 R
0 63 V
5504 663 M
0 -63 V
0 4200 R
0 63 V
6117 663 M
0 -63 V
0 4200 R
0 63 V
0 -4200 R
63 0 V
-63 827 R
63 0 V
-63 828 R
63 0 V
-63 827 R
63 0 V
-63 828 R
63 0 V
-63 827 R
63 0 V
stroke
600 4800 N
600 663 L
5517 0 V
0 4137 V
-5517 0 V
Z stroke
LCb setrgbcolor
LTb
LCb setrgbcolor
LTb
1.000 UP
1.000 UL
LTb
1.000 UL
LT0
1826 1490 M
-1 -8 V
-1 -8 V
-5 -13 V
-12 -19 V
-23 -24 V
-39 -26 V
-53 -26 V
-67 -25 V
-82 -23 V
-97 -20 V
-114 -18 V
-132 -14 V
-152 -11 V
-170 -6 V
-188 -1 V
-90 1 V
4904 241 R
-1 28 V
-1 27 V
-5 44 V
-12 66 V
-23 83 V
-39 94 V
-53 98 V
-69 99 V
-86 101 V
-105 103 V
-130 108 V
-161 113 V
-198 118 V
-245 124 V
-305 128 V
-317 111 V
-326 95 V
-332 82 V
-338 70 V
-341 59 V
-86 13 V
-260 35 V
-180 21 V
-168 18 V
-351 30 V
-352 22 V
-355 15 V
-65 2 V
3665 1490 M
-1 21 V
-1 21 V
-6 34 V
-12 47 V
-24 57 V
-36 63 V
-50 64 V
-62 64 V
-77 65 V
-94 65 V
-114 68 V
-139 69 V
-169 70 V
-205 70 V
-250 70 V
-301 67 V
-334 56 V
-340 42 V
-346 30 V
-350 20 V
-154 5 V
1291 752 R
-6 24 V
-17 24 V
-13 14 V
-12 11 V
-14 11 V
-19 15 V
-27 18 V
-38 24 V
-53 31 V
-75 40 V
-108 52 V
-158 71 V
-229 90 V
799 3846 L
-199 61 V
637 1490 M
37 0 V
36 0 V
74 0 V
147 0 V
294 0 V
368 0 V
233 0 V
135 0 V
368 0 V
367 0 V
368 0 V
368 0 V
233 0 V
135 0 V
368 0 V
367 0 V
368 0 V
368 0 V
233 0 V
135 0 V
368 0 V
110 0 V
-2452 0 R
-1 -14 V
-1 -14 V
-5 -23 V
-12 -34 V
-23 -42 V
-39 -47 V
-53 -48 V
-68 -46 V
-84 -45 V
-101 -42 V
-122 -39 V
-146 -35 V
-172 -31 V
-203 -23 V
2398 993 L
-273 -3 V
-313 8 V
-339 19 V
-345 25 V
-349 30 V
-179 15 V
2331 3254 M
4 23 V
13 21 V
12 14 V
12 11 V
14 11 V
20 15 V
29 20 V
42 27 V
64 41 V
90 63 V
95 83 V
59 84 V
19 63 V
1 18 V
-2 26 V
-9 33 V
-15 30 V
-19 29 V
-25 29 V
-32 29 V
-39 30 V
-47 30 V
-56 31 V
-68 32 V
-82 34 V
-100 35 V
-122 38 V
-151 40 V
-188 43 V
-237 45 V
-302 48 V
-350 45 V
-354 36 V
-7 0 V
stroke
LTb
600 4800 N
600 663 L
5517 0 V
0 4137 V
-5517 0 V
Z stroke
1.000 UP
1.000 UL
LTb
stroke
grestore
end
showpage
  }}%
  \put(508,4386){\makebox(0,0)[r]{\strut{}$b_6$}}%
  \put(508,3890){\makebox(0,0)[r]{\strut{}$b_7$}}%
  \put(508,3393){\makebox(0,0)[r]{\strut{}$b_5$}}%
  \put(508,2566){\makebox(0,0)[r]{\strut{}$b_4$}}%
  \put(508,994){\makebox(0,0)[r]{\strut{}$b_3$}}%
  \put(508,1242){\makebox(0,0)[r]{\strut{}$b_2$}}%
  \put(1887,3145){\makebox(0,0)[r]{\strut{}$P_5$}}%
  \put(2408,3062){\makebox(0,0)[r]{\strut{}$P_4$}}%
  \put(5535,1325){\makebox(0,0)[r]{\strut{}$P_3$}}%
  \put(3910,1656){\makebox(0,0)[r]{\strut{}$P_2$}}%
  \put(1841,1656){\makebox(0,0)[r]{\strut{}$P_1$}}%
  \put(508,1573){\makebox(0,0)[r]{\strut{}$b_1$}}%
  \put(3358,100){\makebox(0,0){\strut{}$s$}}%
  \put(6999,2731){%
  \special{ps: gsave currentpoint currentpoint translate
270 rotate neg exch neg exch translate}%
  \makebox(0,0){\strut{}$u(x^*,y^*)$}%
  \special{ps: currentpoint grestore moveto}%
  }%
  \put(6300,4800){\makebox(0,0)[l]{\strut{} 2}}%
  \put(6300,3973){\makebox(0,0)[l]{\strut{} 1.5}}%
  \put(6300,3145){\makebox(0,0)[l]{\strut{} 1}}%
  \put(6300,2318){\makebox(0,0)[l]{\strut{} 0.5}}%
  \put(6300,1490){\makebox(0,0)[l]{\strut{} 0}}%
  \put(6300,663){\makebox(0,0)[l]{\strut{}-0.5}}%
  \put(6117,400){\makebox(0,0){\strut{} 9}}%
  \put(5504,400){\makebox(0,0){\strut{} 8}}%
  \put(4891,400){\makebox(0,0){\strut{} 7}}%
  \put(4278,400){\makebox(0,0){\strut{} 6}}%
  \put(3665,400){\makebox(0,0){\strut{} 5}}%
  \put(3052,400){\makebox(0,0){\strut{} 4}}%
  \put(2439,400){\makebox(0,0){\strut{} 3}}%
  \put(1826,400){\makebox(0,0){\strut{} 2}}%
  \put(1213,400){\makebox(0,0){\strut{} 1}}%
  \put(600,400){\makebox(0,0){\strut{} 0}}%
\end{picture}%
\endgroup
 

%% file: workers.tex
\begin{algorithm2e}[h]
\dontprintsemicolon
\linesnumbered
\hrule
\medskip
\SetKwFunction{KwPB}{pushBack}
\SetKwFunction{KwInsert}{insert}
\SetKwFunction{KwRemove}{remove}
\SetKw{KwBreak}{break}
\SetKw{KwWhile}{while}
\SetKwFor{Rep}{repeat}{}{}
\SetKwSwitch{Switch}{Case}{Other}{switch}{}{case}{otherwise}{}

\KwIn{none}
\KwOut{none}
\smallskip
\hrule
\smallskip
\Rep{}{
  wait for a message from the boss \;  
    \Switch{the message is a}{
    \Case{data share request}{
      accept the broadcast message from the boss \;
      call MPQswitch to handle the broadcast message \;
    }
    \Case{job}{
      call MPQswitch to run the job \;
      send the result back to the boss \;
    }
    \Case{stop command}{
       stop \;
    }    
    }
}
\medskip
\hrule
\caption{\label{alg2} Pseudo code for the main loop of the workers. The workers get to this loop from the 
{\sf MPQstart} function.}

\end{algorithm2e}

%% file: runjobs.tex
\incmargin{-1em}
\begin{algorithm2e}[h]
\dontprintsemicolon
\linesnumbered
\hrule
\medskip
\SetKwFunction{KwPB}{pushBack}
\SetKwFunction{KwInsert}{insert}
\SetKwFunction{KwRemove}{remove}
\SetKw{KwBreak}{break}
\SetKw{KwWhile}{while}
\SetKwSwitch{Switch}{Case}{Other}{switch}{}{case}{otherwise}{}


\KwIn{input queue}
\KwOut{output queue}
\smallskip
\hrule
\smallskip
\While {input queue is not empty or workers are working}{ 
  \eIf{input queue is not empty and there are idle workers}{
    assign a job to a worker \;
    remove the job from the input queue \;
  }{
    wait for a message from a worker \;
    \Switch{the message is a}{
    \Case{job submission}{
      add the job to the input queue \;
    }
    \Case{job result}{
      \If{result is not empty}{add the result to the output queue \;}
    }
    \Case{task}{
      call MPQswitch to run the task \;
      send the result back to the worker \;
    }
    \Case{info request}{
      send the info back to the worker \;
    }    
    }
  }
}
\medskip
\hrule
\caption{\label{alg1} Pseudo code for the {\sf MPQrunjobs} function. This function is run by the boss.}

\end{algorithm2e}

%% file: MPQueueCommandsTable.tex
\scalebox{1}{
\begin{tabular}{|l|l|}
\hline 
\lstinline!struct Tjob \{int type; string data; \};! & $\vphantom{\int_{\int}^{\int}}$Define job data type.\tabularnewline
\hline 
\lstinline!typedef queue <Tjob> Tjobqueue;! & $\vphantom{\int_{\int}^{\int}}$Define jobqueue data type.\tabularnewline
\hline 
\lstinline!void MPQinit(int argc, char {*}argv[ ]);! & $\vphantom{\int_{\int}^{\int}}$Initialize MPI.\tabularnewline
\hline 
\lstinline!void MPQstart();! & $\vphantom{\int_{\int}^{\int}}$Split the workers from the boss process.\tabularnewline
\hline 
\lstinline!void inline MPQsubmit(const Tjob \& job);! & $\vphantom{\int_{\int}^{\int}}$Submit a job into the currently running
job queue.\tabularnewline
\hline 
\lstinline!template <class T> void ! & $\vphantom{\int^{\int}}$Ask the boss to run a task.\tabularnewline
\lstinline!inline MPQtask(Tjob \& job);! & $\vphantom{\int_{\int}}$\tabularnewline
\hline 
\lstinline!void MPQinfo! & $\vphantom{\int^{\int}}$Get the number of jobs in the running job
queue \tabularnewline
\lstinline{(int \& queuesize, int \& sent);} & $\vphantom{\int_{\int}}$and the number of workers currently working.\tabularnewline
\hline 
\lstinline!void MPQrunjobs ! & $\vphantom{\int^{\int}}$Assign the jobs in \textsl{inq} to the workers
and collect \tabularnewline
\lstinline!(Tjobqueue \& inq, Tjobqueue \& outq);! & $\vphantom{\int_{\int}}$the results in \textsl{outq}.\tabularnewline
\hline 
\lstinline!void inline MPQsharedata(const Tjob \& job);! & $\vphantom{\int_{\int}^{\int}}$Send data to all the workers.\tabularnewline
\hline 
\lstinline!void MPQswitch(Tjob \& job);! & $\vphantom{\int_{\int}^{\int}}$Execute a job.\tabularnewline
\hline 
\lstinline!void MPQstop();! & $\vphantom{\int_{\int}^{\int}}$Release the workers and stop MPI gracefully.\tabularnewline
\hline 
\lstinline!template <class T> string! & $\vphantom{\int^{\int}}$Serialize any variable.\tabularnewline
\lstinline!to\_string(const T \& in);! & $\vphantom{\int_{\int}}$\tabularnewline
\hline 
\lstinline!template <class T> void! & $\vphantom{\int^{\int}}$Deserialize a variable.\tabularnewline
\lstinline!from\_string(T \& out, const string \& str);! & $\vphantom{\int_{\int}}$\tabularnewline
\hline 
\lstinline!\#define LOAD\_DIAGRAM! & $\vphantom{\int_{\int}^{\int}}$Generate load data.\tabularnewline
\hline 
\end{tabular}
}